\documentclass[onecolumn,a4]{report}
\usepackage{pst-all}
\usepackage{epsfig}
\usepackage{psfig}
\usepackage{setspace}
\usepackage{makeidx}
\usepackage{multind}
\newcommand{\be}{\begin{equation}}
\newcommand{\ee}{\end{equation}}
\renewcommand{\=}{~=~}
\makeindex{subject}

\begin{document}

\title{\bf \LARGE Structure studies of conventional and
novel excitations to the continuum in reactions 
with unstable nuclei}
\author{\bf Lorenzo Fortunato \\ ~\\ ~\\
XVI ciclo - Corso di Dottorato in Fisica\\~\\~\\
Universita' degli studi di Padova \\ ~\\
Dipartimento di fisica "Galileo Galilei"  \\~\\~\\
Supervisor:  Prof. A.Vitturi\\}

\begin{figure}
\begin{center}
\epsfig{file=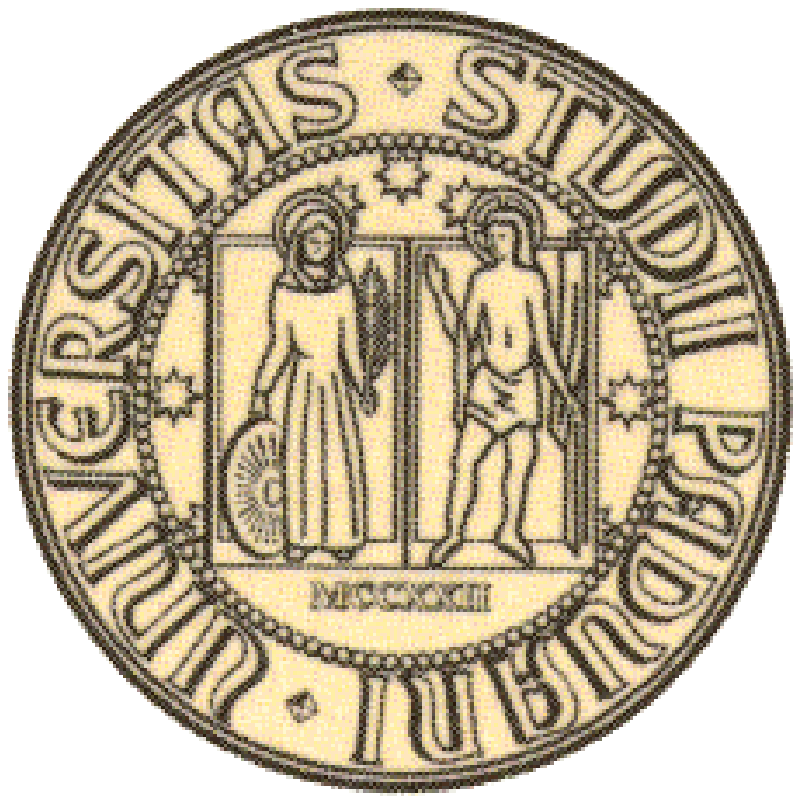,width=2.5cm}
~~~~~~~~~~~~~~~~~~~~~~~~~~~~~~~~~~~~~~~~~~~~~~~~~~~~~~~~
\epsfig{file=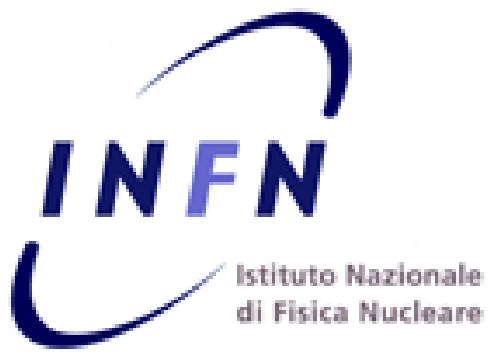,width=3cm}
\end{center}
\end{figure}

\maketitle

\tableofcontents

\newpage
~\\~\\~\\
~\\~\\~\\
~\\~\\~\\
~\\~\\~\\
~\\~\\~\\
\begin{flushright}
{\it To the 'bdot' subroutine \\
(or D02PCF \& D02PVF) \\
and their inventors,\\
with gratefulness and esteem,}
~\\~\\~\\
{\it and also to the Amontillado.}
\end{flushright}

\newpage


~~
\pagebreak

\chapter{Preface}

\begin{quote}
He had a weak point -- this Fortunato -- although in other regards 
he was a man to be respected and even feared. He prided himself on his 
connoisseurship in wine. Few Italians have the true virtuoso spirit. For the 
most part their enthusiasm is adopted to suit the time and opportunity to 
practise imposture upon the British and Austrian MILLIONAIRES. In painting 
and gemmary, Fortunato, like his countrymen , was a quack, but in the matter 
of old wines he was sincere. In this respect I did not differ from him 
materially; I was skilful in the Italian vintages myself, and bought largely 
whenever I could.\\ - {\it The cask of Amontillado 
\index{subject}{amontillado}} - Edgar Allan Poe.
\end{quote}  

\setstretch{1.33}

\section{Introduction and motivations}

We are now living a new era of development of nuclear sciences that has 
expanded  recently thanks to the discovery of the so-called exotic nuclei.
These nuclei lie far from the stability valley, and are also named unstable 
nuclei in contrast 
to the stable ones, which nuclear scientists were used to deal with in the 
past. The Odyssey to reach  the limits of stability, is pushing nuclear 
sciences to new and unexpected discoveries and to a redefinition of its 
scope and methods.  At a variance with other revolutions happened suddenly 
in science, this is a quite slow and step-by-step revolution, that is both 
experimentally and theoretically a scientific challenge and an exciting 
cultural  progress. The description of such systems requires a 
reconsideration of the role of the continuum that increases its importance 
while  moving toward the drip-lines.
In fact the closer a nucleus sits to the drip-lines, the weaker is the binding
energy: in this way only little room is left in the discrete part of the
spectrum for other bound states. Typically if we move from the valley of
stability outward we encounter the situation in which the bound excited states
of the stable nucleus are shifted to the continuum in unstable isotopes, 
forming low-lying resonances. 
The coupling to these states becomes of fundamental
importance for the description of reactions in which weakly-bound nuclei are
involved. At the same time the coupling to non-resonant continuum states 
changes its role becoming more relevant in these systems.

The continuous part of the spectrum also displays other interesting
modifications in exotic nuclei: it is worth mentioning the effect of the
presence of a neutron skin on the excitation of conventional modes 
(as the giant dipole resonance). 

The main aim or {\it file rouge} of the present thesis is to try 
to link many 
different aspects of the complexity of the continuum spectrum in nuclear 
structure and in nuclear reactions  involving both stable and unstable 
nuclei either as a 
as a tool or as the subject of our study. Not only we would like to 
study the continuum in exotic nuclei, but we would also like to tackle the 
issue of  ``exotic continua'', that is to say to address ourselves to a  
number of problems 
regarding the excitation of unusual modes  in stable nuclei, as we will 
mention in the following.
This thesis is organized in four main chapters, each one discussing a 
particular issue.

The first chapter deals with a detailed study of the excitation of double 
phonon giant resonances \index{subject}{double giant resonance}
in stable nuclei. The experimental evidence
for the so-called giant modes in the continuum dates back to the thirties
and spurred the first attempts to describe them theoretically. From the 
other side theorists had come to a precise formulation of the problem in terms
of excitation of phonons in finite quantum systems, thus demanding 
the discovery of two-phonons or many-phonons excitations. 
They were found experimentally (in particular in the case of double 
phonon giant dipole and quadrupole resonances) and
we had developed a method and a computer code to study in a simple way the
dependence on various parameters of the excitation of these modes. 

The following chapter is concerned with the excitation of giant 
pairing vibrations \index{subject}{Giant Pairing Vibration}
in stable nuclei, but excited with an unstable beam. 
Low-lying pairing states are well-known, but the giant pairing mode, that 
was predicted to lie at higher excitation energy, has never been found 
experimentally. Two neutrons transfer (t,p) reactions were studied in 
the past to look for
this excitation, but the results have not been published.
After a structure (RPA or BCS+RPA) study of 
the monopole states in the continuum of the targets, we perform a 
calculation (based on the coupled channel formalism) for two neutron 
transfer reactions with both a conventional ( $^{14}$C) 
\index{subject}{$^{14}$C} and an unconventional beam 
( $^6$He)\index{subject}{$^6$He}, showing that in the latter case 
the excitation of the giant pairing mode is enhanced as a consequence 
of the optimal Q-value that comes 
from the weakly-bound nature of  unstable projectiles.
The main conclusion of this chapter is that an effort should
be undertaken in order to repeat the already tried experiments with
weakly-bound projectiles to identify the giant pairing vibration.

In the third chapter we introduce a new model to study the effects of the 
diffuseness of the nuclear surface and of the presence of the skin on the 
excitation of isovector giant dipole modes in both stable and unstable nuclei. 
Surprisingly, being the model a modification of the Steinwedel-Jensen model, 
we find a strong modification of the predictions for the energy of giant 
dipole resonance, due to the presence of the diffuseness. This is true 
already at the level of nuclei that lie in the stability valley and becomes 
even more effective at the drip-lines.

The last part deals with the problem of the non-resonant nature of the 
low-lying continuum in dicluster nuclei. In the light-mass region of the 
nuclear chart, many nuclei are believed to possess a cluster structure and
a clear distinctions between stability valley and drip-lines is difficult.
In this region many stable nuclei display binding energies comparable to 
drip-line isotopes (though the half-life is a way to make a distinction).
This is 
particularly true for $^6$Li and $^7$Li, \index{subject}{$^6$Li}
\index{subject}{$^7$Li}
respectively formed by an alpha particle plus a deuton or a triton.
Within a simple cluster model, where cluster and core interact via 
an effective Coulomb plus Woods-Saxon nuclear potential, we compute 
many properties of the system under study.  The 
weakly-bound nature of these isotopes is responsible for an enhanced 
excitation of states in the low-lying continuum that confirms older similar 
results obtained in halo nuclei. The contribution of the non-resonant 
continuum is found to mix with the excitation of resonant states (in some 
case being stronger and washing out the lineshape of many states).
This line of research is directly connected with the experimental work on 
breakup reactions of light weakly bound nuclei, that is going on
 at the Laboratori Nazionali di Legnaro. In particular we will
address some comments on the breakup of $^6$Li at the end of the last chapter.

This thesis contains extracts form previously published works in which
the author had participated as well as parts that we intend to publish in 
the future. The list of works, conference proceedings and preprints in 
which the author has been involved and has a connection with the
present thesis is given at the end of the bibliography.

\vspace{0.5cm}
\section{Acknowledgments}
I would like to gratefully thank all the people that helped me during
my Ph.D. in Padova: my supervisor Andrea Vitturi, who shared with me his 
invaluable experience with an almost monastic patience, and all the 
researchers and professors that gave me precious suggestions and teachings 
directly related to the present thesis (as C.H.Dasso, K.Hagino, 
E.G.Lanza, S.Lenzi, C.Signorini, H.M.Sofia, W.von Oertzen, F.Zardi)
or related to other studies that I began during the course of the Ph.D. (as
J.M.Arias Carrasco, F.Iachello and D.J.Rowe).
I would also like to express my gratitude to my Ph.D. colleagues for their help
and for the friendly atmosphere that we have kept. In particular I would like 
to mention, among the other colleagues from which I got a great deal of
informations, suggestions and helps,
M.Mazzocco (for the work on breakup reactions, directly related to 
this thesis) and A.Torrielli (with whom I worked hard to a common project
and from whom I learned a lot of useful stuff). I would also mention
S.Montagnani (for the common interests and exchange of informations).\\
I acknowledge financial support for my research activity 
during this three years from the university of Padova and INFN and the support
for a three-months visit obtained from the University of Sevilla (Spain),
\index{subject}{Sevilla} 
where the first few pages of this thesis were written.

\vspace{0.5cm}~~
\begin{flushleft}
Padova, Italy.\\
Oct. 2003
\end{flushleft}
\begin{flushright}
{\bf Lorenzo Fortunato}
\end{flushright}


\chapter{Double Giant Resonances}

\index{subject}{double giant resonance}

\begin{flushright}
\begin{picture}(70,80)(50,50)
\psset{unit=1.4pt}      
{\red  
\rput(12,60){\small 2ph}
\psline[linecolor=red]{-}(25,60)(55,60)
\rput(70,62){\small GQR$\otimes$GQR}}
\rput(12,40){\small 1ph}
\psline{-}(25,40)(55,40)
\rput(65,42){\small GQR}
\rput(12,20){\small gr}
\psline{-}(25,20)(55,20)
\psline{->}(40,40)(40,60)
\psline{->}(40,20)(40,40)
\end{picture}
\end{flushright}

\section{Introduction.}
Giant Resonances (GR) are considered as one of the most important
elementary modes of excitation in nuclei and have been
studied for more that 50 years: they represented a major discovery
in nuclear sciences and they still stimulate theoretical as well as
experimental developments~\cite{harak}. 

Giant Dipole Resonances \index{subject}{giant dipole resonance}
in nuclei were observed by Bothe and Gentner in 1937
\cite{Bothe}. They observed a broad peak in the spectra of (p,$\gamma$)
reactions around 17 MeV. Subsequently a more systematic investigation of
the energy region between 10 and 25 MeV \cite{Bald} for a larger number of
isotopes was done. A schematic representation of the photoabsorption 
cross-section of a nucleus may be divided in three region: a low-lying region 
under the threshold for nucleon emission where a number of discrete states are
present, a threshold region where states with a non-zero width start to appear
and overlap, and a higher-lying energy region where a broad and huge peak,
followed by lower ones is the major feature of the nucleus. This broad peak
is called Giant Dipole Resonance (GDR) and it has been shown that is a common
feature of nuclei, being almost always present in the photoabsorption spectrum
across the whole table of nuclei (with the exception of the smallest isotopes 
for which it is difficult to identify a GDR).

This giant mode has always a large width (4-7 MeV), being smaller for closed 
shell nuclei and larger for mid-shell isotopes and the integral cross section
exhausts almost completely the Thomas-Reiche-Kuhn (TRK) sum rule.

The first theoretical interpretation was published in 
1948 by Goldhaber and Teller \cite{GT} for the isovector giant dipole 
resonance  in terms of a model in which a rigid proton sphere 
oscillates against a rigid neutron sphere. Some years later Steinwedel and
Jensen \cite{SJ} considered a model in which proton and neutron fluids were 
allowed to oscillate out of phase within a rigid sphere (See chapter 3).

\begin{figure}[!h]
\begin{center}
\begin{picture}(70,80)(50,50)
\psset{unit=2.pt}        
\rput(12,60){\small 2ph}
\psline{-}(25,60)(55,60)
\rput(65,62){\small GQR$\otimes$GQR}
\rput(12,40){\small 1ph}
\psline{-}(25,40)(55,40)
\rput(65,42){\small GQR}
\rput(12,20){\small gr}
\psline{-}(25,20)(55,20)
\psline{->}(40,40)(40,60)
\psline{->}(40,20)(40,40)
\psline{->}(-20,20)(-20,60)
\rput(-25,40){\rotateleft{energy}}
\end{picture}
\end{center}    
\caption{Schematic picture aimed at illustrating how a double giant quadrupole
resonance is built.}
\label{pict}
\end{figure}
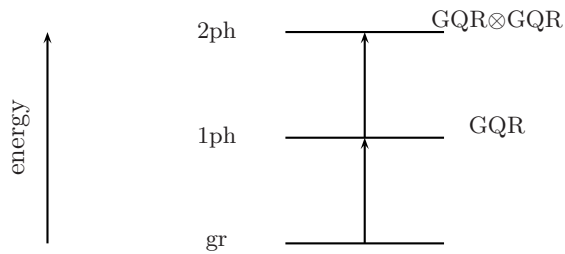

The isoscalar quadrupole resonance \index{subject}{giant quadrupole resonance}
was found (in 1971) 
in the inelastic electron scattering \cite{Pitth} as well as in proton
scattering experiments \cite{Lewis}. Other modes have been identified as the
Giant Monopole Resonance, \index{subject}{giant monopole resonance}
 the Gamow-Teller resonances \index{subject}{Gamow-Teller resonance}, etc. 
that we will not discuss.
They are generally interpreted as harmonic density vibrations of the 
quantum fluids around the equilibrium distribution of the nucleons 
\cite{BroBer}. Within this point of view
one should also expect to observe higher-lying states of the harmonic
spectrum such as, for instance, the two-phonon Double Giant Resonance
(DGR). Double and, in general multiphonon resonances, are seen as excitations
of the second, or higher, phonon on top of the excitation of the first one.
This idea is illustrated in figure \ref{pict} where a double giant quadrupole
resonance is built upon the single-phonon GQR.

\index{subject}{DGQR}

The excitation of higher-lying states is not only limited to phonon of the
same kind, but also the excitation of phonons of different multipolarities
built on top of excited state (both bound and unbound) has been observed 
and discussed.
The existence of the double-phonon excitation of low-lying
collective vibrational states has been known for a long time, but only 
recently the multiple excitations of GR have been
systematically observed (for a complete review see ref.\cite{rev,Chom} or
\cite{harak} and references therein).
The first example of a giant dipole resonance built on a $2^+$ excited state 
has been seen in a (p,$\gamma$) reaction on $^{11}$Be \cite{Ko79}.
In this case however the giant mode is built on a discrete excited state. We 
will instead be concerned, in the following, with giant modes built on 
other giant modes.

\begin{figure}[!h]
\begin{center}
\begin{picture}(360,160)(10,0)
\psset{unit=1.pt}        
\psline{->}(5,0)(5,150)\rput(-2,145){$\sigma$}
\psline{->}(0,5)(360,5)\rput(350,-2){E}
\psline{-}(12,5)(12,40)
\psline{-}(20,5)(20,70)
\psline{-}(22,5)(22,15)
\psline{-}(39,5)(39,50)
\psline{-}(44,5)(44,25)
\psline{-}(59,5)(59,88)
\pscurve{-}(60,5)(63,7)(69,18)(72,20)(75,17)(84,8)(85,7)(88,10)(95,35)(97,37)
(100,34)(108,7)(112,5)
\pscurve{-}(120,5)(125,8)(135,35)(150,90)(160,100)(163,102)(166,101)(180,88)
(200,60)(240,22)(245,30)(260,8)(262,9)(280,12)(290,20)(320,30)(340,20)(360,16)
\rput(30,90){\small discrete}
\rput(100,50){\small continuum}
\rput(200,110){\small giant resonances}
\rput(320,50){\small double resonances}
\end{picture}
\end{center}    
\caption{Schematic picture of the typical energy regions in a photoabsorption 
spectrum. See text. }
\label{pict2}
\end{figure}
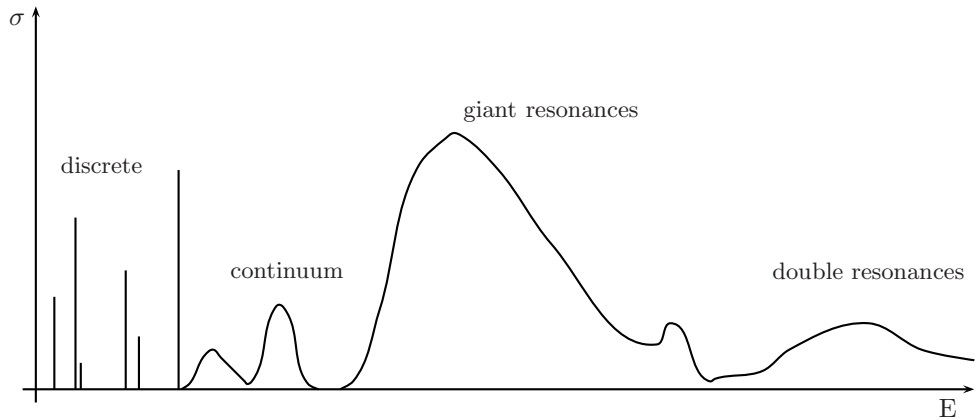 
\index{subject}{DGQR}

\begin{figure}[!h]
\begin{center}
\epsfig{file=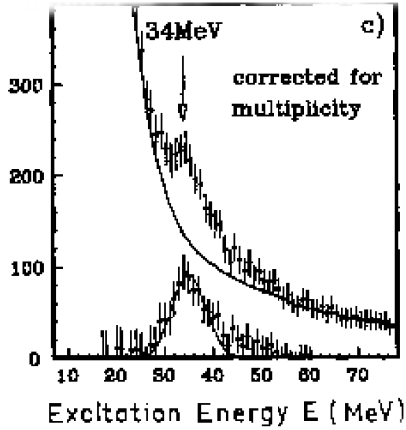,width=0.5\textwidth}
\caption{Experimental inelastic spectrum corrected for the proton multiplicity
for the $^{40}$Ca +$^{40}$Ca reaction at 44 MeV/n. The bump at 34 MeV
corresponds to the double phonon giant quadrupole resonance. From \cite{Chom}.}
\label{40Ca}
\end{center}
\end{figure}

The observation of multiphonon giant resonances raises a number of experimental
difficulties\cite{Frasca}. These modes are to be searched at high 
excitation energy in the continuum where various states with large width 
are overlapped. Moreover their cross section is in general thought to be 
quite small and very selective reactions are demanded in order to yield
appreciable cross sections. In heavy ion collisions at low incident energy 
(where with ``low'' we mean some 
20-50 MeV/n) the inelastic cross-section is dominated by the isoscalar 
resonances because of the strong nuclear interaction. 
The giant quadrupole resonance is the strongest among these kind of 
excitations and we expect that the double phonon excitation built on this 
state will be one of the most favourably seen. This is the case 
in the inelastic spectrum from the $^{40}$Ca +$^{40}$Ca reaction at 44 MeV/n
where the DGQR is seen as a small bump at 34 MeV (see fig. \ref{40Ca}).

\index{subject}{$^{40}$Ca}
The main difficulty is the extraction of the structure from the background, 
whose characteristics are inevitably affected by uncertainties  on the 
correction function.

\begin{figure}[!h]
\begin{center}
\includegraphics[width=1.0\textwidth]{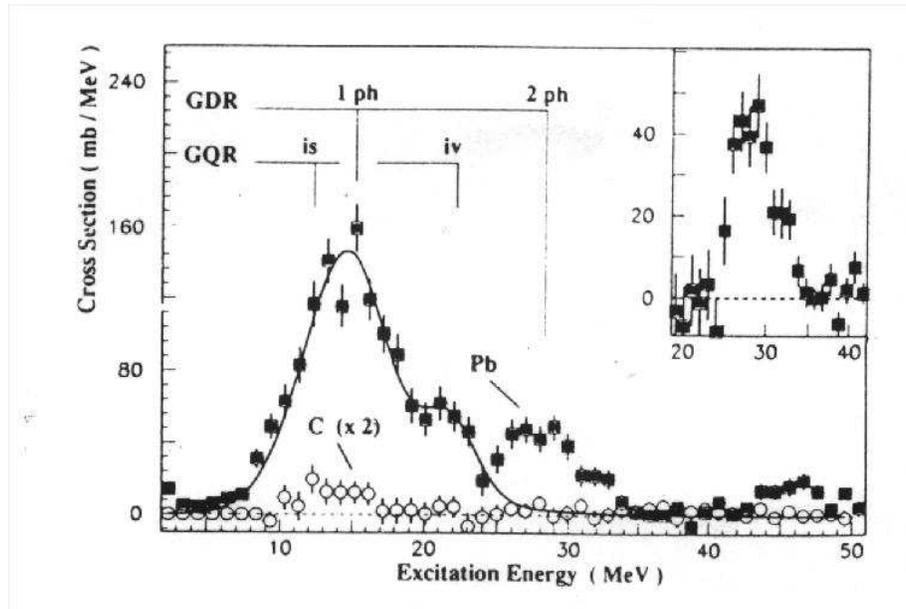}
\caption{Experimental inelastic spectra for $^{136}$Xe on $^{208}$Pb (black 
squares) at 700 MeV/n and on $^{12}$C (white circles, multiplied by 2).
This figure is taken from \cite{Chom}, for more details see \cite{LAND}.}
\label{gsi}
\end{center}
\end{figure}

\index{subject}{GSI}
Interest in this subject has been renewed by recent experiments with
relativistic heavy-ion beams at GSI, where inelastic cross sections for the
excitation of the dipole DGR have been precisely measured. In fig. \ref{gsi}
we report the findings of the LAND collaboration (Large Area Neutron Detector).
A structure at 28 $\pm$ 1 MeV is clearly observed with a width of about 6 $\pm$
2 MeV and it is interpreted as a two phonon state, namely a double-GDR.
The cross section for this state is measured to be 175 $\pm$ 50 mb.

\index{subject}{$^{136}$Xe}\index{subject}{$^{208}$Pb}
\index{subject}{$^{12}$C}

The theoretically calculated cross sections --~when performed within the
framework of the standard harmonic model~-- systematically
underestimate the experimental data by as much as a factor of
two. This unexpected enhancement of the cross sections puts in
evidence shortcomings in either the description of the structure of
the modes or in the formulation of the reaction mechanism. Attempts to
improve over this situation have followed different paths.

Another experiment has been performed for $^{208}$Pb on $^{208}$Pb at 
640 MeV/n \cite{Eml}. In this case the double phonon GDR is found at about 
23.8 MeV with a width of about 6 MeV and a cross section of about 350 mb.
Again this value is a factor of two larger that the prediction.

The microscopic understanding of these resonances, for instance, has
been taken beyond the simple superposition of the 1p-1h configurations
to include couplings to 2p-2h, 3p-3h and/or states of higher
complexity~\cite{wam,pon,ber}. Residual interactions give rise to
anharmonicity in the energy spectrum~\cite{cat} and, also, appreciable
changes in the structure of the wave functions.  Recently, a
systematic study of the anharmonicity in the dipole DGR has been
carried out for several nuclei~\cite{pon2}. This study, based on the
\index{subject}{QRPA}
quasiparticle RPA, has shown an effect of few hundred keV. The
same order of magnitude had been found in ref.~\cite{lan} for
 $^{208}$Pb and $^{40}$Ca.  These effects have been taken into
\index{subject}{$^{208}$Pb}\index{subject}{$^{40}$Ca}
account in macroscopic models that add small anharmonic
contributions~\cite{vol,bor} to the otherwise harmonic hamiltonian in
the presence of an external time-dependent field. Depending on the
magnitude of these anharmonic terms the inelastic cross sections for
the population of the dipole DGR can reach values which are close to
the experimental data. Microscopic calculations in the context of the
RPA approximation,\index{subject}{RPA} have also succeeded in reducing 
the discrepancy
between the experimental data and the theoretical predictions down to
the level of a few per cent~\cite{lan}. Another approach to the
problem that has been examined~\cite{car,wei} exploits the so-called
Brink-Axel hypothesis~\cite{brax}. \index{subject}{Brink-Axel hypothesis}
It also seems possible, through
this formalism, to obtain enhancements in the population of states in
the energy range around the DGR.

In this chapter we investigate the role of the nuclear coupling
in the excitation of GR's and DGR's and its interplay with the
long-range Coulomb excitation mechanism.  Furthermore, we study the
consequences of the spreading of the strength distribution of the
single giant resonance on the inelastic cross section for both the GR
and DGR. These topics have been previously explored in the literature.
In refs.~\cite{cat87,lan2,and} nuclear and Coulomb interactions where
taken into account for medium-heavy nuclei at low bombarding energy
(around 50 MeV/A).  While these studies put in evidence interference
effects between the two excitation mechanisms there was no clear
resolution concerning what could be actually attributed to each of
them.  Also, the role played by the resonances' width on the reaction
cross sections was covered in refs.~\cite{can,ber96,wei}. The
analysis, however, were done only for the case of the Coulomb
excitation mechanism and lead to somewhat ambiguous results.

We will carry on this survey within a simple reaction model that has the
virtue of conforming to the standard treatment of inelastic
excitations which is familiar to many active participants in this
field.  Our original intention was limited to investigate the
qualitative dependence of the probabilities of excitation of the
Double Giant Resonances as a function of several global parameters
such as the excitation energies, bombarding energies, multipolarity,
anharmonicity, width, etc.  In the process of refining the computer
programs we used to obtain these global trends we ended up with a
quite transparent and yet powerful tool that --~we believe~-- can be
useful for the experimentalists to make {\it quantitative} predictions
for measurements in a wide variety of circumstances.  With this very
practical purpose in mind we shall take the width
of the states as a free parameter.  We shall also limit our
calculations to the non-relativistic regime and, for the different
examples, consider the excitation of single- and double-phonon Giant
Quadrupole Resonances.
 
Following this Introduction we describe in Sect. 2.2 the formalism
employed to make our estimates.  Relevant results for the reaction
$^{40}$Ar + $^{208}$Pb are given with illustrations in
Sect. 2.3.  The conclusions that can be inferred from these examples
are also the subject of this Section. Some concluding remarks are left
for Sect. 2.4.
\index{subject}{$^{40}$Ar}
\index{subject}{$^{208}$Pb}

\section{The model}

The excitation processes of the one and two-phonon states are
calculated within the framework of the standard semiclassical model of
\index{subject}{semiclassical model}
Alder and Winther \cite{ald} for energies below the relativistic
limit. According to this model for heavy ion collisions, the nuclei
move along a classical trajectory determined by the Coulomb plus
nuclear interaction. We will explore the energy range from few MeV up
to hundreds of MeV per nucleon. During their classical motion the
nuclei are excited as a consequence of the action of the mean field of 
one nucleus 
on the other. The excitation processes are described according to quantum
mechanics and they are calculated within perturbation theory.

We assume that the colliding nuclei have no structure except for the
presence, in the target, of one and two-phonon states whose energies
are $E_1$ and $E_2=2 E_1$, respectively. For the ion-ion potential we
have used the Coulomb potential for point charged particles and the
Saxon-Woods parametrization of the proximity potential $U_N(r)$ that
are commonly used in heavy ion collisions~\cite{win}.

In the theory of multiple excitations the set of coupled equations
describing the evolution of the amplitudes in the different channels
can be solved within the perturbation theory. We can write the
probability amplitude to excite the $\mu$ component of the one-phonon
state with multipolarity $\lambda$ as

\begin{equation}
a^{(1)}_{\lambda \mu} (t)= (-i / \hbar) 
\int_{-\infty}^{\infty} dt F_{\lambda \mu }  
(r(t),\hat{r}(t))  e^{iE_1 t/\hbar}~,
\label{a1}
\end{equation}
where the integrals are evaluated along the classical trajectories
${\bf{r}}(t)$. In this equation the main ingredient is the coupling
form factor
\begin{equation}
 F_{\lambda \mu }  (r(t),\hat{r}(t))~=~ f_{\lambda}(r)~Y_{\lambda \mu}
(\hat{r})~,
\end{equation}
chosen according to the standard collective model
prescription~\cite{land}. For a given multipolarity $\lambda$ the radial
part assumes the form
\begin{equation}
f_\lambda (r) = {{3 Z_p Z_t e^2}\over {(2 \lambda +1) R_C}}
\beta^C_\lambda \Bigg( {R_C \over r} \Bigg)^{\lambda +1}
- \beta^N_\lambda R_T {d\over dr} U_N(r)~.
\label{ff}
\end{equation}
The deformation parameters $\beta$ determine the strength of the
couplings, and they are normally directly associated with the
$B(E\lambda)$ transition probability.  The expression for the nuclear
component of the form factor is not valid for $\lambda=1$. In this
case the inelastic form factor is obtained from the Goldhaber-Teller
or Jensen-Steinwendel models. The $Z_p$ ($Z_t$) denotes the charge
number of the projectile (target), while $R_C$ and $R_T$ are the
Coulomb and matter radii of the target nucleus.

In a similar way, the amplitude for populating the two-phonon state
with angular momentum $L$ and projection $M$ can be obtained as
\begin{eqnarray}
a^{(2)}_{LM} (t) &=& (1/ \hbar )^2
\sum_{\mu}~ \sqrt{(1 +\delta_{\mu,M-\mu})} \nonumber \\
& \times & \int_{-\infty}^{\infty} dt F_{\lambda,M-\mu}({\bf r}(t))  
e^{i(E_2-E_1)t /\hbar}
\int_{-\infty}^{t} dt' F_{\lambda,\mu}({\bf r}(t'))  e^{iE_1t' /\hbar}
\label{a2}
\end{eqnarray}

Solving the classical equation of motion we can calculate for each
impact parameter $b$ the excitation probability $P^{(1)}(b)$ and
$P^{(2)}(b)$ to populate the single- and the double-phonon
state. These are given by
\begin{equation}
 P^{(1)}(b)~=~\sum_{\mu} |a^{(1)}_{\lambda \mu}(t=+\infty)|^2
\end{equation}
and
\begin{equation}
 P^{(2)}(b)~=~\sum_{L}P^{(2)}_{L}~=~\sum_{LM} |a^{(2)}_{LM}(t=+\infty)|^2~.
\end{equation}
In order to get the corresponding cross sections we have then to
integrate the probabilities $P^{(\alpha)}$'s ($\alpha$ =1,2)
\begin{equation}
\sigma_{\alpha} = 2 \pi \int_0^\infty P^{(\alpha)}(b) T(b) b db \,.
\label{xsec}
\end{equation}
Generally, in Coulomb excitation processes the transmission
coefficient is taken equal to a sharp cutoff function
$\theta(b-b_{min})$ and the parameter $b_{min}$ is chosen in such a
way that the nuclear contribution is negligible. We want to take into
account also the contribution of the nuclear field so in our case T(b)
should be zero for the values of $b$ corresponding to inner trajectory
and then smoothly going to one in the nuclear surface region. This can
be naturally implemented by introducing an imaginary term in the
optical potential which describes the absorption due to non elastic
channels.  Then the survival probability associated with the imaginary
potential can be written as
\begin{equation}
T(b)= \exp\Bigg\{ {2\over \hbar }  \int_{-\infty}^{+\infty} W(r(t)) 
dt \Bigg\} \,,
\label{Tb}
\end{equation}
where the integration is done along the classical trajectory. The
imaginary part $W(r)$ of the optical potential was chosen to have the
same geometry of the real part with half its depth.

The excitation processes of both single and double GR can change
significantly when one takes into account the fact that the strength
of the GR is distributed over an energy range of several MeV. Among
the few standard choices for the single GR strength distribution, we will
assume a Gaussian
shape, with a width $\Gamma_1=2.3 \sigma $ which we will take as a
parameter, of the following form
\begin{equation}
S(E) = {1 \over {\sqrt{2 \pi}\sigma}} \exp \Bigg\{ 
{-(E - E_1)^2 \over {2 \sigma^2}}\Bigg\} \,.
\label{GAU}
\end{equation}
Calculations have been also performed with a Breit-Wigner shape
yielding similar trends. However, the Gaussian form guarantees a
better localization of the response and prevents superposition of the
modes for the largest widths (for a further discussion see
ref.~\cite{ber}). 

To get the cross section to the one-phonon state one then defines a
probability of excitation per unit of energy,
\begin{equation}
 dP^{(1)}(E,b)/dE~=S(E)~\sum_{\mu} |a^{(1)}_{\lambda \mu}(E,t=+\infty)|^2~,
\end{equation}
where the single amplitudes $a^{(1)}_{\lambda \mu}(E,t)$ are obtained
as before, but with a variable energy $E$.  The probability of
exciting the double-phonon state is then obtained by folding the
probabilities of single excitation, in the form
\begin{equation}
dP^{(2)}(E,b)/dE~=~\int dE'~ {dP^{(1)}(E',b)\over dE}~ ~ 
{dP^{(1)}(E-E',b)\over dE}~. 
\end{equation}
The total cross section for one- and two-phonon states can then be
constructed as
\begin{equation}
\sigma_{\alpha} = 2 \pi \int_0^\infty \int_0^
\infty {dP^{\alpha} \over dE}(E,b)~ T(b) \,b \,db
\,dE \, \,.
\label{xsecg}
\end{equation}

Due to the Q-value effect it is clear that one expects a distortion in
the shape of the distribution of the cross section which will favor
the lower part of the distribution in energy.

\section{Results}

\begin{figure}[!t]
\begin{center}
\epsfig{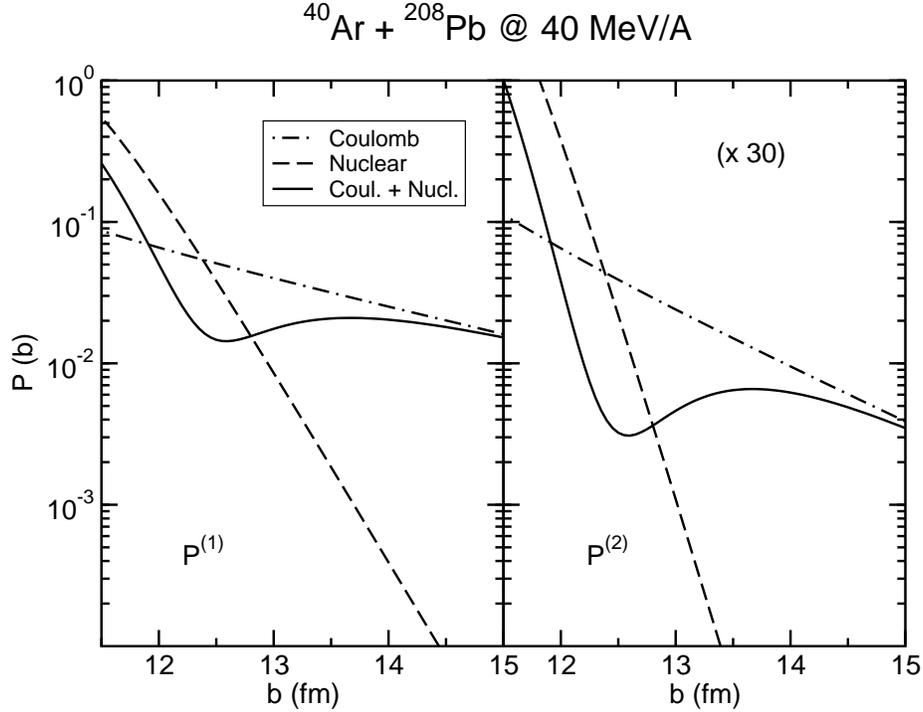}
\end{center}
\caption{Excitation probability vs. impact parameter for the one- (left part)
and two-phonon (right part) states of the GQR in lead for the reaction
$^{40}$Ar + $^{208}$Pb at 40 MeV/A. The Coulomb (dot-dashed line) and
nuclear (dashed) probabilities are displayed as well as the total
(solid line). The curves on the right part have been multiplied by
30.}
\label{PB}
\end{figure}

We show in fig. \ref{PB} the dependence on the impact parameter of the
excitation probabilities for the one- and two-phonon states of the
Giant Quadrupole Resonance \index{subject}{giant quadrupole resonance}
in lead.  The reaction we have chosen for
\index{subject}{$^{40}$Ar}  \index{subject}{$^{208}$Pb}
this illustration is $^{40}$Ar + $^{208}$Pb at a bombarding energy of
40 MeV per nucleon.  The deformation parameters have been chosen equal
$\beta^C = \beta^N = 0.07$, in agreement with the currently estimated
value for the $B(E2)$. The range of impact parameters given in the
figure covers the relevant grazing interval, and in a classical
picture (including both Coulomb and nuclear fields) 
yields scattering angles between $3.4$ and $5.5$ degrees.  In
the strictly harmonic case the probabilities for excitation of the
double-phonon state can of course be constructed from those
corresponding to the single-phonon; they are both explicitly given
here for a matter of later convenience.  Each frame displays a set of
three curves that allows us to compare the individual contributions of
the Coulomb and nuclear fields to the excitation process and put in
evidence a value of $b\approx 12.5$ fm for the maximum (destructive)
interference between the competing mechanisms.

\begin{figure}[!t]
\begin{center}
\epsfig{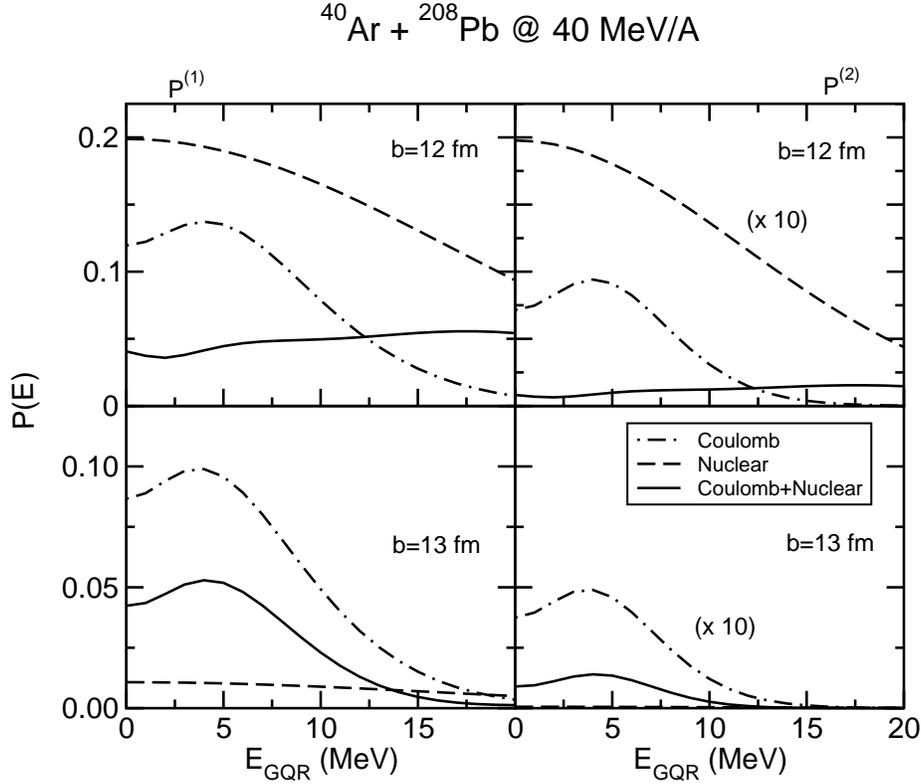}
\end{center}
\caption{ Excitation probabilities as function of the GQR energy, which
is taken as a parameter, for the reaction $^{40}$Ar + $^{208}$Pb at 40
MeV/A. The graphs on the left correspond to the excitation probability
of the single GQR ($P^{(1)}$) while the ones on the right correspond
to DGQR ($P^{(2)}$) and they are multiplied by a factor 10. The
Coulomb (dash-dotted line) and nuclear (dashed) probabilities are
displayed as well as the total (solid line). The upper (lower) figures
correspond to an impact parameter of 12 (13) fm.}
\label{PQ}
\end{figure}

\begin{figure}[!t]
\begin{center}
\epsfig{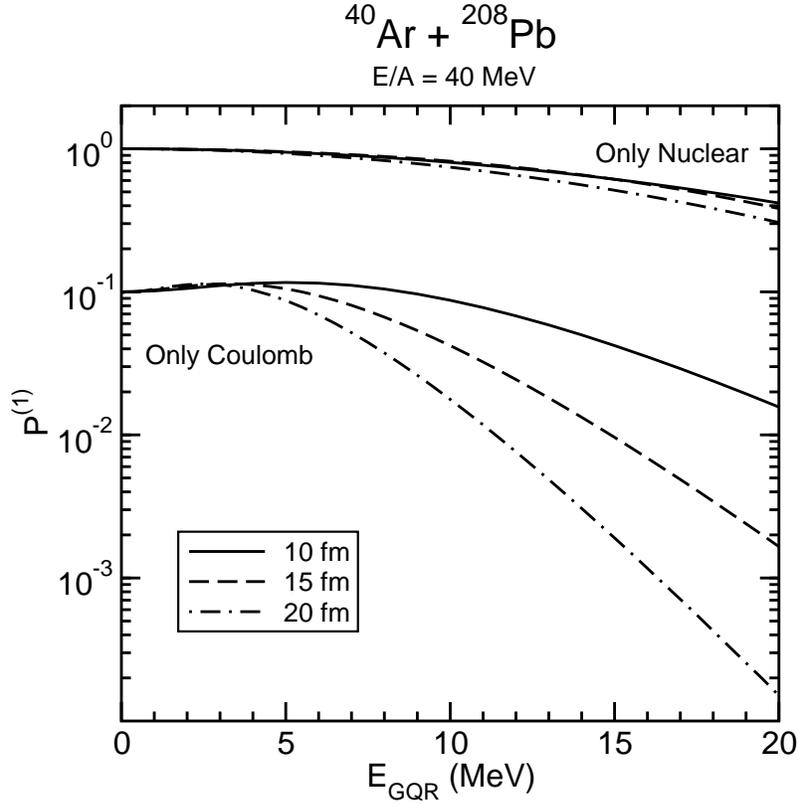}
\end{center}
\caption{ Excitation probability for the single GQR as a function of the
GQR energy for three values of impact parameters. They have been
normalized to their values at $E_{GQR} =0$.  The upper curves
correspond to the excitation probability due only to the nuclear
field. The probabilities calculated only with the Coulomb field are
shown in the lower part of the picture. They have been divided by 10
in order to render the figure readable.}
\label{P1QB}
\end{figure}

\begin{figure}[!t]
\begin{center}
\epsfig{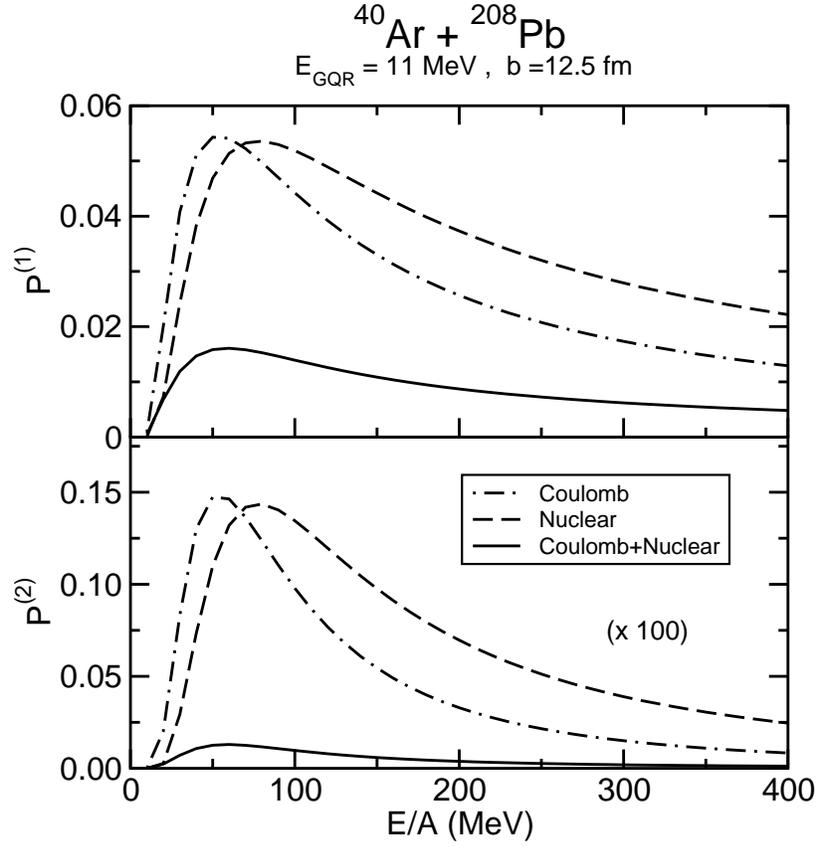}
\end{center}
\caption{Excitation probability as a function of incident energy of the 
one- (upper part)
and two-phonon (lower part) states of the GQR in lead for the reaction
$^{40}$Ar + $^{208}$Pb for an impact parameter of 12.5 fm. The Coulomb
(dot-dashed line) and nuclear (dashed) probabilities are displayed as
well as the total one (solid line). The curves on the lower part have
been multiplied by 100.}
\label{PE}
\end{figure}

\begin{figure}[!t]
\begin{center}
\epsfig{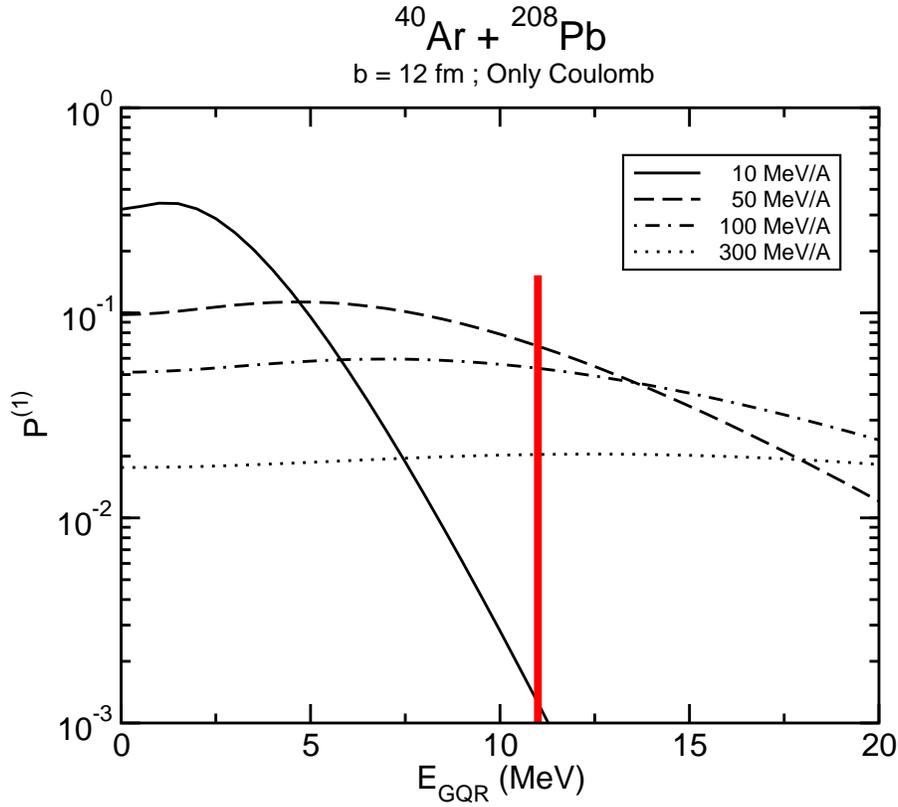}
\end{center}
\caption{Coulomb excitation probability for the one-phonon state 
as a function of the Q-value and for four different incident energies
as shown in the legend. The curves correspond to calculations done for
an impact parameter b=12 fm. The vertical line indicates the GQR
energy for lead.}
\label{P1QE}
\end{figure}

\begin{figure}[!t]
\begin{center}
\epsfig{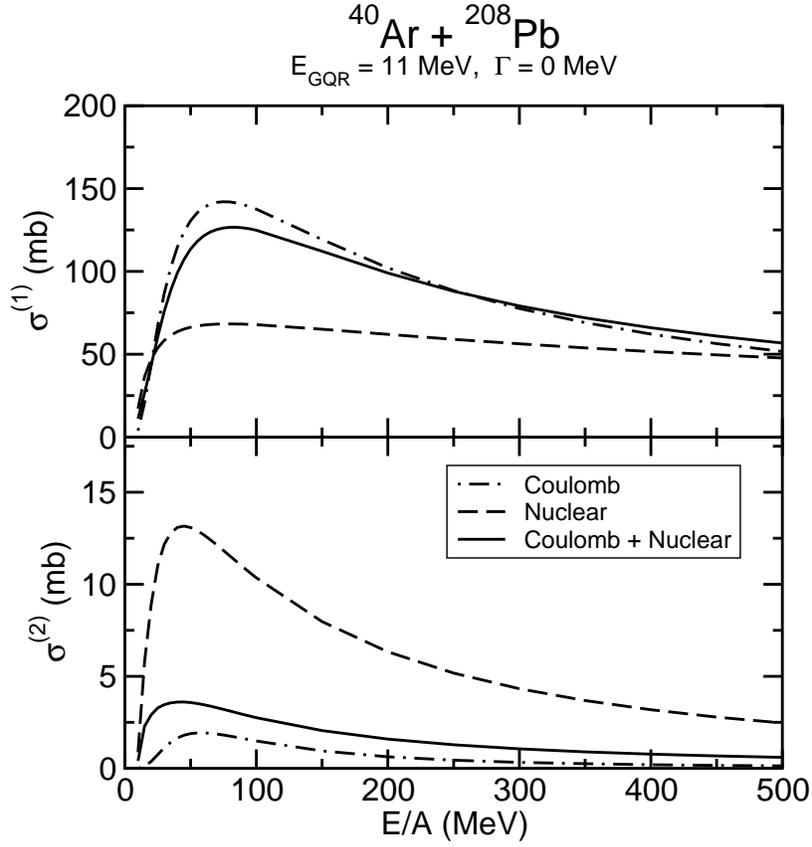}
\end{center}
\caption{Excitation cross section for the GQR (upper figure) and DGQR
state (lower figure) as a function of the incident energy. Again, the
Coulomb (dot-dashed line), nuclear (dashed line) and total (solid
line) contributions are shown.}
\label{XE}
\end{figure}

\begin{figure}[!t]
\begin{center}
\epsfig{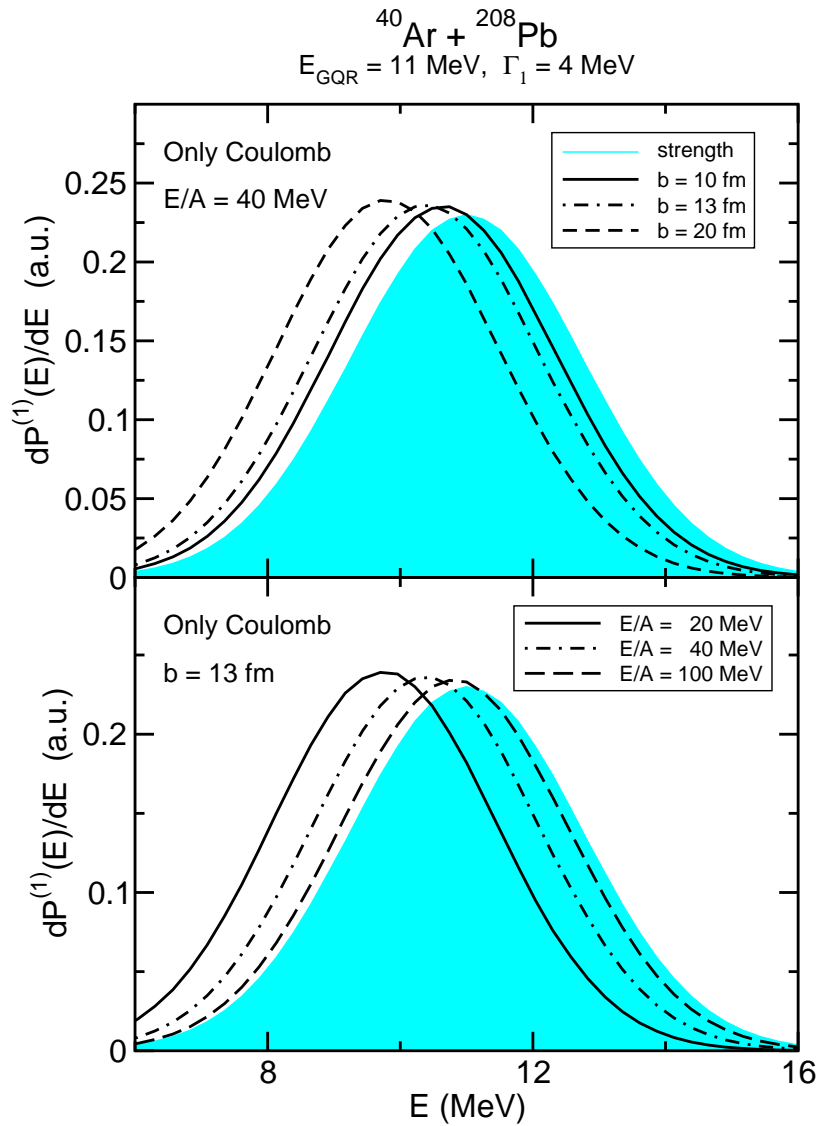}
\end{center}
\caption{Normalized distributions of Coulomb excitation probability for 
the one-phonon state 
for different impact parameters (upper figure) and bombarding energies
(lower figure) as shown in the legend. The shaded area shows the
Gaussian strength distribution used as input in the calculation. The
width has been chosen to be 4 MeV.}
\label{P1W}
\end{figure}

\begin{figure}[!t]
\begin{center}
\epsfig{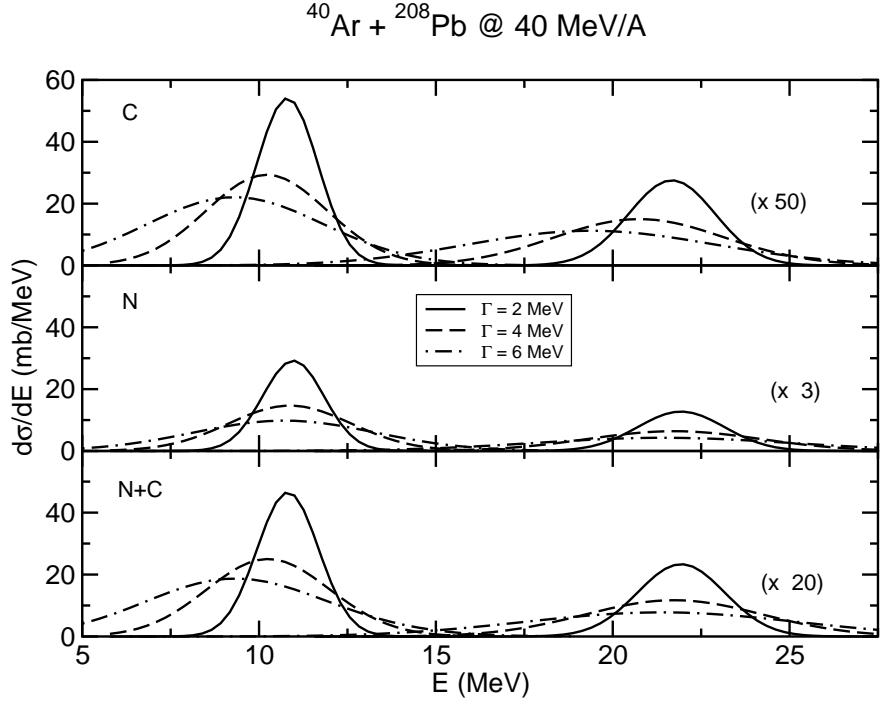}
\end{center}
\caption{Cross section distribution for the reaction
$^{40}$Ar + $^{208}$Pb at 40 MeV/A for three values of the width as
indicated in the legend.
The contributions of the Coulomb (C), nuclear (N) and total (N+C) cross
sections are shown in single graphs. The cross sections for DGQR are
multiplied by the factors reported in the figure.}
\label{XW}
\end{figure}

\begin{figure}[!t]
\begin{center}
\epsfig{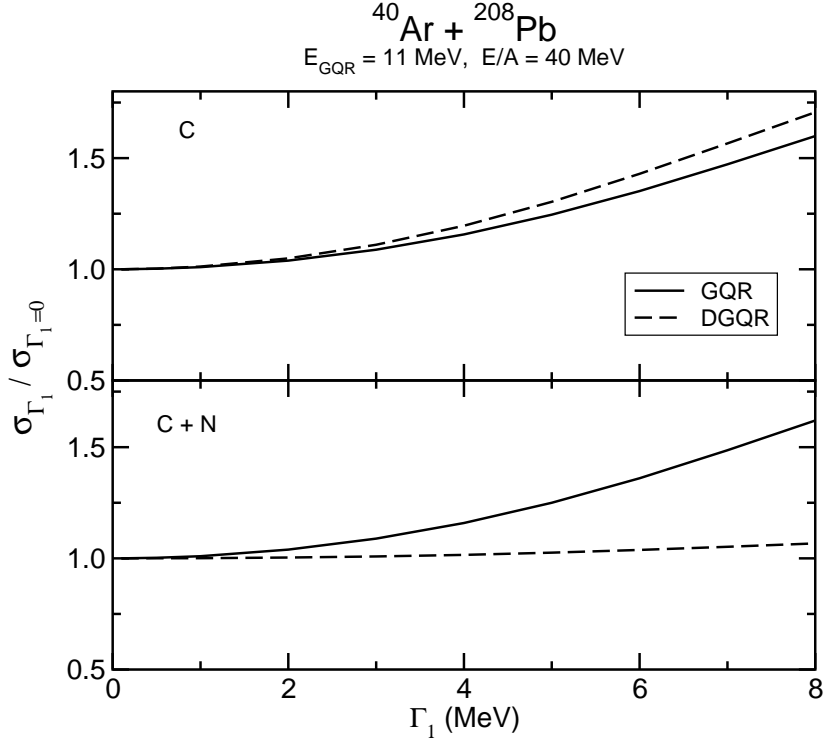}
\end{center}\index{subject}{width}
\caption{Total cross sections for the excitation of GQR (solid line)
and of DGQR (dashed line) as a function of the strength distribution
width $\Gamma_1$ of the single GQR for the reaction $^{40}$Ar +
$^{208}$Pb at $30 MeV/A$. In the graphs we report the cross sections
due to the Coulomb field (upper) and total (lower), each of them
divided by their corresponding value for sharp distribution
($\Gamma_1=0$). We have not reported the nuclear contribution because
the cross sections for both GQR and DGQR do not change appreciably 
when a finite
distribution is assumed.}
\label{XG}
\end{figure}

\
\begin{figure}[!t]
\begin{center}
\epsfig{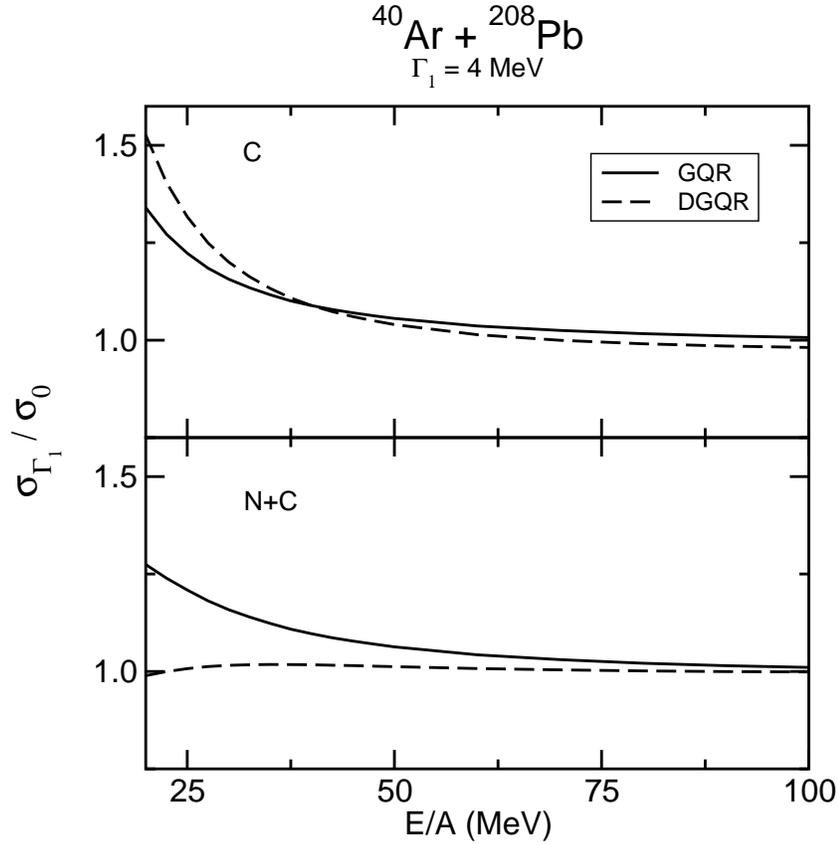}
\end{center}
\caption{Same as fig.~\ref{XG} as a function of the bombarding energy.
The strength distribution width has been chosen equal to 4 MeV.}
\label{XEW}
\end{figure}

\par
We use the same reaction as in fig.\ref{PB} to illustrate the effect of the
reaction Q-values on the transition probabilities.  The dependence of
the single-step inelastic excitation to the one-phonon state of energy
$E_{_{GQR}}$ and the sequential process feeding the double-phonon
state at twice this value are shown in fig.\ref{PQ}.  As before, the three
curves in each frame display the separate contributions of the Coulomb
and nuclear fields and the combined total.  Two values of the impact
parameter have been chosen specifically to
cover a situation of nuclear ($b$=12 fm) and Coulomb ($b$=13 fm)
dominance. The results show --~even in a linear scale~-- a somewhat
moderate dependence with the frequency of the mode.  This is due to
the relatively high bombarding energy chosen in this example, for
which the time-dependence of both the Coulomb and nuclear excitation
fields are quite well-tuned to the intrinsic response.
\par
There is a qualitative difference in the effective collision time for
Coulomb and nuclear inelastic processes that is worth mentioning.  We
refer to the dependence, for a given bombarding energy, of the
excitation probabilities for the one- and two-phonon states on the
impact parameter. Because of the long range of the formfactors the
change of the effective collision time $\tau$ for Coulomb excitation
follows a different law than the one corresponding to the nuclear
inelastic processes. It can be estimated that $\tau_{_C}/\tau_{_N}
\approx A_\lambda\sqrt{b}$, where the proportionality factor
$A_\lambda$ is a monotonically decreasing function of the
multipolarity $\lambda$.  For all multipolarities, however,
$\tau_{_C}$ is larger than $\tau_{_N}$.  It follows from these
arguments that the adiabatic cut-off function that affects the
transition amplitudes for Coulomb excitation varies significantly over
the large range of impact parameters that contributes to this process.
For the nuclear field a favorable matching between effective collision
times and the intrinsic period of the mode applies, on the other hand,
to most of the relevant partial waves.  This can be understood by
examining fig. \ref{P1QB} , where the probability for excitation of the
one-phonon level is plotted as a function of the energy of the mode
for three impact parameters, $b$=10, 15 and 20 fm.  Two sets of curves
are shown, corresponding to Coulomb and nuclear excitation only.  In
both instances the probabilities are normalized to their values for
$E_{_{GQR}}$=0 MeV to emphasize the different character of the
response.  Notice, for instance, that for $E_{_{GQR}}$=20 MeV the role
of the Coulomb field for $b$=20 fm would be effectively quenched by
two orders of magnitude in spite of its long range. (This is of course
{\rm in addition} to the gradual reduction of the transition
amplitudes caused by the slow $r^{-(\lambda+1)}$ dependence of the
couplings.)
\par
We use fig.\ref{PE} to illustrate the dependence of the excitation
probabilities upon the bombarding energy.  For this we take a value of
$E_{_{GQR}}$=11 MeV, close to the actual excitation energy of the
Giant Quadrupole Resonance in lead.  The impact parameter $b$ is set
to 12.5 fm, which provides the condition in which the importance of
the Coulomb and nuclear excitations become comparable.  Of course it
is also the choice of impact parameter that yields the maximum
(negative) interference between the two reaction mechanisms. What we see
is a rapid increase of the probabilities for the one-phonon and
two-phonon levels up to a bombarding energy of about 50 MeV/nucleon.
After that a gradual decline sets in up to about 400 MeV/nucleon, an
energy beyond which a relativistic formalism must be implemented.  The
trend, however, is not to be significantly altered and, in view of
these results one cannot but wonder about the actual need of
exploiting relativistic bombarding energies to probe the excitation of
double-phonon giant resonances in nuclei.  In principle, and entirely
from an adiabatic point of view, the higher the bombarding energy the
better.  Yet, optimal matching conditions reach saturation and one
cannot ignore the fact that, beyond this point, one can no longer
expect a further enhancement of the excitation probabilities. Quite on
the contrary, the interaction time is effectively reduced up to a
point where (as the figure shows) the excitation of the modes becomes
less and less favored (see caption to fig. \ref{P1QE} and, also,
ref.~\cite{ber96}).
\par
In fig.\ref{XE} we display the cross section for the excitation of the GQR
and DGQR as a function of the bombarding energy in MeV per nucleon.
This observable quantity combines the effect of all impact parameters
and the plot puts in evidence a quite interesting feature.  Notice
that at all bombarding energies the population of the one-phonon state
is dominated by the Coulomb formfactors.  At the two-phonon level, on
the other hand, it is mostly the nuclear coupling that determines the
outcome.  To understand the origin of this exchange of roles it may be
helpful to re-examine fig.\ref{PB}.  We have here to pay attention to the
dependence of the ratio between the probabilities for nuclear and
Coulomb excitation in the relevant range of impact parameters, 11-13
fm.  (To this end the display factor of 30 introduced for the case of
the two-phonon state is of no consequence.)  The enhanced logarithmic
slopes for the DGQR resulting from the squaring of the one-phonon
probabilities suffice to give the leading edge to the nuclear
couplings.  This realization has major consequences insofar as the
global properties of the excitation of the GQR and DGQR is concerned.
In fact, the transition probabilities will inevitably reflect the
different characteristics of the reaction mechanism that it is mostly
responsible for the population of one state or the other.
\par
Q-value considerations have such a pronounced effect on the excitation
probabilities that it is clear that they will play an important role
when one takes into account the sizable width of the one- and
two-phonon states.  Suppose that instead of having the total strength
of the mode at a fixed value $E_{_{GQR}}$ we distribute it with the
profile of a Gaussian distribution of width $\Gamma$.  If the energy
of the mode is quite off the optimal Q-value window one should expect
that the distribution of measured cross sections will follow a quite
different law. In fact, whenever the dynamic response in the vicinity
of $E_{_{GQR}}$ is a rapidly changing function of the energy (see, for
instance, fig. \ref{P1QE} for $E$=10 MeV/A) the experimental distribution will
be significantly distorted and shifted toward lower energies.  We
illustrate this aspect in fig. \ref{P1W}, where the distribution of Coulomb
excitation probabilities for the one-phonon state, $dP^{(1)}/dE$, is
shown for different impact parameters and bombarding energies.  In
each frame the shaded curve shows the Gaussian distribution of
strength that is the input to the calculation.  Notice that all
distributions have been normalized in order to emphasize the effect of
interest and to eliminate the over-all dependence on $b$ and $E$
discussed earlier.  As it follows from our considerations one can
easily see that the smaller distortion corresponds indeed to the
smaller impact parameters and/or the larger bombarding energies.
\par  \index{subject}{width}
The distortion of the line profile at the one-phonon level increases
as a function of the width $\Gamma$, as it is clearly seen in fig. \ref{XW},
where reaction cross sections (i.e. the result of an integration over
impact parameters) are shown for a typical value of the bombarding
energy.  For the larger width $\Gamma$=6 MeV the apparent shift of the
distribution is large enough as to place most of the cross section
outside of the initial range set by the Gaussian curve.  The effect
seems to be more noticeable at the two-phonon level, as shown on the
right-hand-side of the figure.  According to our previous discussion,
it is the Coulomb excitation mechanism that contributes most to the
difference between the strength and cross section profiles.
\par
From the energy distributions displayed in fig.\ref{XW} one can calculate
the total one- and two-phonon cross sections, by integrating over the
excitation energy.  The global effect of the finite width is shown in
fig. \ref{XG}, where the total cross sections for different values of the
width are compared with the corresponding values for sharp resonances.
The enhanced excitation in the lower part of the distribution leads to
a global enhancement in the case of the Coulomb field.  As a
consequence, a corresponding enhancement is present in the combined
Coulomb+nuclear case in the one-phonon excitation, which is dominated
by the Coulomb interaction.  On the contrary, being the two-phonon
cross section predominantly due to the nuclear process, no 
appreciable variation
is predicted for this case with finite values of the width.
\par
Since the effect of the width arises from the Q-value kinematic
matching conditions, variations are expected with the bombarding
energy.  In particular one expects that the effects will tend to
vanish at high bombarding energies.  This is illustrated in fig. \ref{XEW},
where the total one- and two-phonon cross sections for $\Gamma$ = 4
MeV are compared to the corresponding values for $\Gamma$ = 0 as a
function of the incident energy.
   
\section{Conclusions and remarks}

We have implemented a simple scheme to calculate the excitation probabilities
for the single and double Giant Resonance as a function of several global
parameters such as excitation energies, bombarding energies, width
etc. We have assumed that the colliding nuclei have no structure
except for the presence, in the target, of one and two-phonon
states.  The excitation processes have been calculated within a
semiclassical model and according to perturbation theory. Since both
nuclear and Coulomb interaction are taken into account the cross
sections are calculated by integrating over all range of impact
parameter with an imaginary potential that takes care of the inner
trajectories. The formalism has been applied to the excitation of 
giant resonances in a typical heavy ion reaction, $^{40}$Ar + $^{208}$Pb. 
In our examples, we have limited our calculation to the giant quadrupole 
resonance.

The role of the nuclear interaction and its interplay with the
long-ranged Coulomb field has been studied. The presence of nuclear
coupling modifies the mechanism excitation of both the GR and the DGR,
the effect being strongly evident in the latter. This has been
ascribed to the difference in the effective collision time which,
together with the qualitative $r$ dependence of the form factors,
produces a different dependence of the transition probabilities on
the reaction Q-value. Hence, the excitation of GR is dominated by the
Coulomb interaction while it is mostly the nuclear coupling which
determines the population of the DGR. 

We have also studied the consequences of the spreading of the strength
distribution of the single giant resonance on the inelastic excitation
of the GR and DGR. Q-value considerations play an important role when
the width of the one- and two-phonon states are considered. Cross
section dependence on both the width of the distribution and the
incident energy has been considered. When compared with the
corresponding values for sharp resonances, the cross sections for GR
and DGR calculated with only the Coulomb field increase as
$\Gamma$ increases. These results are qualitatively similar to the one
obtained in ref.~\cite{wei} where the relativistic Coulomb excitation
of dipole giant resonance (GDR) and double GDR are calculated within a
random matrix theory including the Brink-Axel hypothesis. When the
nuclear interaction is switched on, the enhancement for the single GR
is maintained while the two-phonon cross section presents no variation
with the case of finite value of the width. Also for the dependence on
the incident energies has been found the same trend. This is due to
the fact that the two-phonon cross section is predominantly governed
by nuclear processes.



\newpage
~~\\

\chapter{Giant Pairing Vibrations}

\begin{figure}[!h]
\begin{flushright}
\epsfig{file=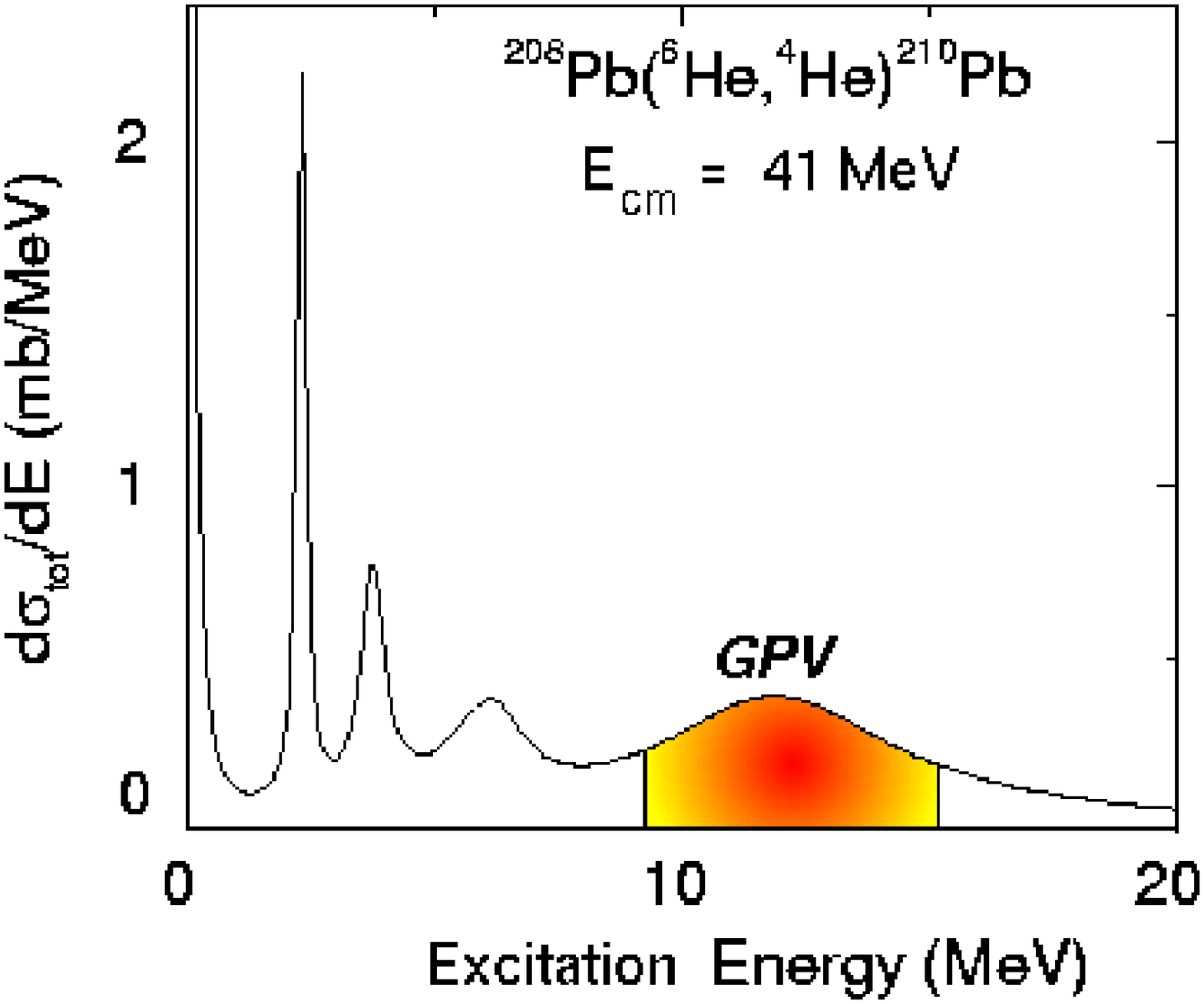,width=.25\textwidth}
\end{flushright}
\end{figure}

\section{Introduction.}
\begin{figure}[!t]
\begin{center}
\epsfig{file=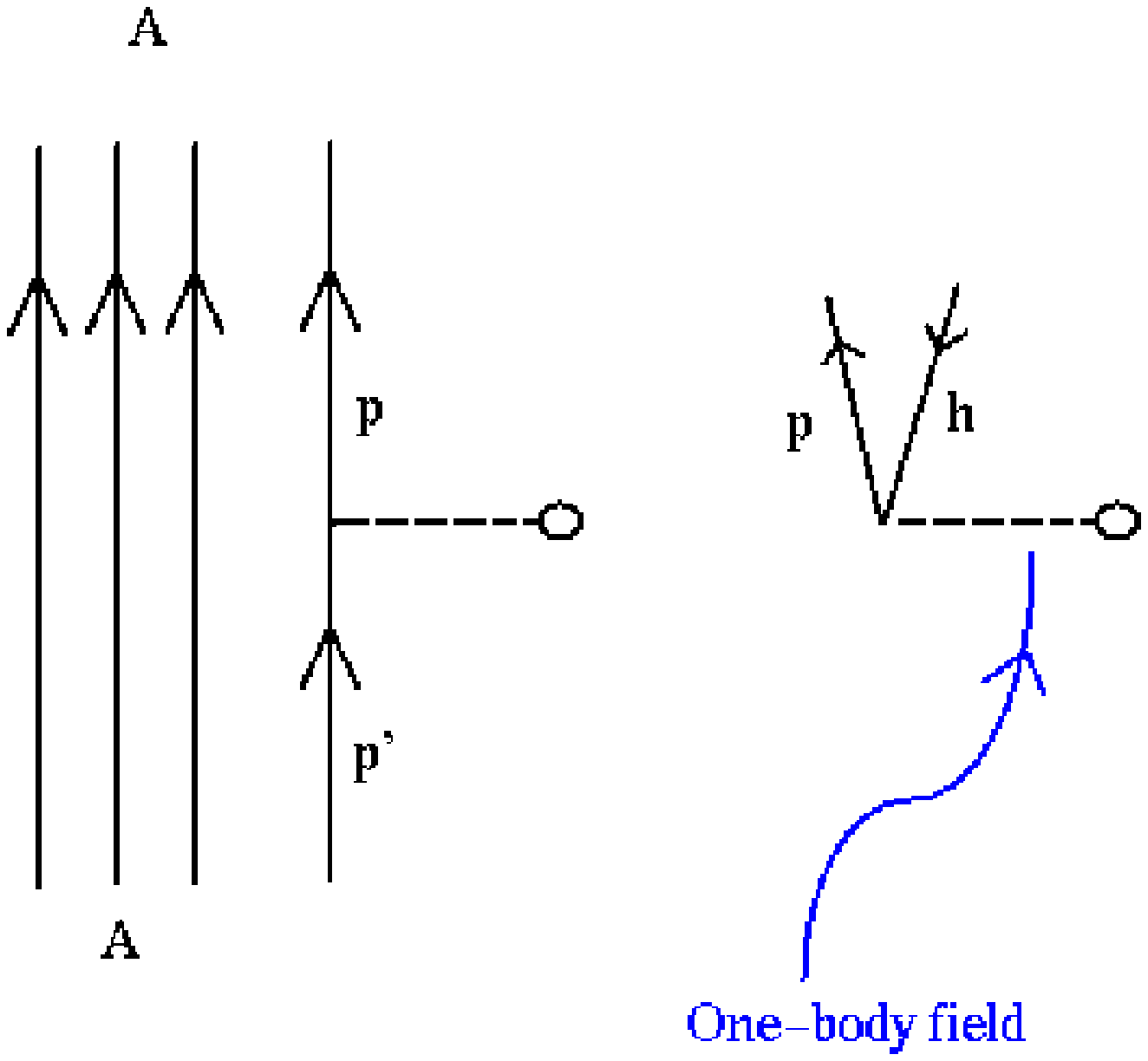,width=.52\textwidth}a
\epsfig{file=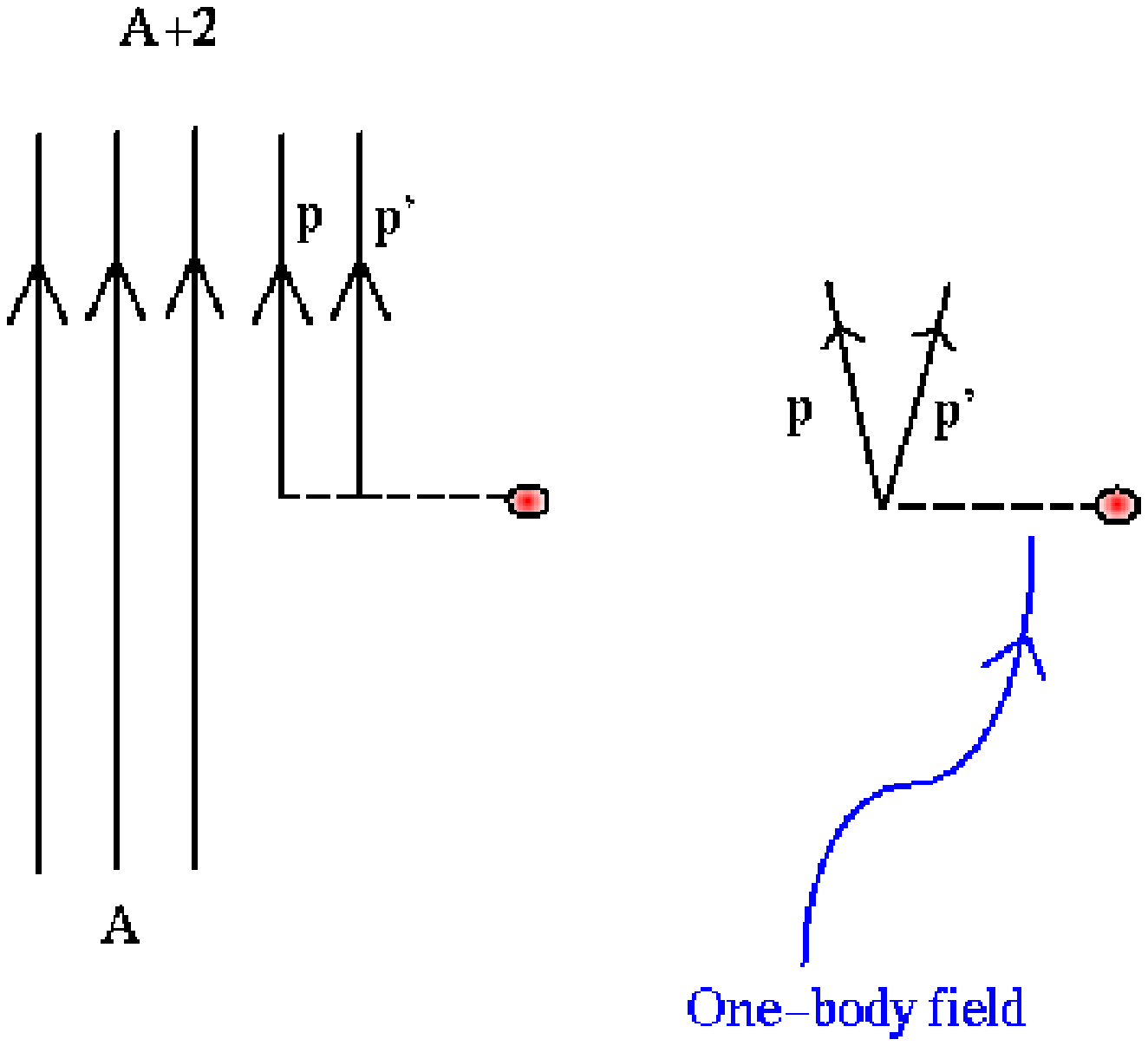,width=.52\textwidth}b
\epsfig{file=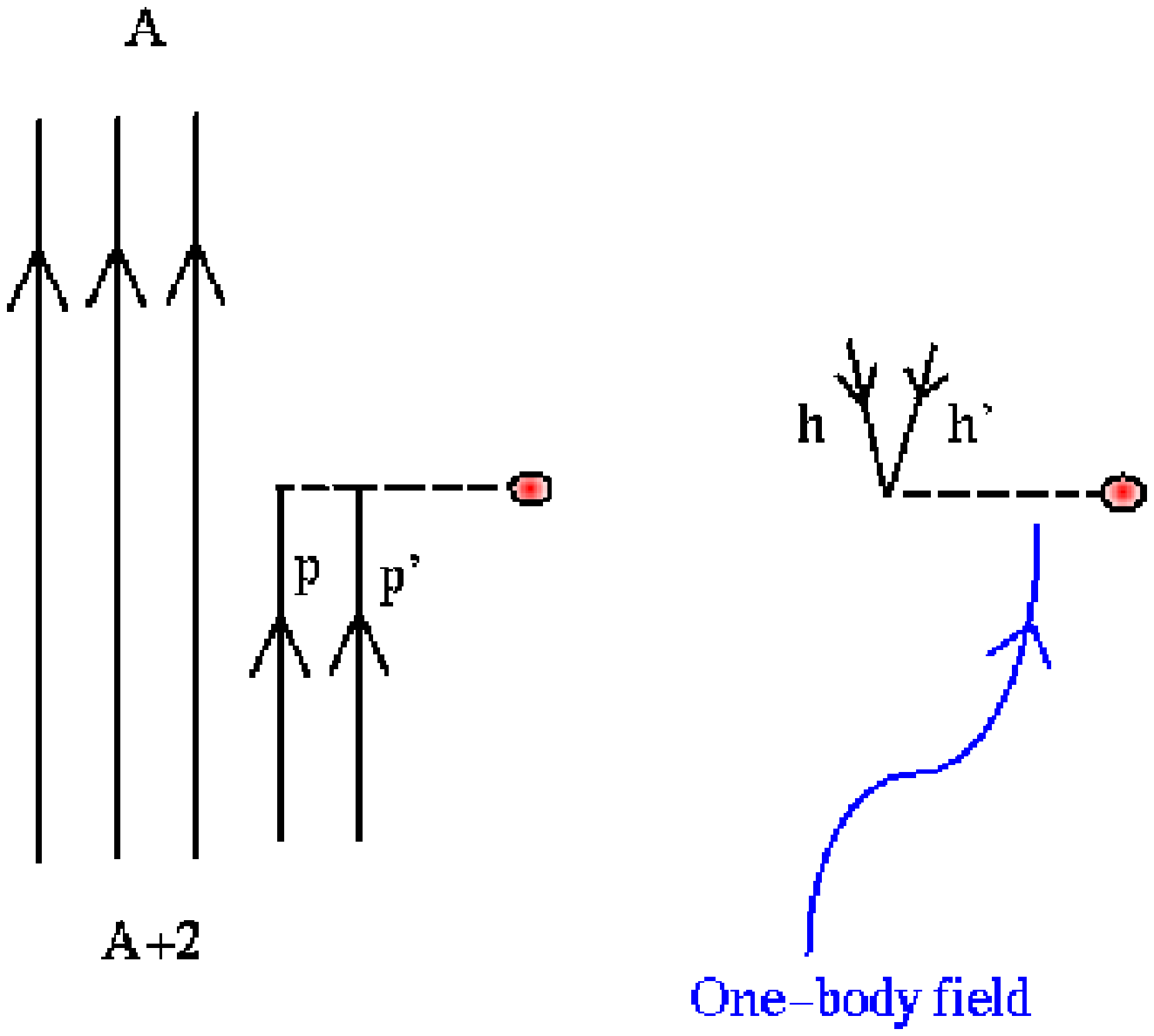,width=.52\textwidth}c
\end{center}
\caption{Formal analogy between particle-hole (a) and particle-particle (b) 
or hole-hole (c) excitations. The one body field that acts in the last two 
cases is the pairing field.}
\label{ph}
\end{figure}

Conventional giant resonances, that have been discussed in the last chapter 
and that will form the subject of the fourth chapter, may be considered as 
collective modes where a significant part
of the constituents of the nucleus oscillates or moves together.
In a microscopic picture giant resonances, corresponding to surface vibrations,
can be viewed as coherent particle-hole excitations to high-lying states,
where the transition matrix elements
receive in-phase contributions from all the various possible configurations.
The amount of collectivity is determined by the values of the matrix
elements of the appropriate transition operator.\index{subject}{pp-RPA}
In this chapter we will concentrate on a collective mode of a
different kind, that nevertheless may be studied exploiting a number
of analogies with the well-known case.
The formal analogy between particle-hole and particle-particle excitations is 
well established as well as the use of a one-body pair field 
(see fig. \ref{ph} for a schematic illustration) 
and may be brought to the concept of high-lying two particle
excitation (populated via two-particle transfer reactions).
In this way the giant mode acquires its collectivity from the
coherent superposition of all the possible two-particle configurations.
This mode is called Giant Pairing Vibration \index{subject}{Giant Pairing 
Vibration} and is created by the action of the
pairing field in the very same way in which low-lying 0$^+$ pairing vibrations 
are encountered in the excitation of closed shell nuclei and their
vicinity. The analogy between shape rotations and vibrations and pairing 
rotations and vibrations may be carried quite far. In closed shell nuclei
strongly enhanced $L=0$ transitions manifest themselves following a vibrational
pattern, in which a pair of transferred particle change the number of phonons
by one, and two types of phonons are present because 
two nucleons can either be added or removed). 
In mid-shell nuclei a series of ground state transitions is seen 
between monopole states that follows a rotational scheme.
While the low-lying pairing states have been sistematically seen, the
corresponding giant mode is still awaiting an experimental confirmation.
Certainly these states are embedded in the continuum and the large background 
produced by other states, with all possible
multipolarities, makes difficult their identification in experiments.
\index{subject}{pair field}

\begin{figure}[!t]
\begin{center}
\epsfig{file=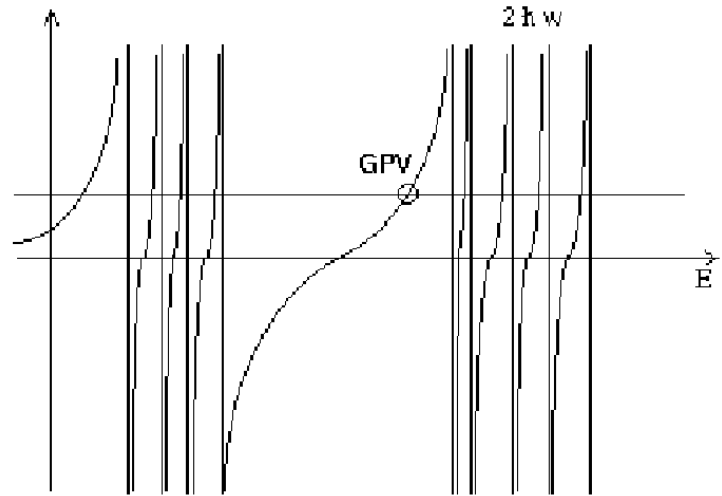,width=0.8\textwidth}
\end{center} 
\caption{Raw picture of the dispersion relation. The two bunches 
of vertical lines represent the unperturbed energy of a pair of
particles placed in a given single particle energy level.
The graphical solutions of the secular equation are the intersection of the 
horizontal line ($1/G$) with the curves. The GPV is the collective 
state relative to the second major shell.}
\label{fig1gpv}
\end{figure}

Large efforts have been recently dedicated to the study of
different aspects of reaction mechanism in collisions induced
by weakly-bound radioactive beams.  The long tails
of the one-particle transfer form factors due to the
weak binding, associated with the possibility of unusual
behaviour of pairing interaction in diluted systems, has raised
novel interest in the possibility of studying the pair field via
two-particle
transfer processes with unstable beams \cite{VOV}. On the other hand,
in transfer reactions induced by weakly bound projectiles 
on stable targets, the Q-values for the low-lying states will
be very large (typically of the order of 10-15 MeV
for the ($^{6}$He,$^{4}$He) stripping reaction).  This will strongly 
hinder these processes \index{subject}{$^6$He}
for reactions where the semi-classical optimum matching 
conditions apply, as it is the case of bombarding energies 
around the Coulomb barrier on heavy target nuclei. Higher bombarding 
energies, where the matching conditions are less stringent, may
on the other hand not be suitable because of large break-up 
cross sections.  The same matching conditions will favour 
instead the population of highly excited states, as the 
Giant Pairing Vibrations (GPV), and the use of Radioactive
Ion Beams (RIB) \index{subject}{Giant Pairing Vibration}
 may therefore become instrumental in offering
the opportunity of studying nuclear structure aspects that are not
usually accessible with stable projectiles.  
These Giant Pairing Vibrations are in fact predicted \cite{BB}
to have strong collective features, but their observation
may have so far failed \cite{WvO} because of large mismatch in reactions 
induced by protons or tritons, at variance to the case of
the low-lying pairing vibrations, which  
have been intensively and successfully studied around 
closed shell nuclei in two-particle transfer reactions \cite{BM}. 
All these 0$^+$
states are associated with vibrations of
the Fermi surface and are described
in a microscopic basis of the shell model as correlated 
two particle- two hole
states.  In the case of the Giant Pairing Vibrations the excitation
involves the promotion of a pair of particles (or holes) in the next
major shell (hence an excitation energy around 2$\hbar\omega$) and
is expected to display a collective pairing strength comparable
with the low-lying vibrations.  
The predicted concentration of strength of a $L=0$ character 
in the high-energy region (8-15 MeV for most nuclei) is understood 
microscopically as the coherent superposition of 2p (or 2h) states in the 
next major shell above the Fermi level. 
With a Random Phase Approximation in mind (or even with a simpler 
Tamm-Dancoff approximation), one may solve the secular problem for
an hamiltonian consisting simply of a kinetic operator. All the possible 
energies obtained by placing two particles in the obtained single-particle 
energy level scheme  may be called unperturbed energies. Once a pairing 
interaction (with constant strength, to fix the ideas) is added to this 
hamiltonian the solution of the secular equation may be drawn and the 
corresponding dispersion relation may be depicted as in Fig. 
(\ref{fig1gpv}). The Giant Pairing Vibration is the collective mode that
is seen in the energy gap between the first and the second major shell.
 
Also in the case of superfluid \index{subject}{superfluid}
systems in an open shell the system is expected to display a collective 
high-lying state, that in this case collects its strength from the
unperturbed two-quasiparticle 0$^+$ states with energy 2$\hbar\omega$.
To investigate this possibility we made estimates of cross sections
to the Giant Pairing Vibrations in two-particle transfer reactions,
comparing the cases of bound or weakly-bound projectiles.
\index{subject}{$^{14}$C}
As examples we have considered the case of ($^{14}$C,$^{12}$C), from 
one side, and the case of ($^{6}$He,$^{4}$He) as representative of 
a reaction induced by a weakly-bound ion.  As targets, we have chosen
the popular cases of the lead and tin regions (so considering
both ``normal'' and ``superfluid'' nuclei).
To perform the calculation,
we will first evaluate the response to the pairing 
operator in the RPA, including both the 
low-lying and high-lying pairing vibrations.  As a following step 
we will then construct two-neutron transfer form factors, using
the ``macroscopic'' model for pair-transfer processes. Finally,
estimates of cross sections will be given using standard DWBA 
\index{subject}{DWBA}
techniques. As we will see, in the case of the stripping reaction
induced by $^{6}$He, the population of the GPV is expected to display
cross sections of the order of a millibarn, dominating over the
mismatched transition to the ground state.

This chapter is organized as follows.
In the next section we discuss the theoretical formalism used for normal
and for superfluid nuclei. In section 3.3 we recall the basics
aspects of the macroscopic form factors for two-particle transfer reactions 
and  in section 3.4 we display the results of calculations for the 
paradigmatic examples of $^{208}$Pb and $^{116}$Sn.

\section{The pairing response and the Giant Pairing Vibration.}
\index{subject}{pair field}
A simple way of displaying the amount of pairing correlations is in terms of 
the pair transfer transition densities \cite{Dal}. These are defined as matrix 
elements of the pair density operator connecting the ground state in nucleus 
$A$ with the generic $0^{+}$ state $|n\rangle$ in nucleus $A\pm 2$, namely
\begin{equation}
\delta\rho_{P} (r)= \langle n| \hat{\rho}_{P} |0\rangle,
\end{equation}
where the generalized density operator is given by
\begin{equation}
\hat{\rho}_{P} (r)= \sum_{\alpha}{\sqrt{2j+1}\over{4\pi}} R_{\alpha}(r)
R_{\alpha}(r)
([a_{\alpha}^{\dagger}a_{\alpha}^{\dagger}]_{00} + 
[a_{\alpha}a_{\alpha}]_{00}).
\end{equation}
Here $ R_{\alpha}(r)$ are the radial wave functions of the 
$\alpha = \{ nlj \}$ level and the sum runs over both particle and hole levels.
The pair transfer strength to each final state can be obtained from the 
corresponding pair transfer transition density by simple quadrature, namely
\begin{equation}
\beta_{P} = \int 4\pi r^{2} \delta\rho_{P} dr.
\end{equation} 

For normal systems around closed shell the strong L=0 transition follows a 
vibrational scheme, 
where the correlated pair of fermions (pairing phonon) change by one 
\cite{RPA}.
In this case, there are two types of phonons associated with the stripping and 
pick-up  reactions. The two-particle collective state is called "addition" 
pairing phonon while the two-holes correlated state is known as "removal" 
pairing phonon. From a microscopic point of view the two kind of 
phonons, corresponding to the $(A\pm 2)$ nuclei 
can be described in terms of the two-particle (two-hole) states of the
Tamm Dancoff Approximation(TDA) or in a better way by a Random Phase 
Approximation (RPA). \index{subject}{RPA}
We start from a hamiltonian with a Monopole Pairing interaction:
\begin{equation}
H=\sum_{j} \epsilon_{j} a^{\dagger}_{j}a_{j} - G4\pi P^{\dagger}P,
\label{p1}
\end{equation}
where
\begin{equation}
P^{\dagger}=\sum_{j_{1} \le j_{2}} {M(j_{1},j_{2}) \over 
\sqrt{1+\delta_{j_1j_2} }}
\left[ a^{\dagger}_{j_{1}}a^{\dagger}_{j_{2}}\right]_{00}.
\label{p2}
\end{equation}
Here the $a^{\dagger}_{j}$ creates a particle in an orbital $j$, where $j$ 
stands for all the needed quantum numbers of the level. 
The constant $G$ is the strength of the pairing interaction and the 
coefficients 
$M(j_{1},j_{2})$ are:
\begin{equation}
M(j_{1},j_{2})={\langle j_{1}||f(r) Y_{00}(\theta,\phi)||j_{2}\rangle \over
\sqrt{1+\delta_{j_1j_2}}},
\label{p3}
\end{equation}
where the detailed radial dependence of $f(r)$ is taken to
be of the form $r^L$ and in our case is a constant since we are dealing 
only with $L=0$ states.
The pairing phonons, that is to say the quanta associated with the two 
collective modes of the above hamiltonian, are defined for closed shell 
nuclei as:
\begin{equation}
\begin{array}{lll}
|n,2p\rangle = \Gamma_{n,2p}^{\dagger}|0\rangle_{RPA} =&~&~ \\
\qquad = \Bigl( \sum_{k} X_{n}(k) [a^{\dagger}_k a^{\dagger}_k]_{00} + 
\sum_{i} Y_{n}(i) [a_i^{\dagger} a_i^{\dagger}]_{00}\Bigr) 
\mid 0 \rangle_{RPA} &~&~\\ &&\nonumber\\
|n,2h\rangle = \Gamma_{n,2h}^{\dagger}|0\rangle_{RPA} =
\Gamma_{n,2p}|0\rangle_{RPA} &~&~
 \nonumber\\
\qquad = \Bigl( \sum_{i} X_{n}(i) [a_i a_i ]_{00} +  
\sum_{k}Y_{n}(k) [a_k a_k ]_{00}\Bigr) \mid 0 \rangle_{RPA}&~&~ ,\\
\label{p4}
\end{array}
\end{equation}
where $k (i)$ stands for levels above (below) the Fermi level. 
The index $j$ runs over both particle and hole levels.
We have indicated with $\mid 0\rangle_{RPA}$ the correlated RPA vacuum. 
It represents the ground state with respect to the boson annihilation
operator $\Gamma_{n,2h}^\dagger \mid 0 \rangle_{RPA} =0$.
The definitions of $X_n$ and $Y_n$ (called
forward and backward amplitudes) are the standard ones and
come from the solution of the RPA equation. 
They may be found in \cite{RPA}. Within this model the pair transfer 
strength associated with each RPA state is microscopically given by
\begin{equation}
\beta_{Pn} = \sum_{j} \sqrt{2j+1} [X_{n}(j) + Y_{n}(j)] .
\label{p5}
\end{equation}
In Fig. 1a we display the predicted pairing response in the case of 
$^{206}$Pb, 
namely two-neutron holes with respect to the double magic $^{208}$Pb.
The set of single-particle levels that has been used in the RPA calculation, 
was obtained using the 
spherical harmonic oscillator levels with corrections due to the 
centrifugal and spin-orbit interactions \cite{MJ}
\begin{equation}
{E\over{\hbar\omega}}= N+{3\over 2} - \mu \left(l(l+1)-{{N(N+3)}
\over 2}\right) + K 
\end{equation}
$$ K = \left\{  \begin{array}{ll} 
-\kappa l & \mbox{  for  } j=l+1/2 \\
-\kappa(l+1) & \mbox{  for  } j=l-1/2
\end{array} \right. ,$$
where $\hbar\omega=41 A^{-{1\over 3}}$, $A$ is the mass number of the
nucleus, $N$ is the principal quantum number and $ j, l$ are the
total and orbital angular momentum quantum numbers, respectively. The
quantities $\kappa$ and $\mu$ are parameters chosen to obtain the best 
fit for each nucleus \cite{Da}.
We have included in the calculation all  the single-particle levels starting 
from $N=0$ up to 10. 
This set is expected to be good enough for our calculation of the Giant 
Pairing Resonance, except for the levels around the Fermi surface. 
In the lead region we prefer to use experimental values for the 
shells just above and below the Fermi surface \cite{GLN}.
The Figure shows, in addition to the strong collectivity associated with the 
ground state transition, a strong collective state with about half of the 
g.s. strength at high excitation energy, around 16 MeV, which can be 
interpreted as the Giant Pairing Vibration. Similar situation is shown in 
Fig. 1b for the corresponding two-neutron addition states in the $^{210}$Pb.
Again one may interpret the strength at about 12 MeV as associated 
with the giant mode. 
Note that in both addition and removal cases,
the contribution of the backward amplitudes to the wavefunction is found 
to be roughly equivalent to 5-10\%  in the ground state, while in the GPV
this contribution reduces to less than 1\%.
\begin{figure}[!t]
\begin{center}
\epsfig{file=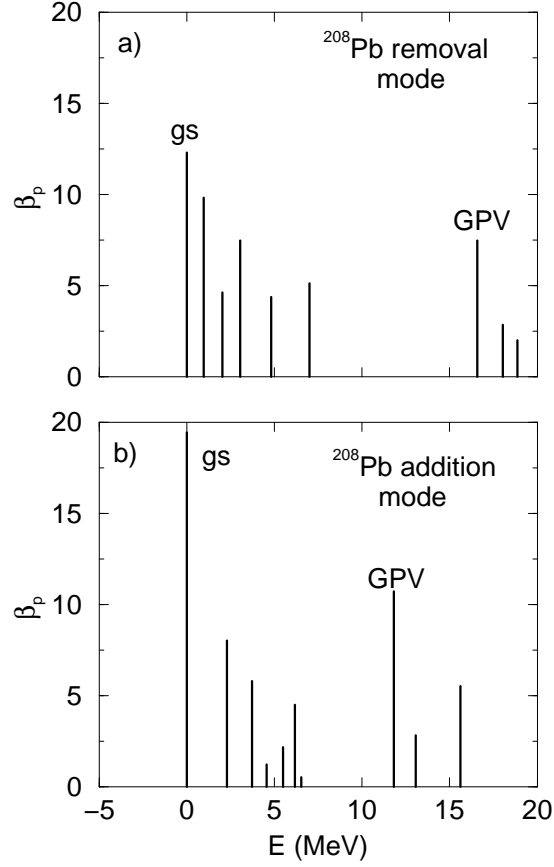,width=.6\textwidth}
\end{center}
\caption{ Pairing response for the removal (upper panel) and addition 
(lower panel) mode in $^{208}$Pb. The ground state transition and the 
candidate for GPV are evidenced.}
\end{figure}

We consider now the case of superfluid spherical-nuclei. 
In this case we make a \index{subject}{BCS}  \index{subject}{superfluid}
BCS transformation of the hamiltonian defined in Eq. [\ref{p1}] changing from 
particle to quasiparticle operators, introducing the usual occupation 
parameters. We start from a single-quasiparticle  
Hamiltonian plus a two-quasiparticle interaction corresponding to the residual
$H_{22} + H_{40}$ of the pairing force
$$
H = \sum_{j} E_{j} \alpha^{\dagger}_{j} \alpha_{j} +
2 \pi G \sum_{j_{1}j_{2}} M(j_1,j_1) M(j_2,j_2) \cdot $$
$$
\cdot \Bigl\{ (U^{2}_{j_{1}} U^{2}_{j_{2}} + V^{2}_{j_{1}} V^{2}_{j_{2}})
[\alpha^{\dagger}_{j_{1}}\alpha^{\dagger}_{j_{1}}]_{00} [\alpha_{j_{2}}
\alpha_{j_{2}}]_{00}-  U^{2}_{j_{1}} V^{2}_{j_{2}}
[\alpha^{\dagger}_{j_{1}}\alpha^{\dagger}_{j_{1}}]_{00}
[\alpha^{\dagger}_{j_{2}}\alpha^{\dagger}_{j_{2}}]_{00} $$
\begin{equation}
-  V^{2}_{j_{1}} U^{2}_{j_{2}}[\alpha_{j_{1}}\alpha_{j_{1}}]_{00} 
[\alpha_{j_{2}}\alpha_{j_{2}}]_{00}\Bigr\} ,
\label{p7}
\end{equation}
where
\begin{eqnarray}
\alpha^{\dagger}_{j} &=& U_{j} a^{\dagger}_{j} - V_{j} a_{\bar j}\\
U^{2}_{j} &=& {1\over {2}} \left(1 + 
{{\tilde \epsilon_{j}}\over E_{j}}\right)\\
V^{2}_{j} &=& {1\over 2} \left(1 - {{\tilde \epsilon_{j}}\over E_{j}}\right).
\label{p8}
\end{eqnarray}
The energies $E_{j} = \sqrt{{\tilde \epsilon}_{j}^{2} + \Delta^{2}}$ are the 
quasi-particle energies, and ${\tilde \epsilon_{j}} = \epsilon -\lambda$ 
are the single-particle energies with respect to the chemical potential
$\lambda$  and $\Delta$ is the BCS gap. 
As usual we have defined $a_{\bar j} \equiv a_{\bar{jm}}=(-1)^{j-m} a_{j,-m}$.
 
For superfluid systems the addition and removal RPA phonons cannot be treated 
separately. The dispersion relation, that relates the strength of the 
interaction with the energy-roots of the RPA, becomes a two by two determinant.
From the RPA equations:
\begin{equation}
\Gamma^{\dagger}_{n} ~=~ \sum_{j} \left( X_{n}(j) [\alpha^{\dagger}_{j} 
\alpha^{\dagger}_{j}]_{00} + Y_{n}(j)[\alpha_{j}\alpha_{j}]_{00}\right)
\label{p9uno}
\end{equation}
\begin{equation}
 \left[ H,\Gamma^{\dagger}_{n}\right] ~=~ \omega_{n} \Gamma^{\dagger}_{n},
\label{p9}
\end{equation}
\begin{equation}
\left[ \Gamma_{n'},\Gamma^{\dagger}_{n}\right] ~=~ \delta_{nn'}
\end{equation}
we can obtain the following factors
\begin{eqnarray}
x &=& \sum_{j_{1} \le j_{2}} |M(j_{1}j_{2})|^{2} \left[\frac{U^{2}_{j_{1}} 
U^{2}_{j_{2}}}{E_{j_{1}}+E_{j_{2}}-\omega_{n}} + \frac{V^{2}_{j_{1}} 
V^{2}_{j_{2}}}{E_{j_{1}}+E_{j_{2}}+\omega_{n}}\right]\nonumber \\
~\\
y &=& \sum_{j_{1} \le j_{2}} |M(j_{1}j_{2})|^{2} \left[\frac{V^{2}_{j_{1}}
V^{2}_{j_{2}}}{E_{j_{1}}+E_{j_{2}}-\omega_{n}} + \frac{U^{2}_{j_{1}}
U^{2}_{j_{2}}}{E_{j_{1}}+E_{j_{2}}+\omega_{n}}\right] \nonumber \\
~\\
z &=& \sum_{j_{1} \le j_{2}} |M(j_{1}j_{2})|^{2} (U_{j_{1}}V_{j_{2}}
U_{j_{2}}V_{j_{1}}) \nonumber \\
~&~& \left[\frac{1}{E_{j_{1}}+E_{j_{2}}-\omega_{n}} + 
\frac{1}{E_{j_{1}}+E_{j_{2}}+\omega_{n}}\right] ,
\label{p10}
\end{eqnarray}
and the dispersion relation is in this case:
\begin{equation}
\left|\matrix{(1 - 4 \pi G x) &4 \pi G z\cr 4 \pi G z 
&(1 - 4 \pi G y)\cr}\right| = 0 .
\label{p11}
\end{equation} 
From this determinant the following relation is obtained 
\begin{equation}
4\pi G\= \Bigl[ {(x+y)\over 2}\pm \sqrt{{(x+y)^2\over 4}+z^2} \Bigr]{1\over 
(xy-z^2)}
\end{equation}
It can be shown that $\omega = 0$ is solution of that equation and correspond 
to the Goldstone boson corresponding to the breaking of the number of particle 
symmetry. 
Once we have obtained the energies $\omega_n$ of the different RPA roots, 
we can write the components of the RPA phonon in the form:
\begin{eqnarray}
 X_{n}(j,j) &=& \frac{4\pi G M(j,j)}{E_{j}+E_{j}-\omega_{n}}
\left(U_{j}^{2} + V_{j}^{2}\frac{4 \pi G z}{(1-4 \pi G y)}\right)
\Lambda_n \nonumber\\
 Y_{n}(j,j) &=& \frac{4\pi G M(j,j)}{E_{j}+E_{j}+\omega_{n}}
\left(U_{j}^{2}\frac{4 \pi G z}{(1-4 \pi G y)} + V_{j}^{2}\right)\Lambda_n ,
\nonumber \\
~
\label{p12}
\end{eqnarray}
where $\Lambda_n$ is determined by normalizing the phonon corresponding to the
$n-$th root of the RPA. The normalization condition reads
\begin{equation}
\sum_{j } [X_{n}^{2}(j) -Y_{n}^{2}(j)] = 1 .
\label{p13}
\end{equation}
Finally, we can obtain for each state $n$ the pairing strength parameter 
$\beta_{P}$ with the following formulae:
\begin{eqnarray}
\beta_{P}(2p) = \sum_{j} \sqrt{2j+1} \langle n|[a^{\dagger}_{j}
a^{\dagger}_{j}]_{00}|0\rangle = \nonumber \\
= \sum_{j} \sqrt{2j+1} [U^{2}_{j} X_{n}(j) + V^{2}_{j}Y_{n}(j)]\nonumber ,\\
\beta_{P}(2h) = \sum_{j} \sqrt{2j+1} \langle n|[a_{j}a_{j}]_{00}|0\rangle 
= & \nonumber \\ 
=\sum_{j} \sqrt{2j+1} [V^{2}_{j} X_{n}(j) + U^{2}_{j}Y_{n}(j)] .
\label{p14}
\end{eqnarray}
From the two equations above one recovers the four contribution to 
formula (\ref{p5}) by putting $U=0$ and $V=1$ when $j$ is below the Fermi level
and by putting $U=1$ and $V=0$ when $j$ is above.
The predictions of the pairing strength distribution for the superfluid system 
$^{116}$Sn are shown in the two panels of Fig. 2.
For the calculation we have used the single-particle levels from Ref. 
\cite{US}. 
These last ones have been proved to give good results in BCS calculations 
using 
a pairing strength $G=g/A$, where $g\simeq 20 MeV$. We assume that the rest of 
the levels have occupation probability 1(0) if they are far below(above) the
Fermi surface. The change of the single particle energies around the Fermi 
surface has been done, in both cases, taking care of keeping the 
energy-centroids of the exchanged levels in the same position.  
The figure clearly shows the occurrence of high-lying strength which can be 
associated with the Giant Pairing Vibration. Note that,with respect to the 
case of $^{208}$Pb, there is a minor fragmentation of the strength both in 
the low-lying and in the high-lying energy region.
\begin{figure}[!t]
\begin{center}
\epsfig{file=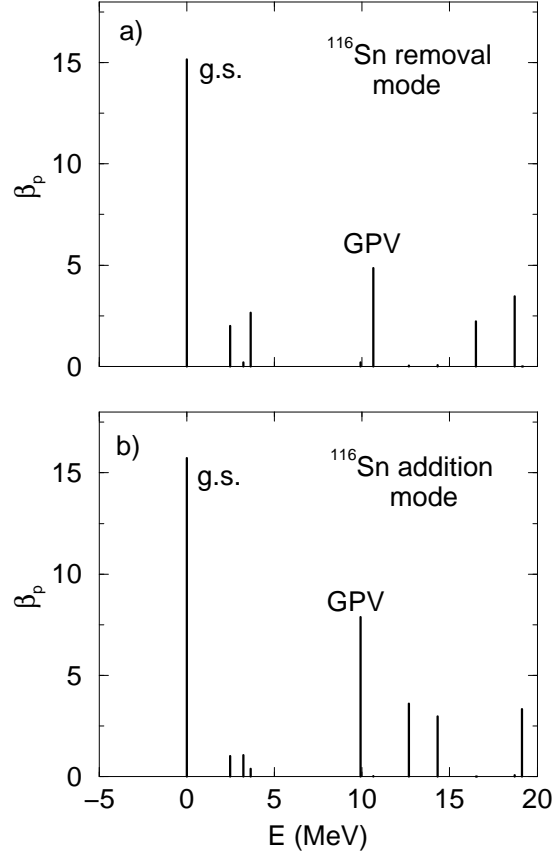,width=.6\textwidth}
\end{center}
\caption{Pairing response for the removal (upper panel) and addition 
(lower panel) mode in $^{116}$Sn. The ground state transition and the 
candidate for GPV are evidenced.}
\end{figure}

 We also report in Fig. (3) 
a number of analogous results for other commonly studied targets 
with the aim of giving some indications to experimentalists on the reasons 
why we think that lead and tin are some of the most promising candidates.
We have studied two isotopes of calcium with closed shells.
Even if the absolute magnitudes of the $\beta_{\rm P}$ is lower,
it is worthwhile noticing that some enhancement is seen in the more 
neutron-rich $^{48}$Ca with respect to $^{40}$Ca. \index{subject}{$^{40}$Ca}
\index{subject}{$^{48}$Ca}
An important role in this change is certainly due to the different shell 
structure of the two nuclei as well as to the scheme that we implemented 
to obtain the set of single particle levels. The latter is responsible
for the collectivity of the removal modes in both Ca isotopes and also 
for the difficulty in finding out a collective state in the addition modes.  
We display also results for $^{90}$Zr
\index{subject}{$^{90}$Zr} where the strength is much more 
fragmented and the identification of the GPV is more difficult.
In the work of Broglia and Bes estimates for the energy of the pairing 
resonance are given as $68/A^{1/3}$ MeV and $72/A^{1/3}$ MeV for normal
and superfluid systems respectively. Our figures follow roughly 
these prescriptions based on simple arguments (and much more grounded
in the case of normal nuclei) as evident from Table \ref{tabe1}.
  
\begin{table}[!h]
\begin{center}
\begin{tabular}{|c|c|c|}
\hline
\raisebox{0pt}[13pt][7pt]{Nucleus} & Our calculation  & Broglia \& Bes 
estimate \\ 
\hline
Sn & 12.68 MeV & 14.76 MeV \\
Pb & 11.81 MeV & 11.47 MeV \\
\hline
\end{tabular}
\end{center}
\caption{Comparison of position of GPV between our calculation and the Broglia
and Bes estimate.} 
\label{tabe1}
\end{table}

\begin{figure}[!h]
\begin{center}
\epsfig{file=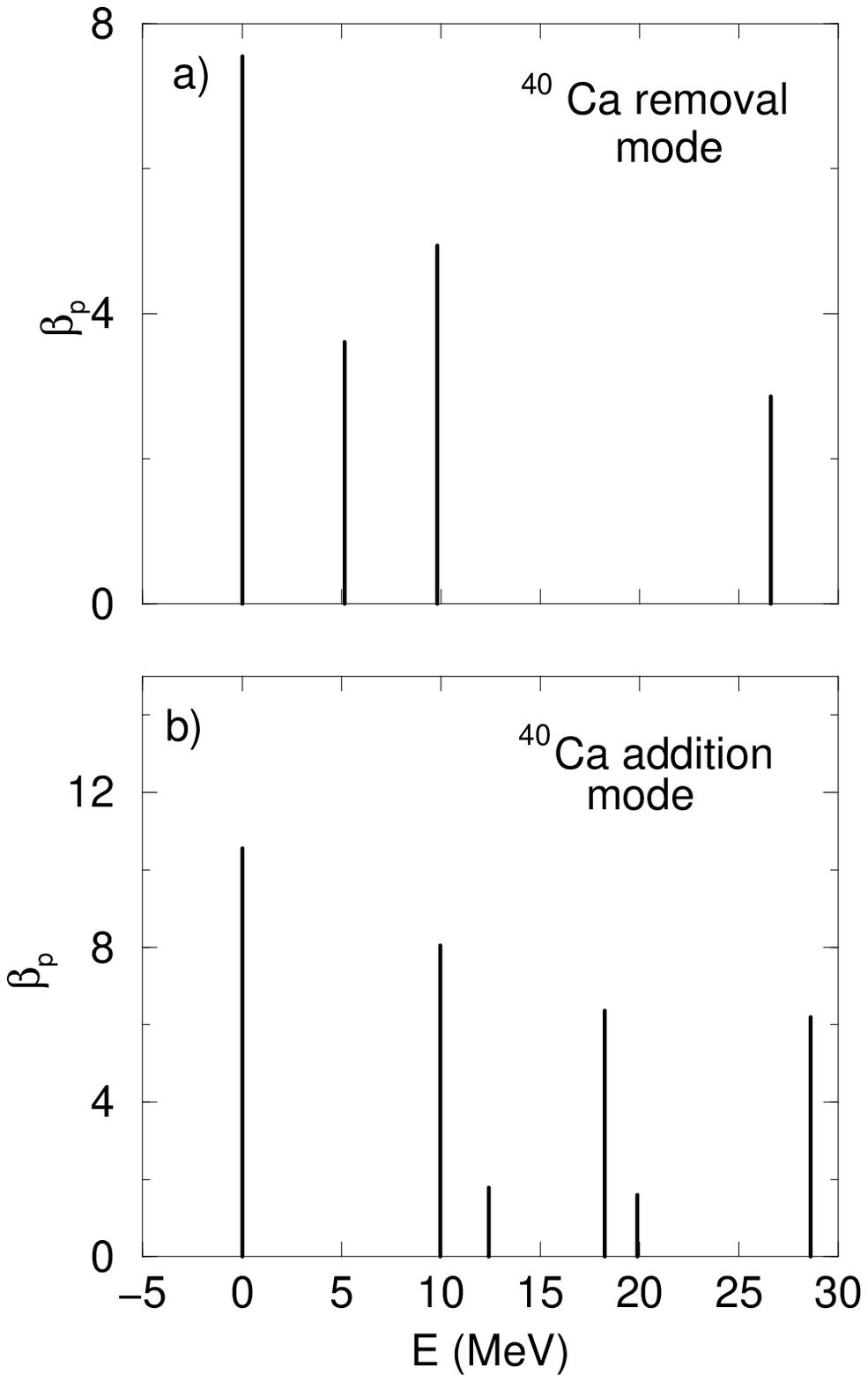,width=0.43\textwidth}
\epsfig{file=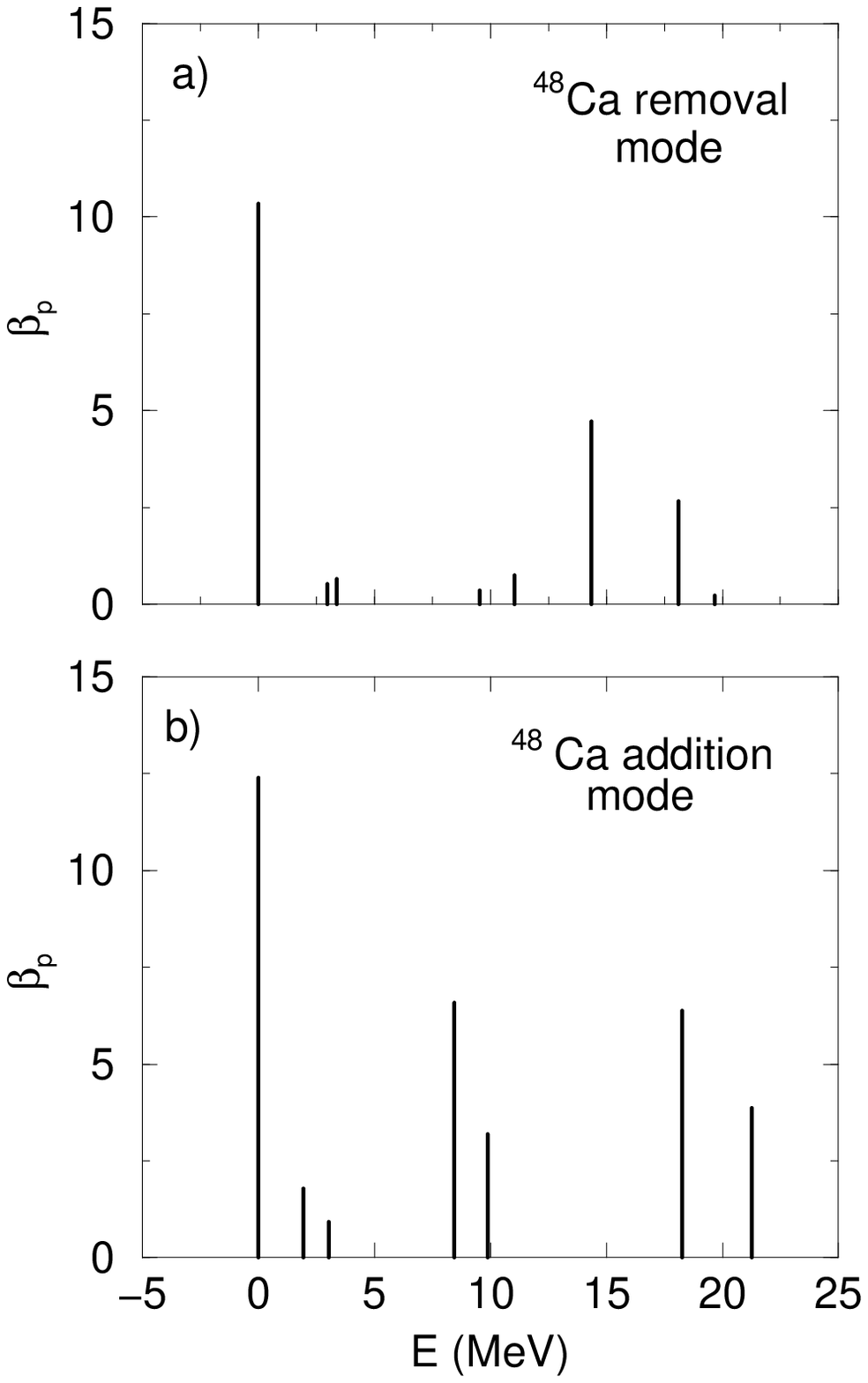,width=0.43\textwidth}
\epsfig{file=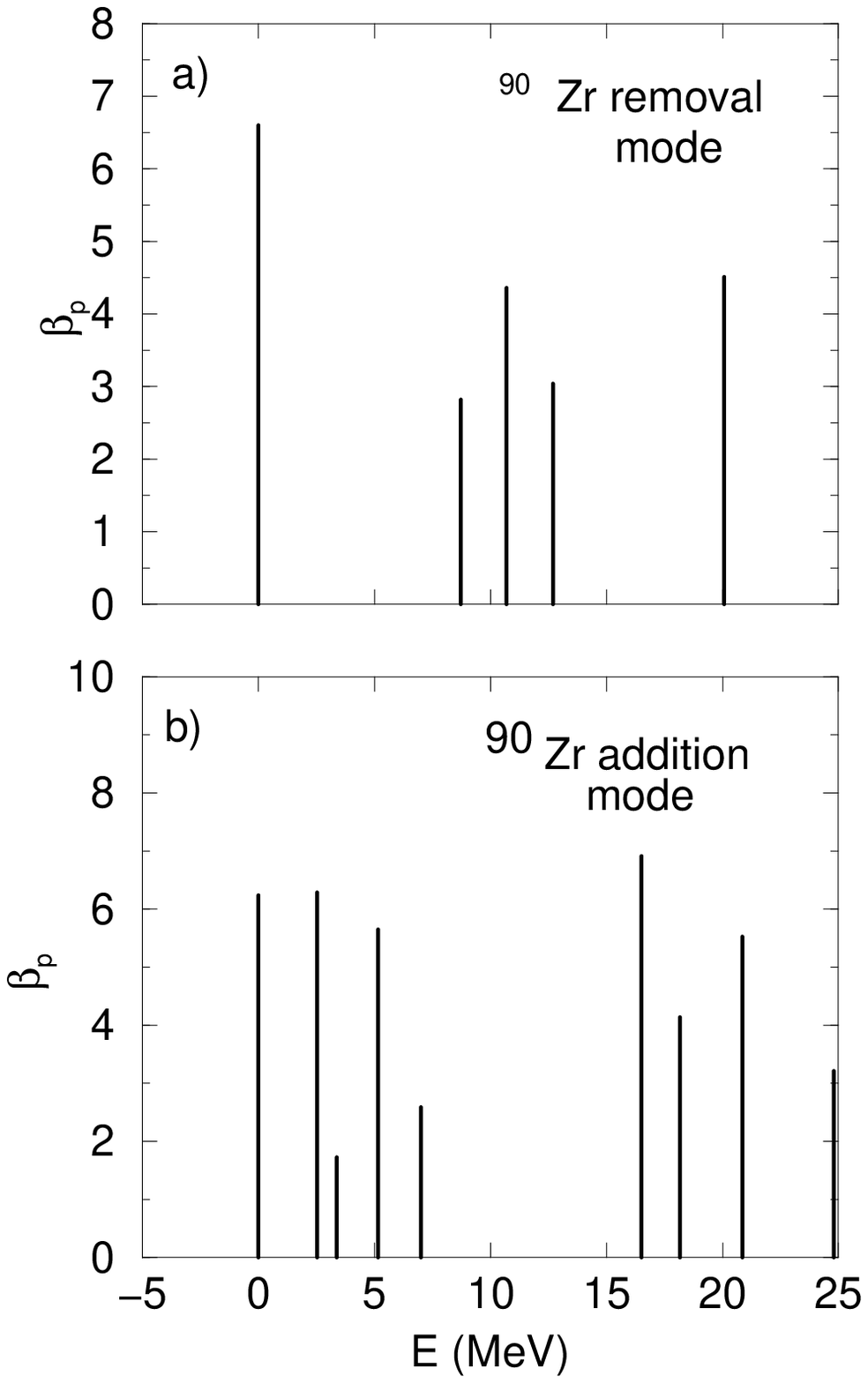,width=0.43\textwidth}
\end{center}
\caption{Pairing response for removal and addition mode in $^{40}$Ca,
 $^{48}$Ca and $^{90}$Zr.}
\label{fig3gpv}
\end{figure}

\subsection{Energy-weighted sum rule} \index{subject}{energy-weighted sum 
rule (pairing)}
Before turning to macroscopic model we want to remind that some
attempts to introduce sum rules for two-particle transfer reactions have
been tried until the formulation of sum rules in terms of elementary modes
of excitation of the target alone \cite{DaLi}. Introducing the operator
\be
\hat F^\dag \= \sum \langle \alpha \mid F \mid \alpha \rangle 
\bigl[ a_{\alpha}^\dag,a_{\alpha}^\dag \bigr]_0
\ee
and its hermitian conjugate, where $F=\sum f(r_k)Y_{00}(\theta_k,\phi_k)$
is a particle transfer monopole field, we can find an energy-weighted sum 
rule from the expectation value of a double commutator of the hamiltonian 
with $\hat F\= (\hat F+\hat F^\dag)/2$ (an hermitian combination of the 
operators above that conserves the number of particles only as an average).
The sum rule reads:
$$
S\= 2\langle 0\mid \bigl[\hat F_H,\bigl[\hat H,\hat F_H \bigr] \bigr]
\mid 0\rangle \=
$$
$$
\= \sum(E_{n+}-E_0)\mid \langle n_+\mid \hat F^\dag \mid 0 \rangle \mid^2+
$$
\be
+\sum(E_{n-}-E_0)\mid \langle n_-\mid \hat F^\dag \mid 0 \rangle \mid^2
\ee
where $E_{n\pm}$ are the energies of the states in the systems with
mass $(A\pm 2)$ and $E_0$ is the energy of the reference state in the 
starting $A$ system.

\index{subject}{macroscopic model for \\2n transfer}
\section{Macroscopic form factors for two-particle \\transfer reactions.}
\begin{figure}[!b]
\begin{center}
\includegraphics[width=0.7\textwidth]{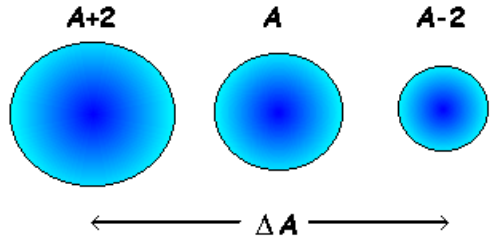}
\end{center}
\caption{Pictorial representation of an oscillation in mass
between different $L=0$ states. The macroscopic variable is the difference
in mass from the initial mass partition.}
\label{DA}
\end{figure}

The description of the reaction mechanism associated with the transfer 
of a pair of particles in heavy ion reactions has always been a rather 
complex issue.  In the limit in which the field responsible for the 
transfer process is the one-body field generated by one of the partners 
of the reactions, at least for simple configurations the leading order 
process is the successive transfer of single particles.  
In this framework the collective  
features induced by the pairing interaction arise from the coherence 
of different paths in the intermediate (A+1 , A--1) channel due to the 
correlation present in the final (A+2) and (A--2) states.  The actual 
implementation of such a scheme may turn out not to be a simple task, 
due to the large number of active intermediate states, and the use of 
a simpler approach may be desirable.  This is offered, for example, 
by the ``macroscopic model'' for two-particle transfer reactions, that 
parallels the formalism used to describe the inelastic excitation  
of collective surface modes. \index{subject}{macroscopic model for pairing}
The starting point of the 'macroscopic model' for two particle transfer
reactions is to push further the analogy of the vibrations of the nuclear
surface
with the 'vibrations' across different mass partitions. If one imagine an
idealized space in which a discrete coordinate (the number of particles of
the system) labels different sections of the space, it is plausible to give
an interpretation of pairing modes as back and forth oscillations
in the number of particles, as in Fig. \ref{DA}.
The role of macroscopic variable in this game
is played by the quantity $\Delta A$, that is the difference in mass from
the initial mass partition. 
The fundamental idea of the macroscopic model for the inelastic excitations 
is to interpret the superposition of particle-hole excitations as representing
 a state of collective motion in which the systems deviates from its spherical
equilibrium shape. 
In that case, as an alternative to the (more correct) 
microscopic description based on a superposition of particle-hole 
excitations, one has traditionally resorted to collective form factors  
of the form \cite{Sch} 
\begin{equation} 
F_{\lambda}(r)~=~\beta_{\lambda}R{dU \over dr} ,
\end{equation} 
in terms of the radial variation of 
the ion-ion optical potential U induced by the surface vibrations,  
with the strength parameter 
$\beta_{\lambda}$ obtained from the strength of the B(E$\lambda$)  
transition.  
To generalize these concepts to the pair transfer processes we need to
fulfill a number of important requirements. A couple of generalized 
particle-particle transition densities must be introduced to deal with 
addition and removal reactions ($\delta \rho^+_P$ and $\delta \rho^-_P$).
An interpretation of these in terms of operator with a one-body character 
should be given in order to be effective.

In the case of the pair transfer, the   
vibration is the fluctuation of the Fermi surface with respect to the 
change in the number A of particles, and the corresponding form 
factor $F_{P}$ is assumed to have the parallel form \cite{Dal}
\begin{equation} 
F_{P}(r)~=~\beta_{P}{dU \over dA} ,
\end{equation} 
in terms of the ``pairing deformation'' parameter $\beta_{P}$  
\index{subject}{pairing deformation}
associated with that particular transition, defined in the previous  
section. The assumption of simple scaling law between nuclear radius R and  
mass number A allows to rewrite the two-particle transfer form 
factor into an expression which is formally equivalent to the one  
for inelastic excitation, namely 
\begin{equation} 
F_{P}(r)~=~{\beta_{P} \over 3A}~R~{dU \over dr} .
\end{equation} 
This formalism has been successfully applied to quite a number of two-particle 
transfer reactions \cite{DP,DV}.  As in the case of inelastic excitations, 
macroscopic 
collective form-factors may in some cases only give a rough estimate 
to the data, requiring more elaborate microscopic descriptions.  Nonetheless, 
the use of simple macroscopic form factors is of unquestionable usefulness 
in making predictions, in particular in cases, as the one we are discussing,  
where experimental data are not yet available and estimates are needed in 
order to plan future experiments.

\section{Applications: estimates of two-neutron \\transfer cross 
sections.} 
 
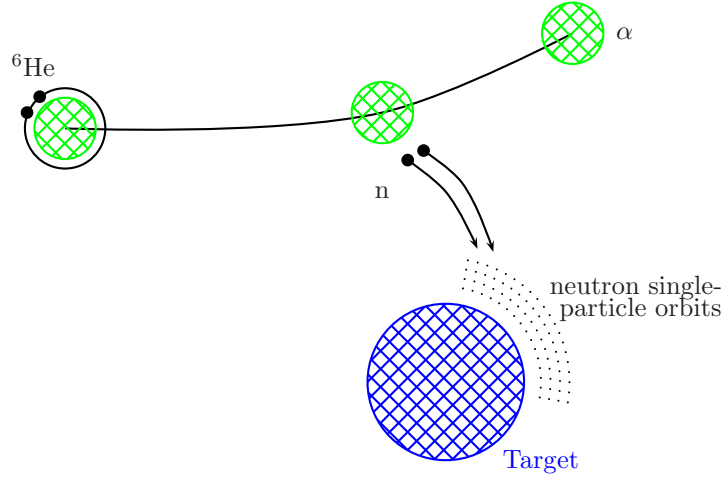
\begin{figure}[!h]
\begin{flushright}
\begin{picture}(250,180)(40,20)
\psset{unit=1.2pt}     
\rput(170,5){\blue Target}
\pscircle[linecolor=blue,fillstyle=crosshatch,hatchcolor=blue](140,30){25}
\pscurve(20,110)(120,115)(180,140)
\psarc[linestyle=dotted](140,30){30}{350}{80}
\psarc[linestyle=dotted](140,30){36}{350}{80}
\psarc[linestyle=dotted](140,30){33}{350}{80}
\psarc[linestyle=dotted](140,30){39}{350}{80}
\rput(200,60){neutron single-}
\rput(200,53){ particle orbits}
\rput(10,130){$^6$He}
\pscircle[linecolor=green,fillstyle=crosshatch,hatchcolor=green](20,110){10}
\pscircle(20,110){13}
\pscircle*(8,115){2}\pscircle*(12,120){2}
\pscircle[linecolor=green,fillstyle=crosshatch,hatchcolor=green](120,115){10}
\pscircle*(128,100){2}\pscircle*(133,103){2}
\rput(120,90){n}
\pscurve{->}(128,100)(140,90)(150,72)
\pscurve{->}(133,103)(145,93)(155,71)
\pscircle[linecolor=green,fillstyle=crosshatch,hatchcolor=green](180,140){10}
\rput(195,140){ $\alpha$}
\end{picture}
\vspace{5mm}
\end{flushright}   
\caption{Scheme of the two neutron transfer reaction ($^6$He,$^4$He).
The neutrons are transferred to single-particle orbits of the target.}
\label{reactsch}
\end{figure}

In order to evidence the possible role of unstable beams in the study 
of high-lying pairing states, we compare in this section two-particle transfer
reactions induced either by a traditionally available beam (e.g. the 
($^{14}$C, $^{12}$C)) or by a more exotic beam (e.g. the reaction 
($^{6}$He, $^{4}$He)).  As a target, we have considered the two cases of 
$^{208}$Pb and $^{116}$Sn, as representative cases of normal and superfluid  
systems in the pairing channels.  
\index{subject}{$^{14}$C}\index{subject}{$^6$He}
\index{subject}{$^{208}$Pb}\index{subject}{$^{116}$Sn}
A typical reaction scheme is shown in fig. \ref{reactsch} where 
in a pictorial way the phenomenon is illustrated. During the process the 
neutrons are transferred from the projectile to neutron single-particle states
of the target, leaving an $\alpha$ particle (or $^{12}C$) 
in the exit channel. \index{subject}{matching factor}

For the semiclassical description to be valid, the Sommerfeld parameter needs
to be larger tan unity. This is certainly the case here where the product
of the charges of the colliding nuclei are big and the c.m. energy is around
the Coulomb barrier. In this case the transfer cross-section may be 
factorized in the product of the scattering cross-section, of the transfer 
probability 
and of a quantal correction factor. This correction factor, or matching 
factor, is important whenever the orbits of the initial and final systems 
(or the donor and acceptor, in the case of a transfer process) have 
differences in the variables that characterize the scattering orbit
\cite{BPW,vOe2,VOV}. The matching factor may be written in the following way
\be
F(Q,L)\= exp \bigl[ -c_1^2(\Delta Q-c_2\Delta L) \bigr]
\ee
where the energy matching condition and the tranferred momentum 
matching condition are taken into account via $\Delta Q$ and
$\Delta L$, respectively. Since here we are dealing with transitions
between monopole states the transferred angular momentum is zero, 
$\Delta L \=0$, and thus the highest transition 
cross-section will be obtained if the quantal correction factor is the
higest possible (equal to unity) and therefore the optimum Q-value for this 
kind of process is $\Delta Q\=0$.

In both cases, we have considered the full 
pairing L=0 response, e.g. all transitions to 0$^+$ states in 
$^{210}$Pb and $^{118}$Sn, as described in Sect 2.
The Q-values corresponding to the transitions to the ground-states and to the 
GPV states are displayed in Table \ref{tabe2}.
\begin{table}[!h]
\begin{center}
\begin{tabular}{|c|c|c|}
\hline
~~&~~\raisebox{0pt}[13pt][7pt]{$^{14}\mbox{C} \rightarrow ~^{12}\mbox{C}$}
~~&~~$^{6}\mbox{He} \rightarrow ^{4}\mbox{He}$\\ \hline
$^{116}\mbox{Sn} \rightarrow ~^{118}\mbox{Sn}_{gs}$~~& 3.15 MeV & 15.298 MeV\\
$^{208}\mbox{Pb} \rightarrow ~^{210}\mbox{Pb}_{gs}$~~& -4 MeV & 8.148 MeV\\
$^{116}\mbox{Sn}\rightarrow ~^{118}\mbox{Sn}_{GPV}$~~& -6.746 MeV & 5.402 MeV\\
$^{208}\mbox{Pb} \rightarrow ~^{210}\mbox{Pb}_{GPV}$~~& -15.81 MeV 
& -3.662 MeV\\
\hline
\end{tabular} 
\end{center}
\caption{Q-values for ground-state and GPV transitions. 
The target (column) and projectile (row) are specified.}
\label{tabe2}
\end{table}  
\begin{table}[!h]
\begin{center}
\begin{tabular}{|c|c|c|}
\hline
~~&~~\raisebox{0pt}[13pt][7pt]{$^{14}\mbox{C} \rightarrow ~^{12}\mbox{C}$}
~~&~~$^{6}\mbox{He} \rightarrow ^{4}\mbox{He}$\\
\hline
$^{116}\mbox{Sn} \rightarrow ~^{118}\mbox{Sn}_{gs}$~~& 19.4 mb & 0.4 mb\\
$^{208}\mbox{Pb} \rightarrow ~^{210}\mbox{Pb}_{gs}$~~& 15.3 mb & 1.8 mb\\
$^{116}\mbox{Sn}\rightarrow ~^{118}\mbox{Sn}_{GPV}$~~& 0.14 mb & 2.4 mb\\
$^{208}\mbox{Pb} \rightarrow ~^{210}\mbox{Pb}_{GPV}$~~& 0.04 mb & 3.1 mb\\
\hline
\end{tabular}
\caption{Cross-sections for ground-state and GPV transitions obtained with
the DWBA code Ptolemy. The target (column) and projectile (row) are specified.}
\end{center}
\label{ta3}
\end{table}       

\begin{figure}[!h]
\begin{picture}(280,170)(0,-10)
\psset{unit=1.2pt}     
\psline{-}(10,10)(40,10)
\psline[linestyle=dashed]{-}(10,40)(40,40)
\psline[linestyle=dotted]{-}(40,40)(60,40)
\psline[linestyle=dashed]{-}(60,40)(90,40)
\psline{<->}(25,10)(25,40)
\rput(25,-2){ $^6$He}
\rput(35,25){ $\sim$ 1 MeV}
\rput(106,40){$\alpha$ + 2n}

\psline{-}(170,10)(200,10)
\psline[linestyle=dotted]{-}(200,10)(250,10)
\psline[linestyle=dotted]{-}(260,105)(260,120)
\psline{-}(220,60)(250,60)
\psline{-}(220,100)(250,100)
\psline[linestyle=dotted]{-}(235,105)(235,120)
\psline{<->}(210,10)(210,60)
\rput(185,-2){ $^{208}$Pb}
\rput(235,-2){ $^{210}$Pb}
\rput(245,35){ some MeV}
\rput(260,60){ $0^+_{gs}$}
\rput(260,100){ $0_1^+$}
\end{picture}
\caption{Energy balance for the two subsystems. The initial state and 
the final ones are displayed.}
\label{subsystems}
\end{figure}
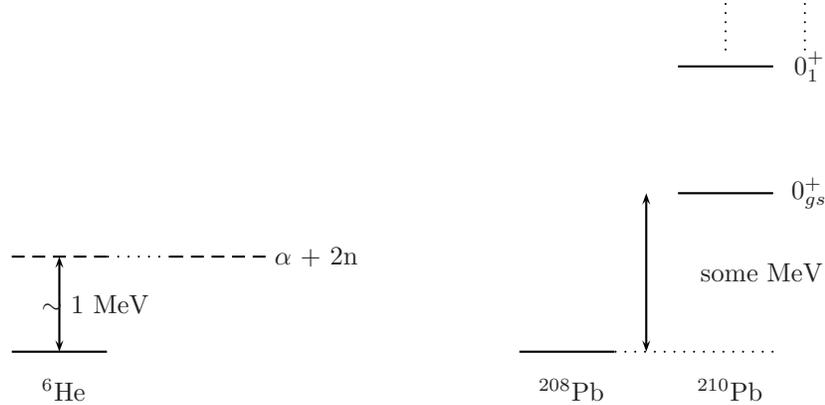

Let us consider in greater detail the energy balance in one case for 
illustative purpose. The projectile and target subsystems are displayed in fig.
\ref{subsystems} where the initial and final configurations are seen.

About 1 MeV is needed for the projectile to 'break', but the lowest 
$0^+$ states of the target are some MeV higher that the ground state. The
remaining energy needed is subtracted from the relative motion kinetic energy.
The total energy balance is depicted in the following figure \ref{totenergy} 
for completeness.
We have taken as an example a lead target but the same considerations apply to
tin or other targets, but with different Q-values as we discussed in the tables
above. Fig. \ref{totenergy} makes immediately clear that in the case of a 
conventional (well bound) projectile, the initial bound state would lie 
at much lower energy thus demanding for a greater energy to release its 
neutrons.

\begin{figure}[!h]
\begin{center}
\begin{picture}(170,220)(30,0)
\psset{unit=1.3pt}     
\psline[linecolor=blue]{-}(10,10)(40,10)
\psline[linestyle=dotted]{-}(10,40)(120,40)
\rput(25,-2){\blue $^6$He}
\rput(105,-2){\red $^{210}$Pb}
\rput(156,40){$\alpha$ + 2n + $^{208}$Pb$_{gs}$}
\rput(140,30){\blue $\underbrace{~~~~~~~~~~~~~~~}_{^6 He}$}
\rput(163,52){\red $\overbrace{~~~~~~~~~~~~~~~~~~}^{^{210}Pb}$}
\psline[linecolor=red]{-}(90,100)(120,100)
\psline[linecolor=red]{-}(90,160)(120,160)
\rput(130,100){\red $0^+_{gs}$}
\rput(130,160){\red $0_1^+$}
\psline[linecolor=green]{|->}(60,10)(60,100)
\psline[linecolor=green]{|->}(70,10)(70,160)
\end{picture}
\end{center}
\caption{Total energy balance for the complete system. The kinetic energy
borrowed from the relative motion is shown as a green arrow while the 
projectile and the target states are in blue and red.}
\label{totenergy}
\end{figure}
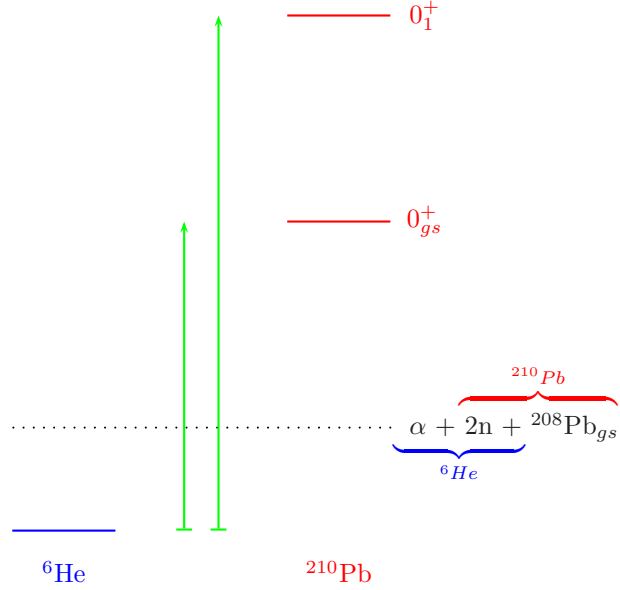

For each considered state the two-particle transfer cross section has been 
calculated on the basis of the DWBA (using the code Ptolemy \cite{COM}) 
\index{subject}{DWBA} \index{subject}{RPA}
employing the macroscopic form factor described above, with a strength 
parameter as resulting from the RPA calculation. 
For the ion-ion optical potential, the standard parameterization 
of Akyuz-Winther \cite{Aky} has been used for the real part, with an 
imaginary part with 
the same geometry and half its strength.  
In all cases, the bombarding energy has been chosen in order to correspond, 
in the center of mass frame, to about 50\% over the Coulomb barrier. 
\begin{figure}[!t]
\begin{center}
\epsfig{file=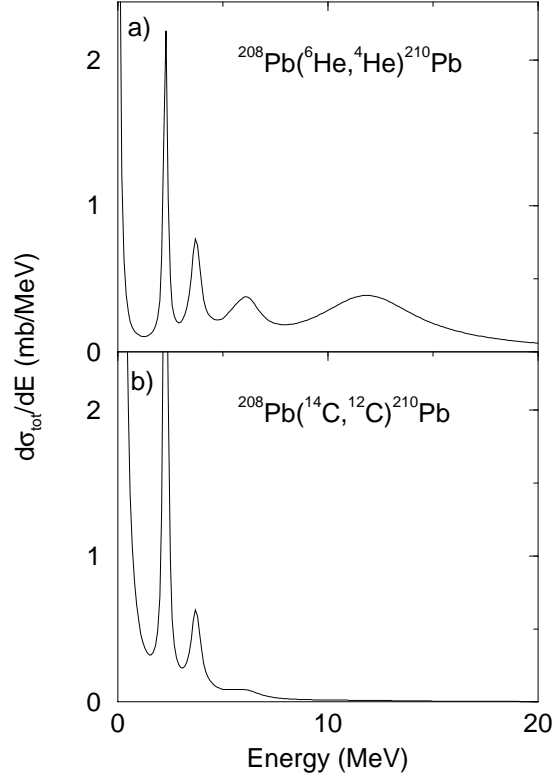,width=.6\textwidth}
\end{center}
\caption{ Differential cross-sections as a function of the excitation energy
 for the two reactions : a) 
$^{208}$Pb($^{6}$He,$^{4}$He)$^{210}$Pb,
 and b) $^{208}$Pb($^{14}$C,$^{12}$C)$^{210}$Pb. See text for details.}
\end{figure}
The angle-integrated L=0 excitation function is shown in Fig. 3b as a 
function of the excitation energy E$_x$ for the  
$^{208}$Pb($^{14}$C,$^{12}$C)$^{210}$Pb reaction at E$_{cm}$=95 MeV.  For 
a more realistic display of the results, the contribution of each 
discrete RPA state is distributed over a lorentzian with $\Gamma$= k E$_x^2$,
with $k$ adjusted to yield a width of $4 MeV$ for the giant pairing vibration.
This could seem rather arbitrary since there is no reason for an {\it a priori}
assignment of this quantity. We have been brought to this simple prescription
because other collective states (of different nature) lying in the same 
energy region display similar values for their width, and it is reasonable to
assume some rule to narrow the low-energy states and to broaden the 
high-energy ones.

As the figure shows, the large (negative) Q-value associated with the region 
of the GPV (see Table 1) completely damps its contribution, and the excitation 
function is completely dominated by the transition to the ground state and the 
other low-lying states.  The situation is very different for the 
$^{208}$Pb($^{6}$He,$^{4}$He)$^{210}$Pb reaction at E$_{cm}$=41 MeV, 
whose excitation function is shown in Fig. 3a. In this case the weak binding 
nature of $^{6}$He projectile leads to a  
mismatched (positive) Q-value for the ground-state transition  
(Q$_{gs}$= 8.148 MeV), favouring the transfer process to the high-lying 
part of the pairing response.  In this case the figure shows that, 
in spite of a smaller pairing matrix element, the 
transition to the GPV is of the same order of magnitude of the 
ground-state transfer (1.8 mb for g.s. and 3.1 mb for the GPV). 
Note that a total cross section to the GPV region of 
the order of some millibarn should be accessible with the 
new large-scale particle-gamma detection systems.
\begin{figure}[!h]
\begin{center}
\epsfig{file=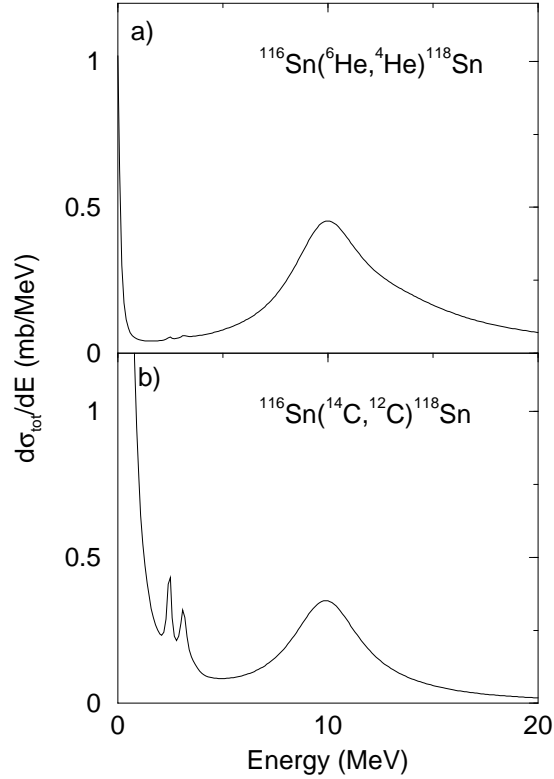,width=.6\textwidth}
\end{center}
\caption{ Differential cross-sections as a function of the excitation energy
for the two reactions : a) 
$^{116}$Sn($^{6}$He,$^{4}$He)$^{118}$Sn,
and b) $^{116}$Sn($^{14}$C,$^{12}$C)$^{118}$Sn. The comparison between
the GPV and the ground-state  clearly shows the different strength.
 Notice the different vertical scale with respect to figure 3.  }
\end{figure}

A similar behaviour is obtained in the case of a tin target.
In Figs. 4a and 4b the corresponding excitation functions for the  
 $^{116}$Sn($^{14}$C,$^{12}$C)$^{118}$Sn reaction (at E$_{cm}$=69 MeV) 
and the $^{116}$Sn($^{6}$He,$^{4}$He)$^{118}$Sn reaction (at E$_{cm}$=40 MeV) 
are compared. Now the transition to the GPV dominates over the ground-state 
transition when using an He beam ( 0.4 mb for g.s. and 2.4 mb for the GPV). 
From a comparison with the RPA strength distributions of Fig. 1 and 2 one can 
see that the giant pairing vibrations is definitely favoured by the use of an 
$^{6}$He beam instead of the more conventional $^{14}$C one, because the 
transition to the ground-state is hindered, while the GPV is enhanced 
(or not changed), because of the effect of the Q-value.

\section{Conclusions.}\index{subject}{$^6$He}
The role of radioactive ion beams for studying different features of 
the pairing
degree of freedom via two-particle transfer reactions  is 
underlined. A $^6$He beam may allow an experimental study of high-lying 
collective pairing states, that have been theoretically predicted,
but never seen in measured spectra, because of previously unfavourable 
matching conditions.
The modification in the reaction Q-value, when passing from $^{14}$C
 to $^6$He, 
that is a direct consequence of the weak-binding nature of the latter 
neutron-rich nucleus, is the reason of the enhancement of the transition 
to the giant pairing vibration with respect to the ground-state. 

The final achievements for the four reactions studied in detail are presented 
in the last two figures. It is worthwhile 
noticing that in the case of Pb there is a considerable gain in using 
unstable beams, while in Sn is much less evident. One sees the need 
for unstable helium when compares the magnitude for the pairing resonance
in the right a) and b) panels with the peak at zero energy: in the first 
panel the transition to the ground state is extremely hindered.

A $^6$He beam is currently available (or it will be available in the very
near future) in many radioactive ion beams facilities around the world and 
the calculations that we have presented could allow a planning for future
experiments aimed to study the not yet completely unraveled role of pairing
interaction in common nuclei, 
using exotic weakly bound nuclei as useful tools. 

Projectiles with neutron excess display favourable conditions for
multi-pair transfer because of their large radial extensions. Since large
neutron excess is usually connected with a low binding energy Q-value 
considerations indicates that these reactions are suitable to populate
states at higher energy, thus exciting high lying pairing vibrations.
The Q-value for the reaction $^{206}Pb (^{8}He,^4He) ^{210}Pb$, for example,
is around $+20$ MeV, which gives optimal conditions to populate the GPV.



\chapter{Extension of the Steinwedel-Jensen model}

\begin{figure}[!h]
\begin{flushright}
\epsfig{figure=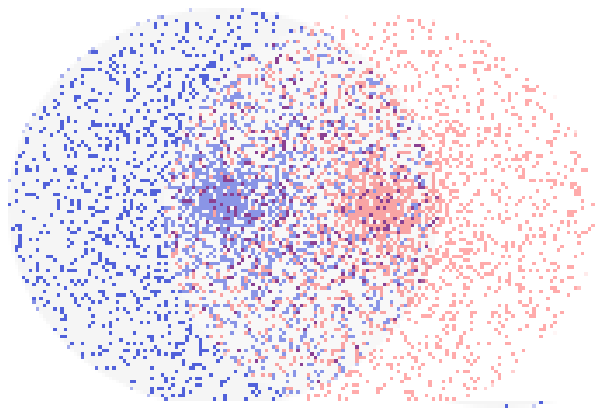,width=.2\textwidth}
\end{flushright}
\end{figure}

\section{Introduction}
The most important role played by exotic nuclei is to force the nuclear 
scientific community to test its ideas within the borders of a broader 
new realm. We have thus discovered 
that the extrapolations of the theories that are working pretty well inside
the stability valley may fail as long as nuclei with a large asymmetry are 
taken into consideration and a number of truly new phenomena arises in this 
region of the nuclear chart. At the driplines the presence of halos and 
neutron skins and the effects of the pairing interaction are believed to 
increase their importance, modifying many observables. This is also true 
for collective features and especially for the Isovector Giant Dipole 
Resonance (GDR). This collective mode represents the most important feature
of the continuum and is very often considered an important point for
the full understanding of the structure of nuclei.
Work in this direction has been pursued for example by Van Isacker and 
collaborators \cite{VanI}, who studied the effect of a neutron skin on the 
excitation of E1 and M1 collective states by means of en extension of the 
Goldhaber-Teller model, finding a lowering of the average energy of these 
modes  that they estimate to be about 5\% and a fragmentation of strength.
Microscopic HF+RPA calculations had shown that the value of the centroid of the
energy distributions in neutron-rich nuclei is invariably smaller than the 
corresponding value in normal nuclei \cite{cat22,Io7}.
Lipparini and Stringari \cite{Lipp}, instead, had modified the 
Steinwedel-Jensen model to include surface effects and interaction current 
terms constructing an energy functional to derive the symmetry energy and 
polarizability as well as sum rules. Their model is nevertheless rather 
complicated and we will propose a simpler alternative to take into
account surface effects in a straightforward way.
A preliminary discussion of some of these topics may be found in \cite{SteMon}.

Migdal \cite{Migd} was the first to derive a simple 
power law for the dependence of the energy of the giant dipole 
resonance upon the atomic mass $A$. The proposed formula 
was $24 A^{-1/3}(a_s Z/A)^{1/2}~MeV$, where 
$a_s$ is the coefficient of the symmetry term of the 
Bethe-Weizs\"acker mass formula.  \index{subject}{Goldhaber-Teller}
Goldhaber and Teller (GT) \cite{GT} assumed the oscillation of
a rigid proton sphere against a rigid neutron sphere with
sharp surfaces, ending in a dependence of the type $A^{-1/6}$.
We refer to the following section for a brief discussion.
Shortly afterwards Steinwedel and Jensen (SJ) \cite{SJ}, developing another 
idea proposed in the cited work of Goldhaber and Teller, derived a 
formula for the oscillation \index{subject}{Steinwedel-Jensen}
of proton and neutron liquids inside a common fixed 
spherical boundary. Their model, also called hydrodynamic or
acoustical, gave the prediction $A^{-1/3}$.
All these models were thought to be promising in the early
stage of the study of nuclear collective phenomena, but with
the growing amount of experiments on various atomic species they were 
negatively tested on many data   
\cite{BF}, and it was found that a good description of the
general trend is achieved with a dependence of $A^{-0.23}$.
We should mention that another model, called the droplet model, 
has been developed \cite{MS}. It encompasses the basic assumptions
of the two models and, although the physical interpretations of the two
approaches remain incompatible to a large extent, 
it gives accurate predictions, reproducing the empirical power law.

Even if all these models are very well accepted, we felt that it was 
worthwhile looking at this problem with a simple model that nevertheless
is capable to go beyond the actual approaches including surface effects.

The purposes of the present chapter are:
\begin{itemize}
\item{to review and comment the predictions of some well-known models about
the energy of the GDR when one moves from the prescription of $N=Z$ nuclei,
showing their different trends.}
\item{to set up a new class of models based on the extension of
the Steinwedel-Jensen model with the 
aim of describing situations in which the nuclear surface is not sharp.
In this class of models the density distribution is assumed to be of Fermi 
type and the region around the surface is divided in $n$ slices or steps where
the density is taken to be constant. This case turns out to be solvable. 
When $n$ is sufficiently large the smooth function is approximated pretty 
well.}
\item{to analyze the outcomings of this new class of models, namely to
study qualitatively the dependence upon the diffuseness of the nuclear 
surface and upon the presence of a skin, showing that
they predict sizable decrements in the energy of the GDR even at the level of
$N=Z$ nuclei and that this is especially effective at the dripline.}
\end{itemize}
The diffuse surface may reduce the energy of the mode to a large extent 
(up to 20\%).
The presence of the skin also decreases the energy of the modes, but is 
effective only as long as the diffuseness is kept small. 
When the diffuseness is taken into account the effect of the skin is not 
bigger than a 10\%.

\section{GT and SJ models at the driplines} \index{subject}{Goldhaber-Teller}
The Goldhaber-Teller model predicts the energy of the giant resonance to be
(see \cite{Grei} for a detailed and simple derivation)
\be
\hbar \omega \= \hbar \sqrt{3 a_s \over 4 \epsilon m} \sqrt{A^2\over ZNR}
\simeq {45 MeV\over A^{1/6}}
\ee
where $R=r_0 A^{1/3}$, $a_s$ is the asymmetry energy, $m$ is the mass of a
nucleon and $\epsilon$ is a somewhat arbitrary parameter
that is fixed to be $2 ~fm$. Since the two spheres are displaced the energy
required is linear in the separation distance. This is certainly a bad 
approximation for very small separations where the symmetry energy must have
a quadratic dependence. Goldhaber and Teller assumed a quadratic dependence 
at small separations fitted to join the linear dependence at some fixed point
$\epsilon$.  It is worthwhile to notice that the formula is usually 
approximated to its second form (valid only at $Z=N$) and that very often the
coefficient is fitted from the data and taken to be $33 MeV$. We have preferred
to use a common value for $a_s$ to give a purely theoretical prediction.   

\begin{figure}[!t]
\epsfig{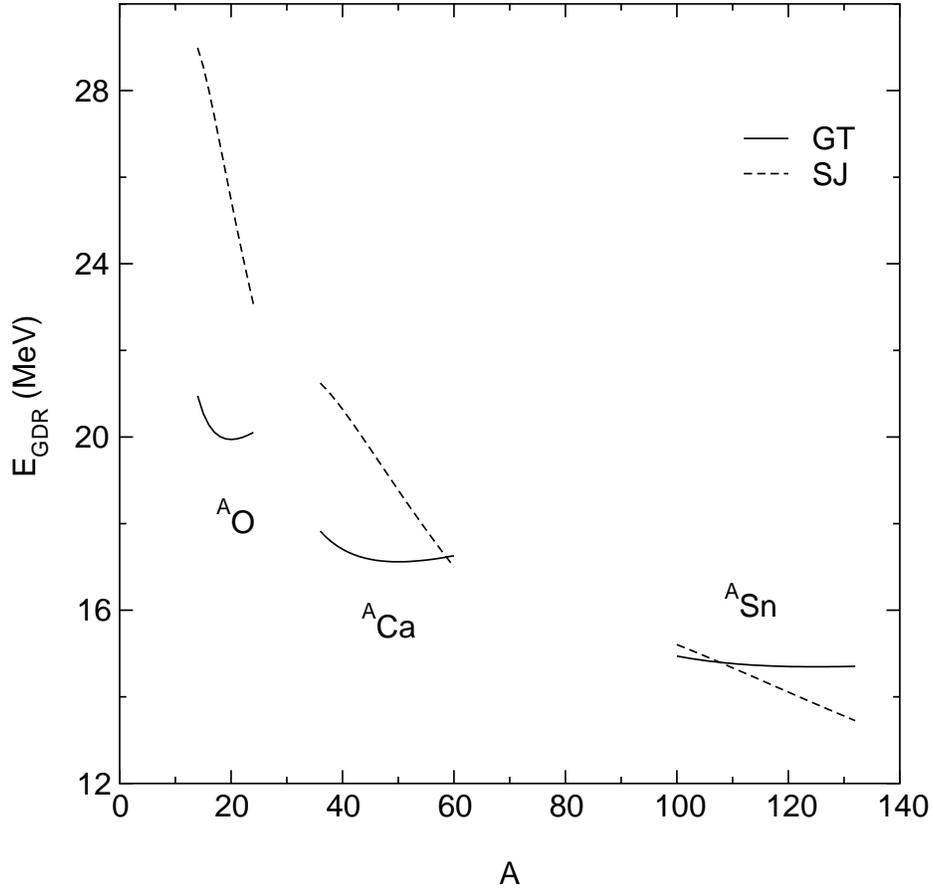}
\caption{Comparison of predictions of Goldhaber-Teller (solid) and 
Steinwedel-Jensen (dashed) models (considering the full formulae 
with $a_s=20MeV$) for the energy of the GDR for three isotopic chains   
(oxygen, calcium and tin).}
\label{GT_SJ}
\end{figure}
\begin{figure}[!t]
\begin{center}
\epsfig{file=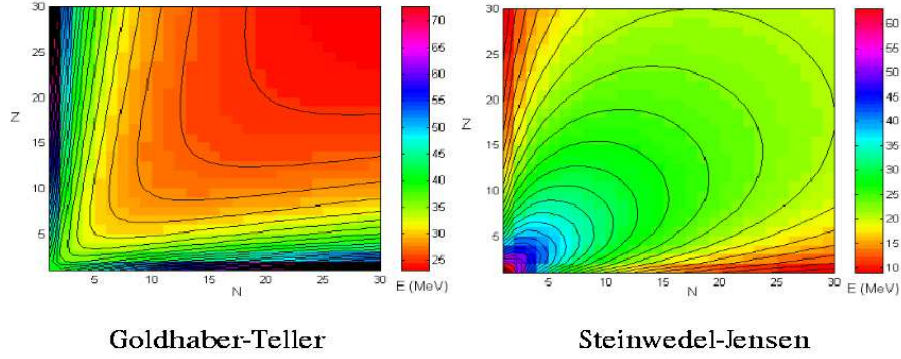,scale=0.4}
\end{center}
\caption{Comparison of predictions of Goldhaber-Teller (left) and 
Steinwedel-Jensen (right) models (considering the full formulae 
with $a_s=20MeV$) in the $(N,Z)$ plane. This is figure is due to
S.Montagnani \cite{SteMon} (pay attention
to the slightly different colouring scheme).}
\label{GT_SJ_2}
\end{figure}

\index{subject}{Steinwedel-Jensen}
The hydrodynamical or acoustical model of Steinwedel and Jensen takes a 
step function as a parameterization of nuclear densities
\be
\rho_{n,p}(r)\=\left\{
\begin{array}{cc}
\rho_{p,n}^0 & ~\mbox{if}~r\leq R_{p,n}\\
~&~\\
0&~\mbox{if}~r>R_{p,n}
\end{array}   \right.
\ee
where the radii $R_{n,p}$ of the two distributions and the  densities of 
the internal region, $\rho_{p,n}^0$, are taken as constants. This is a 
very crude approximation. The derivation of the energy is quite
straightforward \cite{Grei} and leads to the following expression 
\be
\hbar \omega \= \hbar \sqrt{8 a_s Z N\over m A^2} {z_1\over r_0 A^{1/3}}
\simeq \sqrt{4ZN\over A^2} {76.5 MeV\over A^{1/3}}
\ee
where $z_1=2.08$ is the first zero of the derivative of the spherical Bessel 
function with $\ell\=1$.

The second formula must be analyzed: usually one takes only the second factor  
because the square root reduces to 1 when $N=Z$. The remaining part agrees to 
a good extent with old data. 

We  notice however that the two formulae derived above lead
to very different results when one extrapolates to the driplines. This is 
especially relevant nowadays since exotic beams are available and large sets
of data are expected from future experiment. 

In fig. \ref{GT_SJ} we display a comparison between the full behaviour of the 
GT and SJ models for three chains of different isotopes: oxygen, calcium and
tin. The change of the number of neutron (and total mass) is responsible for 
the very different slope predicted by the two models. The differences are 
impressive not only for the general trends but also for the magnitude of the 
energy of the isovector giant dipole mode (30$\%$ in the worst case).
In fig. \ref{GT_SJ_2}, for each point in the $(N,Z)$ plane, the 
corresponding energy of the Giant Dipole Resonance is shown using different
colours for different magnitudes. The opposite behaviour of GT and SJ models
is clearly seen either for stable and unstable nuclei: not only the predictions
along the diagonals are different in magnitudes, but also the surfaces 
display different curvatures while moving toward regions with excess of 
one of the two type of nucleons (borders of the square).

\section{The extension of the SJ model}
The nuclear surface is not sharp. It is diffuse and 
the Fermi distribution is known to be an efficient way to parameterize the 
nuclear proton and neutron densities:
\be
\rho_t(r)\= {\rho_t^0\over 1+e^{(r-R_t)/a_t}}
\ee
where $t={n,p}$ is an index that indicates neutrons or protons respectively,
$R_t$ and $a_t$ are the radius and diffuseness  of the 
density distributions of neutrons and protons, while the saturation 
values are $\rho_t^0$.
One may wonder that the effect of the diffuseness should be
small for light nuclei and even negligible in the case of heavy nuclei. 
Our aim is to show in a qualitative way  that it is indeed very effective in 
changing the predicted energy of the GDR within the Steinwedel-Jensen model, 
already at the level of $N=Z$ nuclei. We define the total density 
distribution has the sum of the density of the two species:
\be
\rho_0(r)\=\sum_{t} \rho_t(r)
\ee
It is worthwhile to insist here on the fact that every well-behaved 
distribution is equally treatable with the method that we are going to
 explain. We have decided to deal with the Fermi distribution because
the degree of approximation that it furnishes is very good.

We now give a criterion to create a subdivision of 
the interval over which the density distribution is  defined with
the purpose to approximate it with a step function.
The procedure that we adopt consists of the following points:
\begin{itemize}
\item{We choose a convenient number $n$ of steps that we wish to use
as an approximation.
Performing the calculations a number of times with increasing $n$, 
we will show that the value of the energy of the giant dipole mode will 
converge to a finite constant value.
}
\item{We define a region around the surface in such a way that the point at 
which the density is one half of the value found at $r=0$ is taken as the 
'center' of the surface and we take a spherical crust whose thickness is
such that the external radius always is a percentage of the inner density
(10\% or 5\% will be taken for simplicity).}
\item{The surface region is then divided in $n$ equally spaced intervals, and
in each interval is taken a constant average density. This is also done in 
the case of the inner interval. In this way we have defined a step function 
that is an approximation to our original density distribution. We have
$r_0$ in the origin, $n$ radii $r_j$, with $j=1,n$, that 
divide two adjacent internal intervals and finally 
the external radius $r_{n+1}$.}
\end{itemize}
Now we have delineated a way to split the density distribution in a 
number of intervals. Depending on the number of intervals the outcoming
model will be called a n-steps SJ model. Obviously the 0-step SJ model reduces
exactly to the SJ model when the parameters $R_t$ and $a_t$ of the two 
distributions are equal.

\begin{figure}[!t]
\vspace{0.5cm}
\begin{center}
\epsfig{file=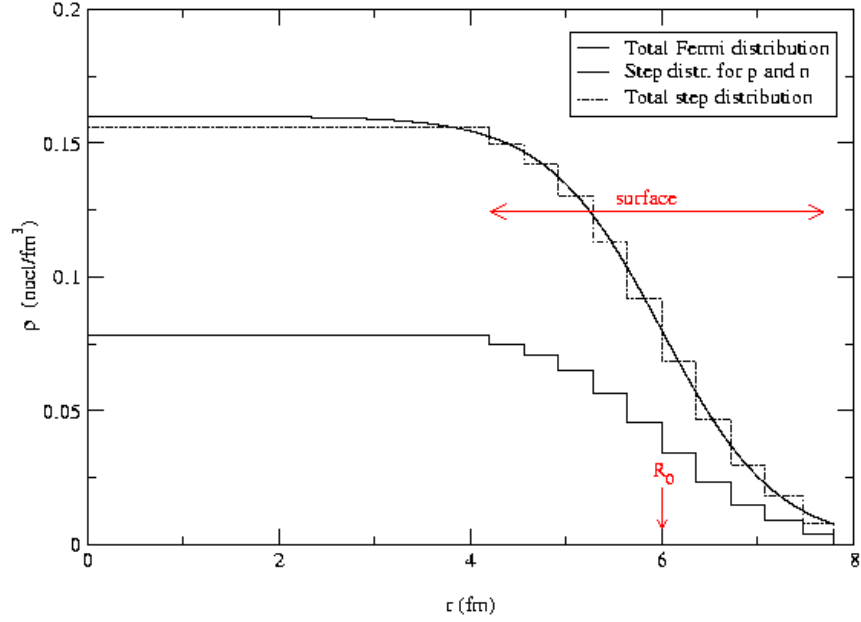,scale=0.8}
\end{center}
\caption{Fermi distribution function (solid line) and 10-steps SJ 
distribution function (dashed for n or p and dot-dashed for their sum). 
The radii $r_j$ correspond to the jumps of the step function.
When the number of partitions in $r$ is increased, the approximation of the 
step function to the Fermi distribution becomes more effective.}
\label{slices}
\end{figure}

We consider a continuous distribution 
replacing the number of particles $(N,Z)$ of each subdivision with the 
constant density in that interval.
A plot of a typical Fermi distribution with the 10-steps function is displayed
in fig. \ref{slices} to illustrate the way in which we approximate the 
density profile and the subdivision of the surface region.
The higher is the number of intervals that we choose, the better is the 
approximation of the Fermi distribution.
Thus in the limit $n \rightarrow \infty$ the step function tends exactly
to the Fermi distribution (cut at some external radius).
The fact that we are cutting the distribution, whose tail extends to infinity,
at a given point may introduce problems. In fact whenever we increase the 
external radius the value of the energy decrease. To fix the ideas and to give
a qualitative trend we have to make a reasonable recipe, defining the surface 
thickness as the region where the density drops from 90\% to 10\% of the inner 
value. This is a common prescription \cite{Jone} when one is dealing with
a Fermi type parameterization of the nuclear surface.

Since the magnitude of the effect depends on the way in which one cuts
the distribution we have repeated these calculation several times, adopting
different strategies and finding always a qualitative agreement in the results.

Insofar we have always made a distinction between neutrons and protons 
for the sake of generalization, but, since the two distributions may have 
different parameters $R$ and $a$, the resulting step functions may differ in 
the set of ${r_j}$'s. To simplify our model we take as a guide the 
distribution whose saturation density is higher and we derive only one set
of intervals. In each interval the two distributions of protons and neutrons
are defined accordingly to the average value between the two extremes of
the interval. This is expected to bear no consequence on the final results,
especially when the number of steps is made big.

Now we introduce, as a straightforward generalization of the 
Steinwedel-Jensen model, a system of $n+1$ 
\footnote{There are $n$ steps in the surface plus the inner interval, 
for a total of $n+1$ different slices.} space and time dependent 
equations that 
describes the variations of the nuclear densities, ${\rho_t}^{(j)} 
(\vec r,t)$, within each interval as small density oscillations,
$\varepsilon^{(j)}(\vec r,t)$, of proton fluid against neutron 
fluid with total fixed densities ${\rho^0_t}^{(j)}$:
\be
{\rho_{t}}^{(j)} (\vec r,t)\= {\rho^0_t}^{(j)} \pm \varepsilon^{(j)}(\vec r,t) 
~~, \forall j=0,\cdots,n-1
\ee
where plus and minus signs refer to the two different isotopic species.
Clearly in every interval the density fluctuation $\varepsilon$
of protons is opposite to the density fluctuation of neutrons. 
Furthermore one may think to more complex fluctuations that involve the 
exchange of material across different intervals. This is also included
in our model because the requirement of constant total densities implies
that all the fluctuations change accordingly to this request. 
The conservation of each kind of 'particles' in each interval is assured by 
the fact that the volume integral of $\varepsilon$ is zero.

In each interval we calculate kinetic and potential symmetry energies in 
the standard way (see for example \cite{Grei}):
$$
T\= {m\over 2} \int d^3r ~ \bigl( \rho_{p}^{(j)}{{\bf v}^{(j)}_p}^2+
\rho_{n}^{(j)}{{\bf v}^{(j)}_n}^2 \bigr)\=$$
\be
{m\over 2} \int d^3r ~ \bigl( \rho_{0}^{(j)}{{\bf V}^{(j)}}^2+
\rho_{red}^{(j)}{{\bf v}^{(j)}}^2  \bigr)
\ee
where ${{\bf v}_{p,n}({\bf r},t)}^{(j)}$ are the flow velocities of protons
and neutrons fluids in each interval. Transforming to relative and center of 
mass velocities leads to the second formula, where the first term may be
eliminated requiring that the nucleus as a whole would remain at rest 
(${\bf V}^{(j)}=0$, $\forall j$) and where the reduced density in the 
$j-$th interval is given by 
\be
\rho_{red}^{(j)}\= {{\rho_p^0}^{(j)}{\rho_n^0}^{(j)} \over {\rho_0}^{(j)}}
+{{\rho_p^0}^{(j)}-{\rho_n^0}^{(j)} \over {\rho_0}^{(j)}} {\varepsilon^{(j)}} -
{{\varepsilon^{(j)}}^2\over \rho_0^{(j)}}.
\ee
In the following only the first term will be kept. The potential energy
reads instead
\be
V\= {4a_s\over {\rho_0}^{(j)}} \int d^3r ~{\varepsilon^{(j)}}^2
\ee
The variation of the lagrangian $L=T-V$, under the assumption of 
hydrodynamical irrotational flow ($\nabla \times v =0$), reads
$$
\delta \int dt L\=0\=$$
\be
\= \int dt \int d^3 r \Biggl({m\over 2}{{\rho_p^0}^{(j)}{\rho_n^0}^{(j)} 
\over {\rho_0}^{(j)}} {{\bf v}^{(j)}}^2 - {4a_s\over {\rho_0}^{(j)}}
{\varepsilon^{(j)}}^2  \Biggr)
\ee 
Defining the velocity and its variation as a function of the 
displacements, exploiting the continuity equation (at zero order in 
$\varepsilon$) and integrating by parts where necessary,
we come up to a system of coupled differential equations for the 
displacements of the following form (see \cite{Grei} for details, p. 200): 
\be
\left\{
\begin{array}{ccc}
{1\over u_0^2}{\partial^2\varepsilon^{(0)}\over \partial t^2 } &\=& 
\nabla^2\varepsilon^{(0)}   \\
\cdots&\=&\cdots \\
{1\over u_j^2}{\partial^2\varepsilon^{(j)}\over \partial t^2 } &\=& 
\nabla^2\varepsilon^{(j)}   \\
\cdots&\=&\cdots
\end{array}
\right.
\label{syst}
\ee 
where we have used the continuity equation at each border within different 
intervals and the propagation speed of density waves $u_j$ inside each 
interval is given by
\be
u_j\=\sqrt{8a_s{\rho_n^0}^{(j)}{\rho_p^0}^{(j)}\over 
m ({\rho_n^0}^{(j)}+{\rho_p^0}^{(j)})^2 }
\label{speed}
\ee
The solution of the system of wave equations (\ref{syst}) gives in general
a system of coupled equations of 
linear combinations of spherical Bessel $j(z)$ and Neumann $n(z)$ functions 
with $\ell\=1$ and proper coefficients that we write as:
\be
y_j(x,r)\equiv \alpha_j j(k_jr)+\beta_jn(k_jr)
\ee
with $k_0\equiv x$ and hence $k_j\=u_0k_0/u_j\=u_0x/u_j$, $\forall j\=0,\cdots
, n-1$. This last equivalence follows from the fact that the 
wavevectors $k_j$ are related by 
\be
k_0u_0\=k_1u_1\=\cdots\=k_ju_j\=\cdots
\ee
Using the definitions (\ref{speed}) we may choose only one independent 
$k_j$, that is the final goal of the solution of the system, 
since the energy of the Giant Dipole Resonance is  $E_{GDR}\=\hbar k_ju_j$.
The solution in the innermost interval contains only the function that is 
regular in the origin (i.e. $\beta_{0}\=0$), 
while in all the other intervals they are both present.
The solutions in two adjacent intervals must connect up, together with their 
derivatives, and the solution in the outermost sector must have a null 
derivative at the outer radius $r_0$. No condition is explicitly required
for the value of the solution at the outer radius.
All these conditions set up a system of $2n+1$ equations in $2n+2$ variables
\be
\left\{
\begin{array}{cc}
   \left.
   \begin{array}{c}
    y_0'(x,r)\mid_{r_{n+1}}\=0\\
   \end{array} 
   \right\}  & at~~~r_{n+1}\\
     \cdots\\
   \left. 
   \begin{array}{c}
    y_{j+1}(x,r)\mid_{r_{j}}\=y_{j}(x,r)\mid_{r_{j}}\\
    y_{j+1}'(x,r)\mid_{r_{j}}\=y_{j}'(x,r)\mid_{r_{j}}\\
   \end{array}  
   \right\} & at~~~r_{j}\\ 
\end{array}
\right.
\label{syst-sol}
\ee 
where $j=1,\cdots, n$. The system above may be easily solved: given
the coefficients in the innermost interval ($\beta_0=0$ and $\alpha_0$
arbitrary, may be fixed by normalization, but this is not relevant for the 
present purpose), one may express the coefficients in the $j-$th interval
by recursion:

\be
\left\{
\begin{array}{c}
\beta_j\={{y'_{j-1}(r)j(k_{j}r)/k_j-y_{j-1}(r)j'(k_{j}r)}\over 
{n'(k_{j} r)j(k_{j}r)-n(k_{j}r)j'(k_{j}r)}} \\
~~~\\
\alpha_{j}\={y_{j-1}(r)-\beta_j n(k_{j}r) \over j(k_j r)} \\
\end{array}
\right.
\ee

The system above may be easily solved by means of a
numerical FORTRAN routine by looping over the wavevector 
(the $(2n+2)^{th}$ variable) and looking for the point where the derivative 
of the solution in the outermost interval changes its sign, while crossing 
zero. This sets the value of the energy of the isovector giant dipole mode.

\section{The effect of the diffuse surface}
The first issue we want to discuss is the effect of the diffuseness
of the nuclear surface.
It is worth mentioning that whenever the diffusivity is very small the 
result of the calculations agrees with the Steinwedel-Jensen prediction.
At all practical purposes when $a$ is $1/100$ or lower there should be no
difference between the Fermi distribution and a sharp distribution.
 
We kept constant the saturation densities ($\rho_p^0=\rho_n^0=0.080$ 
nucleons/fm$^3$) and the mass of the system ($N=Z=20$)
and we studied the dependence of the energy of the giant dipole resonance 
on a common diffuseness parameter $a_p\=a_n$ finding a non-negligible 
reduction of the energy as one can see from table \ref{ta1}.

\begin{table}\vspace{5mm}
\begin{center}
\begin{tabular}{c|c|c}
a (fm) & $E_{GDR}$ (MeV) & \% \\ \hline\hline
0.0&24.23&\\ \hline
0.01&24.13&0.4\%\\
0.1&22.99&5.1\%\\
0.2&21.90&9.6\%\\
0.3&20.99&13.4\%\\
0.4&20.21&16.6\%\\
0.5&19.58&19.2\% 
\end{tabular}
\end{center}
\caption{Energy of the giant dipole resonance as a function of the 
diffuseness for $^{40}$Ca. The first value is the prediction of the 
Steinwedel-Jensen model, while the others are the outcome of the 
presently discussed model (see text). The distribution is integrated 
until the 10\% of the inner density is reached.}
\label{ta1}
\end{table}

The first row in this table is separated from the other because is the
prediction of the pure Steinwedel-Jensen model with a rectangular distribution 
with the parameters given above. 
This calculation has a qualitative character, because we changed the radii in
order to keep constant the mass of the system.
The first column is the diffusivity, the second is the energy of the 
mode. When increasing the diffusivity we have to lower the radius of the 
center of the surface because, otherwise, the total 
mass and charge of the nucleus vary. We have thus found, for each case, the 
radius that maintain the total mass fixed to the original
values and the effect of lowering on the energy of the giant 
dipole mode may amount to a lowering of about 
20\% with respect to the Steinwedel-Jensen prediction.

It is found, by fitting the data in this and other cases, that the dependence 
on the diffusivity is parabolic.

The first conclusion that one may get is that the effect of diffuseness
is very strong already at the level of $N=Z$ nuclei, or along the 
stability valley and may be expected to be even stronger for unstable nuclei.

\section{The effect of the skin} \index{subject}{skin}
One of the most fascinating issues concerning the novel properties of very
neutron- (or proton-) rich nuclei is the presence of a neutron- 
(or proton- ) skin.
It is commonly believed, thanks to electron and hadron scattering experiments
, that along the stability valley the matter distribution (radial density) of
protons and neutrons have almost the same spatial extension, although different
absolute value of the saturation density. That is to say the proton
distribution 'follows' the neutron distribution due to the strong pn 
interaction. This is certainly true for light and medium mass nuclei and
partly fullfilled by heavier nuclei, but it is no more true when one moves 
away from stable systems.

\begin{figure}[!t]
\vspace{0.5cm}
\begin{center}
\epsfig{file=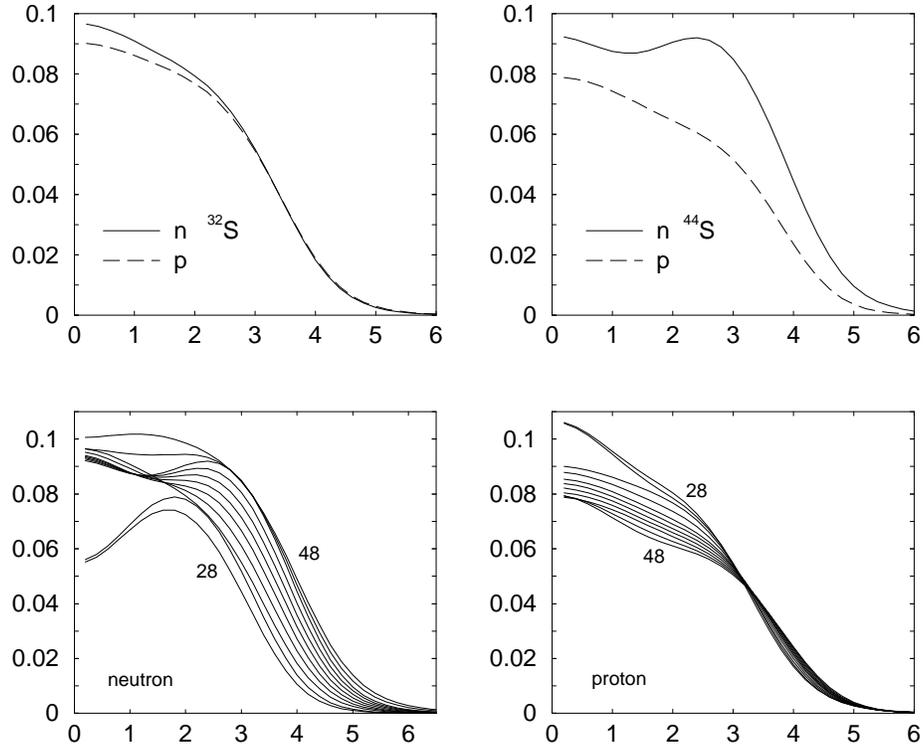,width=1.\textwidth}
\end{center}
\caption{Hartree-Fock predictions for radial density distributions of 
protons and neutrons in various isotopes of sulphur. The stable case 
(left,upper panel) has no skin while the neutron-rich isotope (right, 
upper panel) shows an abundance of neutron in the superficial region. 
The evolution of the density profiles with the number of particles is 
illustrated in the lower row. The vertical scales are in nucleons/fm$^3$, 
while the horizontal scales are in fm.}
\label{sulfu}
\end{figure}

\begin{figure}[!t]
\vspace{0.5cm}
\begin{center}
\epsfig{file=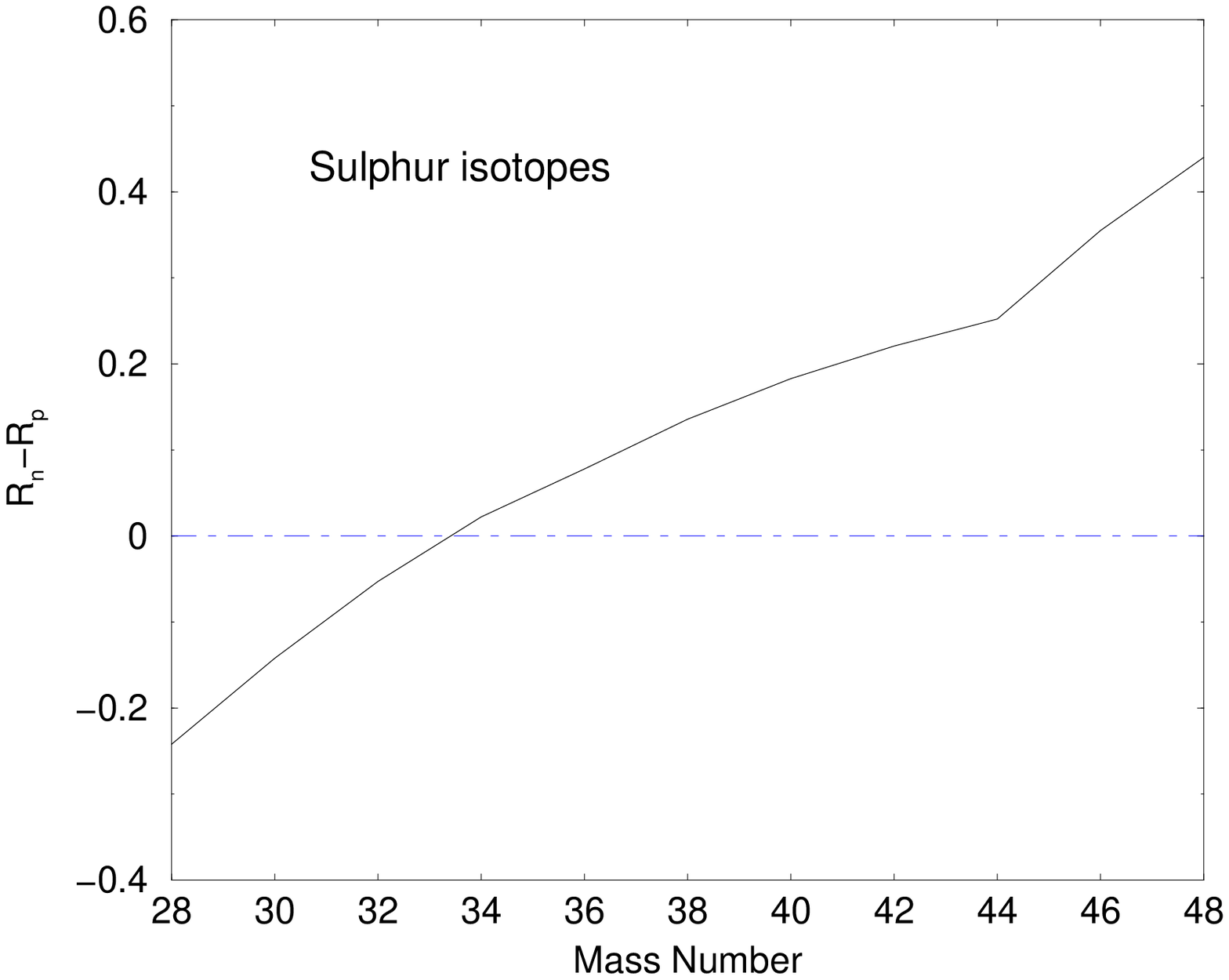,width=0.85\textwidth}
\end{center}
\caption{Extension of skin (in fm) versus mass number in sulphur isotopes.
(HF calculations with Skyrme force). Negative values correspond to proton 
skins, while positive ones correspond to neutron skins.}
\label{skn}
\end{figure}

In fact whenever the number of one of the two kinds of nucleons is by far 
exceeding the number of the others, many calculation and experiments indicate
that the above picture is no more valid and that there is a superficial region
with abundance of one of the two species. We do not attempt to resume here
the extensive literature on this topic, but we refer the reader to some of the
works in this field as \cite{skin,cat22,Das2,Das1,Cata,Io7}.
We prefer instead to present as an example the results of simple 
Hartree-Fock (HF) calculations with \index{subject}{Hartree-Fock}
a common Skyrme type interaction to give an argument in support of the 
theoretical predictions of neutron skin.
In fig. \ref{sulfu} a comparison between density distributions of protons
and neutrons in various isotopes of sulphur is made and the corresponding
differences between the neutron and proton radii are drawn in fig. 
\ref{skn}. It is seen that, with this approach, a skin of a fraction of fm
is envisaged even for a relatively light nucleus as sulphur.
\index{subject}{sulphur}

We address now the problem of the effect of the neutron (or proton) skin 
on the excitation of the GDR in our schematic model increasing only one 
of the radii of the two distributions and keeping fixed and equal 
the central densities. In the following we are not interested in making 
accurate predictions on the real value of the energy of the Giant Dipole Mode,
but we only want to discuss the effect of the presence of skin: for this
reason the parameters in the table do not necessarily correspond to an 
integer value for the masses.
We repeated these calculations with different values of 
the diffuseness to shed light on an interesting fact.
When the diffuseness is small ($a=0.01$ fm) and the two distributions 
are almost box-like the effect of the skin is appreciable: 
the energy of the GDR drops of about 2-6 \% with respect to the case with 
no skin as shown in table \ref{ta2}. 

\begin{table}[t]
\vspace{5mm}
\begin{center}
\begin{tabular}{c|c|cl}
$R_p$ (fm) & $R_n$ (fm) &$E_{GDR}$ (MeV)\\ \hline
3.0&3.0&27.362&\\
3.0&3.1&27.354&(0.03\%)\\
3.0&3.2&26.370&(3.6\%)\\ \hline
5.0&5.0&18.166&\\
5.0&5.1&17.796&(2.0\%)\\
5.0&5.2&16.914&(6.9\%)
\end{tabular}
\end{center}
\caption{Energy of the giant dipole resonance as a function of the neutron 
skin (difference between the two radii), having fixed 
$\rho_t^0 = 0.080$ nucl/fm$^3$ and $a_t=0.1$ fm.}
\label{ta2}
\end{table} 
\index{subject}{skin}

In the same table it is also seen, by comparing the relative values of the 
$R=3$ fm and $R=5$ fm, that this lowering due to the skin is less 
effective in system with smaller masses.

\begin{table}[t]
\vspace{5mm}
\begin{center}
\begin{tabular}{c|c|cl}
$R_p$ (fm) & $R_n$ (fm) & $E_{GDR}$ (MeV)\\ \hline
4.0&4.0&18.621&\\
4.0&4.1&18.393&\\
4.0&4.2&18.119&\\
4.0&4.3&17.541&(5.4\%)\\\hline
6.0&6.0&13.385&\\
6.0&6.1&13.248&\\
6.0&6.2&13.111&\\
6.0&6.3&12.927&(3.4\%)
\end{tabular}
\end{center}
\caption{Energy of the giant dipole resonance as a function of the neutron 
skin (difference between the two radii), having fixed 
$\rho_t^0 = 0.080$ nucl/fm$^3$ and $a_t=0.5$ fm.}
\label{ta3}
\end{table}

A more appropriate value of the diffusivity ($a\=0.5$ fm) is used in the 
calculations displayed in table \ref{ta3} to understand the effect of 
the skin. Notice that now the value of the energies are in general smaller 
than in the previous case because of the effect of the diffuseness, 
as discussed in the previous section.
In this case the additional lowering due to the presence of the skin 
is still sizable (3-6 \%) and roughly in agreement with
the estimate made in \cite{VanI}.

The interplay between the effect of the presence of a diffuse surface and 
of the skin have opposite trends in the two tables above. To understand it
properly one should compare the reductions in cases where, for instance, the
skin is exactly equal to one diffusivity, or to a fraction of diffusivity, 
and see which are the trends from light to heavy systems. 
This is done in table \ref{ta4} where the densities and diffusivities are fixed
and the energy of the GDR is given as a function of the radii of protons 
distributions for three cases: no skin (first column), skin equal to 1/5 of
diffusivity (second column) and skin equal to diffusivity (third column).
The third case, in which the skin has been taken unrealistically large to 
highlight the effect, has been supplemented with the percentage with
respect to the case with no skin. It is seen that going from light systems to
heavier systems, the effect of the neutron skin becomes less important.

\begin{table}
\vspace{5mm}
\begin{center}
\begin{tabular}{c|c|c|c}
$R_p (fm)$&\multicolumn{3}{|c}{$E_{GDR}$}\\ \hline
&$\Delta r_{skin}/a=0$&$\Delta r_{skin}/a=1/5$&$\Delta r_{skin}/a=1$ \\ \hline
3.0&23.13&22.81&20.66 (10.7\%)\\
4.0&18.62&18.39&16.93 (9.1\%)\\
5.0&15.57&15.39&14.34 (7.9\%)\\
6.0&13.38&13.25&12.43 (7.1\%)
\end{tabular}
\end{center}
\caption{Energy of the giant dipole resonance: trends from light to heavy 
systems with the same ratios of skin and diffusivity ( $\rho_p = \rho_n =0.08$ 
nucl.$/fm^3$ and $a=0.5~fm$). It is seen how the presence of skin is less 
dramatic in heavier systems.}
\label{ta4}
\end{table}

\section{Conclusions}
We have attempted a study of surface effects on the excitation of the Giant 
Dipole mode with the aim of shedding light on the trends associated with
 the presence of a diffuse surface or with the presence of a displacement of
proton and neutron radii (skin).
To this end we have introduced an extension of the Steinwedel-Jensen model
in which the nuclear density, parameterized as a Fermi distribution, is 
sliced in a number of intervals and taken constant inside each interval. 
An exact solution of this model is proposed in terms of an iterative formula 
that may be easily solved.

Surprisingly we have found large reductions in the energy in connection with
the presence of a diffuse surface that may even amount to the 20\% of the 
total. In a similar way,
the effect of the neutron skin has large consequences on the  
distribution with a very small diffusivity, but becomes less remarkable 
whenever a reasonable diffusivity is taken into account. 
This is due to the fact that situations in which the densities of the 
two fluids are very different tend to lower more the energy of the mode. 

The study presented in this chapter is still preliminary in the sense 
that the prescription that we have employed to cut the distribution 
is arbitrary, albeit reasonable. Since the results depend on the recipe
used we can only make qualitative statements:
 both the presence of diffuse surface and skin affects the Giant
Dipole Resonance favouring a marked reduction of the energy of the mode.
These findings confirm the hints that were put forward in the works of 
Van Isacker and collaborators, based on a modification of the Goldhaber-Teller
model and in the works of Catara and others, where a microscopic description,
 based on RPA calcualtions, was adopted. 
The effects discussed here are expected to play a role in those systems 
far from the stability line where the presence of nuclear halos and skin 
are thought to be present.


~~\newpage

\chapter{Break-up of dicluster nuclei}

\begin{figure}[!h]
\begin{flushright}
\epsfig{figure=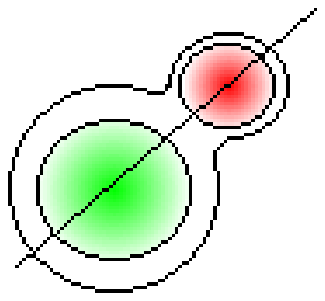,width=.2\textwidth}
\end{flushright}
\end{figure}

\section{Introduction}
A unique feature of nuclear systems along the neutron drip-line
is the concentration of strength at excitation energies just above the 
continuum threshold. 
This concentration of strength is directly measured in breakup reactions, 
but it has strong effects also on other processes, such as elastic 
scattering or sub-barrier fusion reactions.
It has been proved that this peculiar feature is 
associated with the weakly bound nature of most nuclei at the drip-line 
\cite{Das2}.
Within a di-cluster description of a weakly-bound nucleus (where one of the 
cluster may even be a single nucleon), the quantum state
that describes the system lies very close to the threshold for separation into
the two subsystems. The potential is in this case is the 
intercluster potential. The wavefunctions associated with such states (and
hence their distributions of matter) extend to large radii, spreading far 
outside the walls of the potential well (this is valid already at the level
of a square well potential, and it is even more evident for a realistic
potential that has a tail at large radii).
This establishes the opportunity to set a matching between the bound wave 
function and some scattering state in the (low-lying) continuum with
approximately the same wavelength. The resulting electromagnetic response
has a marked concentration of strength in the threshold region.
Besides the intrinsic excitation properties of these systems, one has to 
consider the unusual extension of nuclear coupling
on reactions, that can compete with coulomb excitation and modify inelastic
form-factors and cross-sections in a sizable way \cite{Das1}.

\begin{figure}[!t]
\begin{center}
\includegraphics[height=.85\textheight]{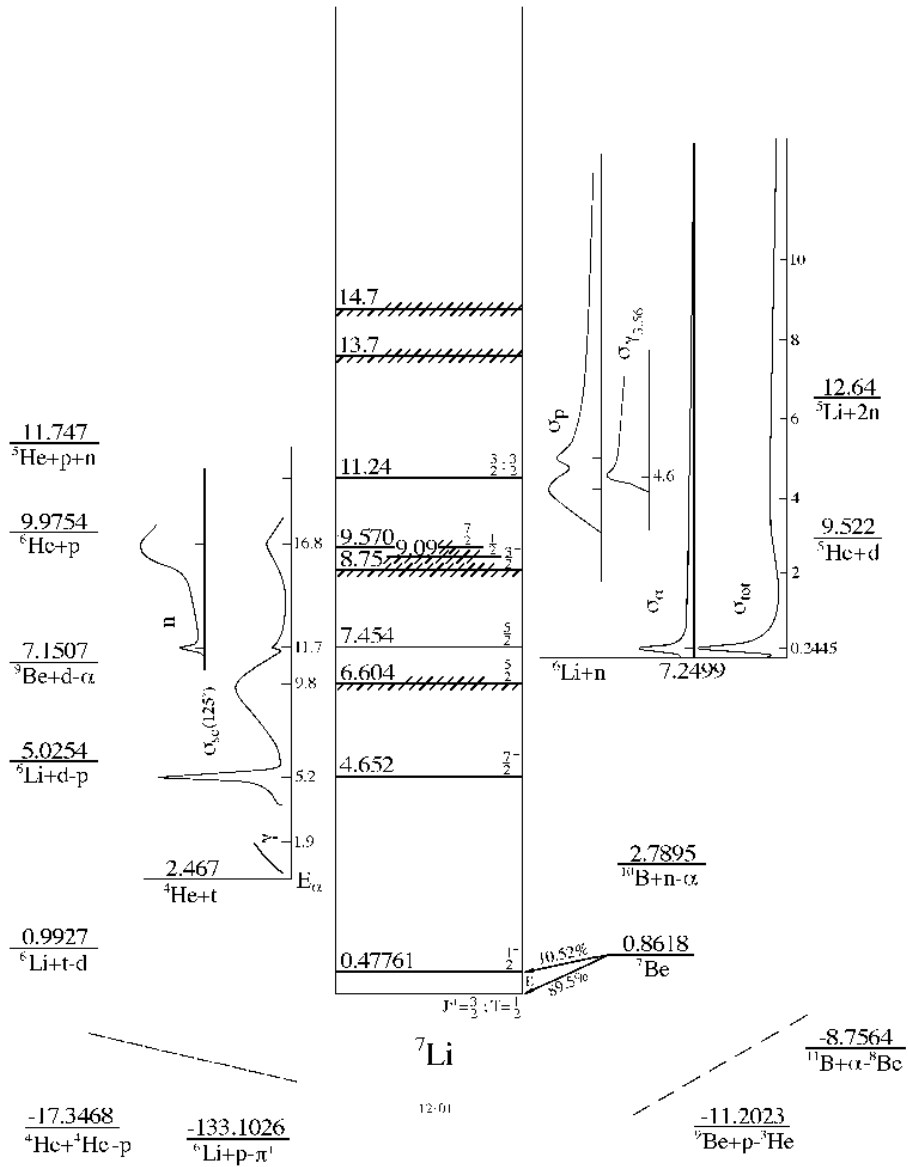}
\caption{Level scheme of $^7$Li from TUNL website \cite{TUNL}. The data
displayed here come from the compilation of Fay Ajzenberg-Selove.}
\end{center}
\label{7Lischeme}
\end{figure}

The picture outlined above finds its simplest application in the case of
single particle halos, where, in a mean field approach,
 it is the last unpaired
nucleon that is responsible for the halo distribution and that, being 
promoted to the continuum single particle states, gives rise to the 
low-lying strength. The precise dipole strength distribution does not depend
on the details of the binding potential, but rather on the value of the 
initial binding energy and on the angular momentum of the initial state,
as well as on the neutron or proton character of the halo state. In all
cases, however, the energy corresponding to the maximum of the strength 
distribution depends linearly on the binding energy \cite{Naga}.
Similarly, the total dipole strength at the threshold depends approximatively
on the inverse of the binding energy and tends therefore to magnify its 
effects as one approaches the drip-lines.
The picture developed so far for single-particle halos can be extended to 
the case of light weakly-bound nuclei within a dicluster model.
We take as an \index{subject}{$^7$Li}
example the case of $^7$Li, whose ground state is well described in terms
of interacting $\alpha$ and triton clusters, which characterize the lowest
continuum threshold (at 2.467 MeV). The basic necessary 
assumption is that also the
excited states, both bound and unbound, are described within the same
dicluster picture. In particular for the bound $1/2^-$ state the relative 
motion has been assumed to be still in the $p$ state, as the ground state,
while for the continuum state the cluster-cluster relative motion can have
all angular momenta. We depict in fig. \ref{7Lischeme} the level scheme of 
$^7$Li (taken from TUNL website \cite{TUNL}).

The simple model for the threshold strength is modified when the system 
displays, in the low-energy continuum, true resonant states in addition to 
the non-resonant part. This is for example the case of $^7$Li which has the
$7/2^-$ and $5/2^-$ states at 4.652 MeV and 6.604 MeV respectively. 
Within the cluster picture these states
correspond to narrow resonances in the relative motion with angular momentum
$\ell =3$. In a proper treatment of the continuum both resonant
and non-resonant contributions arise in a natural way and may have comparable
strengths. {\it Ad hoc} formalisms, which only include either the resonances
or the non-resonant continuum, may therefore be inadequate to describe
the full process.

A preliminary account of the concepts discussed in this article may be found 
in \cite{Io6}.

\section{Status of dicluster systems}
\begin{figure}[!t]
\begin{center}
\epsfig{figure=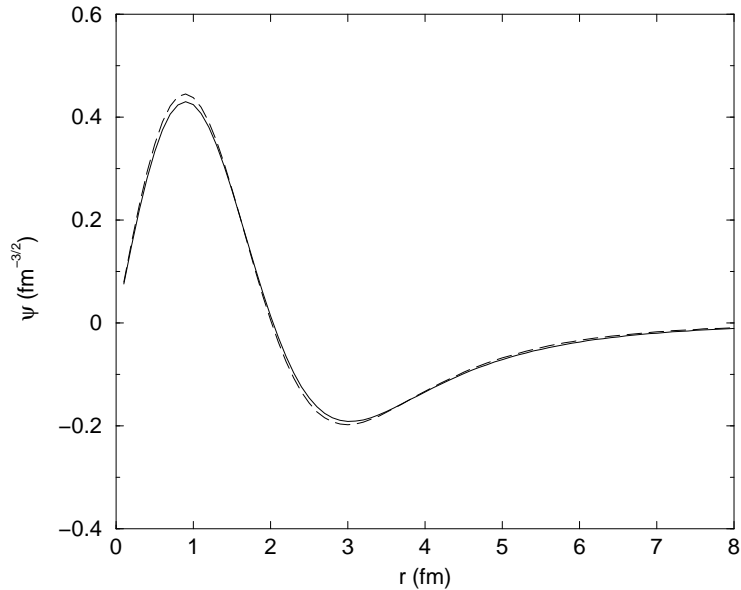,width=0.8\textwidth}
\end{center}
\caption{Wavefunctions for the ground state (solid) and first excited state
(dashed) obtained solving a unidimensional Schr\"odinger equation for the
relative motion, adjusting the depth of the Wood-Saxon potential to obtain 
the correct eigenvalues. }
\label{waves2}
\end{figure}   
Walliser and Fliessbach \cite{Wall} discuss a 
cluster picture for $~^7$Li, in which the constituents of the nucleus
are treated as elementary, that is without internal structure, but not 
necessarily point-like. They obtain considerable agreement with
experimental data and we conform, in principle, to their model. 
The main difference is that we determine the wavefunctions for the 
relative motion of the cluster by solving the unidimensional Schr\"odinger
equation for an effective $\alpha-t$ potential that reads:
\be
V_{\alpha-t}(r)\=V_{coul}(r)+V_{WS}(r)+V_{\bf l \cdot s}(r)
\ee
It contains the coulomb repulsion (corrected at small distances for the 
sphericity of charge distributions), the nuclear Woods-Saxon attractive 
potential and the usual spin-orbit term \cite{Bohr}.
By adjusting the depth of the Woods-Saxon well ($V_{WS}=-74.923$ MeV) and 
the magnitude of the spin-orbit correction ($V_{ls}=1.934$ MeV) we can obtain 
exactly the energy eigenvalues for the two bound states.
The $\alpha$ cluster has spin $0$ while the $t$ cluster has spin ${1\over 2}$.
The angular momentum coupling between the relative motion and the spin of the
triton provides the total angular momenta $({3\over 2})^-$ for the ground 
states with energy $-2.467$ MeV and $({1\over 2})^-$ for the first excited 
state at $-1.989$ MeV \cite{Till}. 
The energy are measured with respect to the $\alpha-t$ break-up threshold.
We give in fig. \ref{waves2} the wavefunctions for the ground state and
for the first excited state to allow a qualitative
comparison with the ones obtained in the paper of Wallisser and Fliessbach 
(for example the radial node is at the same point).
The treatment of the scattering states will be discussed later.

\begin{table}[t]
\begin{center}
\begin{tabular}{|l|c|c|}
\hline
Quantity& This work & Experiments\\ \hline \hline
$<r^2>_{ch}^{1/2} (fm)$ & $2.44$& $2.55 \pm 0.07^{(a)}$ \\
&&$2.39\pm 0.03^{(a)}$\\ \hline
$Q_{el} (fm^2)$& $-3.77$ & $-3.8 \pm 1.1^{(a)}$ \\
&& $-3.4 \pm 0.6^{(a)}$\\
&& $-3.70 \pm 0.08^{(a)}$\\ \hline
$Q_{mat} (fm^2)$& $-3.99$ & $-4.1 \pm 0.6^{(a)}$\\
&&$-4.00\pm0.06^{(b)}$\\ \hline
$B(E2,{3\over 2} \rightarrow {1\over 2}) (e^2fm^4)$& $7.55$ & $8.3 \pm
0.6^{(a)}$\\
&&$8.3 \pm 0.5^{(a)}$\\
&&$7.59 \pm 0.12^{(b)}$\\
&&$7.27 \pm 0.12^{(b)}$\\ \hline
$B(M1,{3\over 2} \rightarrow {1\over 2}) (\mu^2)$& $2.45$ & $2.50 \pm
0.12^{(a)}$\\
\hline
$\Gamma ({7\over2}^-) (keV)$ & $\sim 110$ & $93 \pm 8^{(c)}$\\
$\Gamma ({5\over2}^-) (keV)$ & $\sim 930$ & $875^{+200}_{-100}$\\ \hline
\end{tabular}\\
\end{center}
\caption{Comparison of calculated and experimental quantities taken from: 
(a) Walliser and Fliessbach \cite{Wall},  (b) Voelk and Fick \cite{Voel}, 
(c) Tokimoto et al. \cite{Toki}}
\end{table}

\begin{table}[t]
\begin{center}
\begin{tabular}{|l|c|c|}
\hline 
Quantity& My work & Other works\\ \hline \hline
$<r^2>_{ch}^{1/2} (fm)$ & $2.44$& $2.43^{(a),(b)}$ \\
&&$2.55^{(c)}$\\ \hline
$Q_{mat} (fm^2)$& $-3.99$ & $-3.82^{(a)}$\\ 
&&$-3.83^{(b)}$\\
&&$-4.41^{(c)}$\\ \hline
$B(E2,{3\over 2} \rightarrow {1\over 2}) (e^2fm^4)$& $7.55$ & $7.74^{(a)}$\\
&&$7.75^{(a)}$\\ 
&&$10.57^{(b)}$\\ \hline
\end{tabular}
{\tiny (a) Keeley,Kemper and Rusek, PRC {\bf 60} ('02),
(b) Buck and Merchant JPG {\bf 14} ('88),
(c) Kajino {\it et al.}, PRL {\bf 46} ('81) }
\end{center}
\end{table}

We have set a simple model for $~^7$Li that nevertheless is capable of
a good agreement with experimental observations, as witnessed by the list of
observables in Table 1. Evaluation of charge radius, electric and matter 
quadrupole moments, $B(E2)$ and $B(M1)$ values for transitions between 
the ground state and the first excited state are reported.
These quantities, except the two width, are calculated accordingly 
to the prescriptions given in \cite{Wall}.
These quantities are very sensitive to the wavefunction shape and therefore
provide a reliability test for our approach as far as bound states are 
concerned. 
The two last rows 
in the table refer to the width of the two $f_{7/2}$ and $f_{5/2}$ 
resonances on which we will comment later and have been calculated with the 
purpose to show that this model gives also sensible predictions for the 
continuum states.
We also compare our results with older calculations in the second table, 
showing that an overall agreement is found.

\section{Electromagnetic response}
\index{subject}{electromagnetic response}
So far we have showed that the dicluster picture is able to give 
reasonable results and we would like to  to apply this
model to the calculations of electromagnetic response for the transitions
to continuum states.
\begin{figure}[!t]
\begin{center}
\epsfig{figure=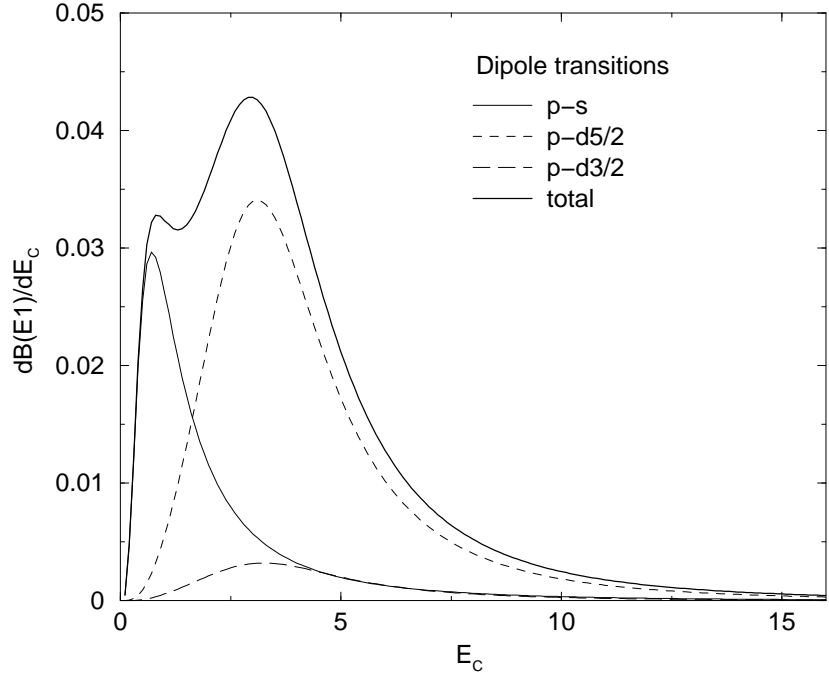,width=0.9\textwidth}
\end{center}
\caption{Differential B(E1) values (in $e^2 fm^2/MeV$) for transitions 
from the ground state to the continuum. 
Energies are in MeV, the different contributions are indicated
in the legend. }
\label{dipole}
\end{figure} 
Starting from the ground state (with $p$ character) we have investigated 
electric dipole transitions to $s$ and $d$ states (in fig. \ref{dipole} we
give the differential reduced transition probability for dipole transitions)  
as well as quadrupole transitions to $p$ and $f$ states (fig. 
\ref{quadrupole}). In the former case the scattering states for even 
multipolarities have been calculated solving again a Schr\"odinger equation 
with the same parameters that have been used to find the bound states.
The same has been done in the case of the $p$-continuum.
In this scheme all the features of the transition are ascribed
to the modification of the character of the angular momentum of the 
relative motion. The clusters are thus frozen in this picture, and their
intrinsic wavefunctions are not modified by the electromagnetic operators.
In the continuum energy region are present a number of resonances that 
deserve some further comments. 
\begin{figure}[!t]
\begin{center}
\epsfig{figure=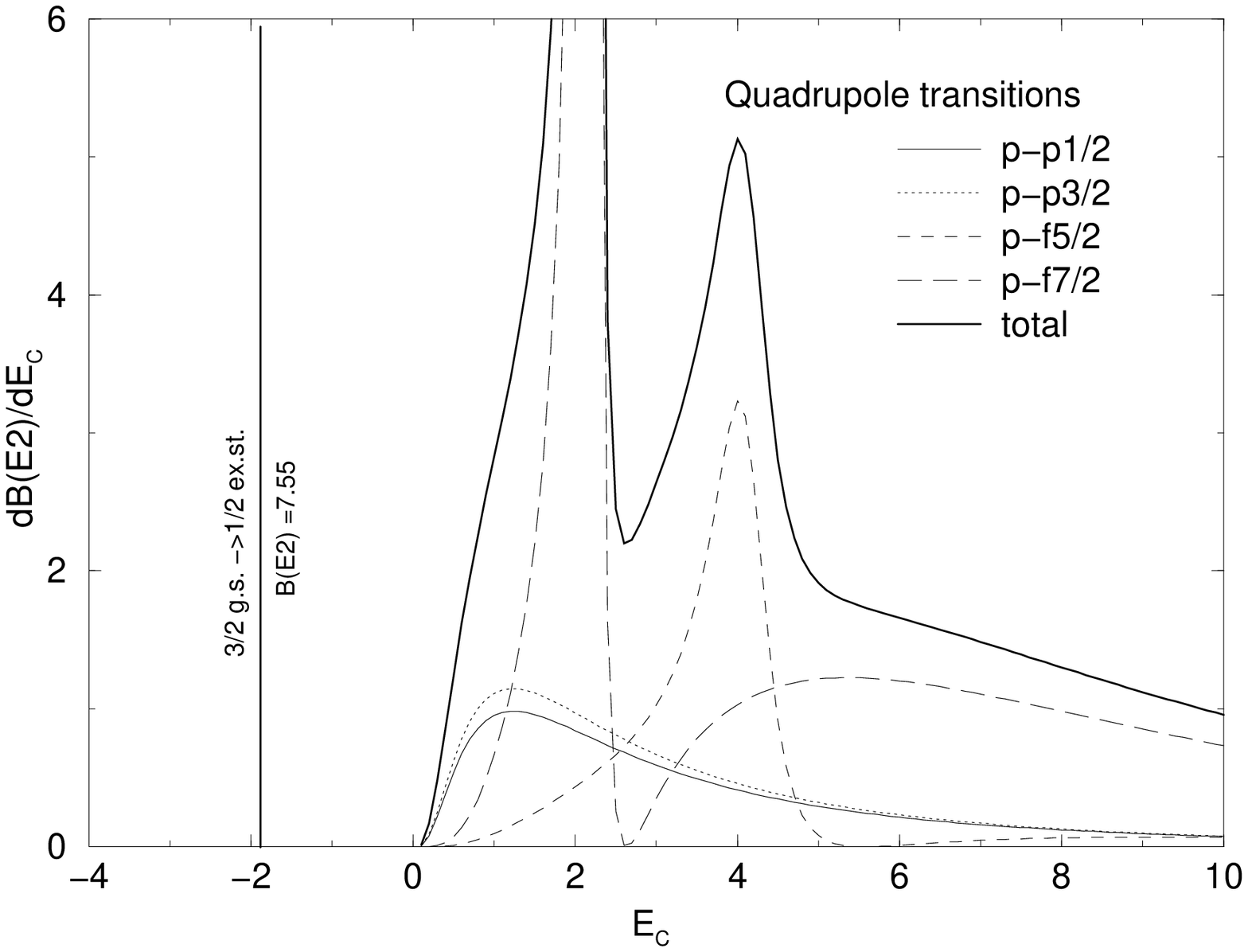,width=0.9\textwidth}
\end{center}
\caption{Differential B(E2) values (in $e^2 fm^4/MeV$) for transitions 
from the ground state to the continuum. 
Besides there is a quadrupole transition to the first excited 
(bound) state, displayed in the figure as narrow peak, whose strength
is indicated on the figure itself. Energies are in MeV, referred to the 
threshold for break-up into the $\alpha-t$ channel.}
\label{quadrupole}
\end{figure} 

\begin{figure}[!h]
\begin{center}
\epsfig{figure=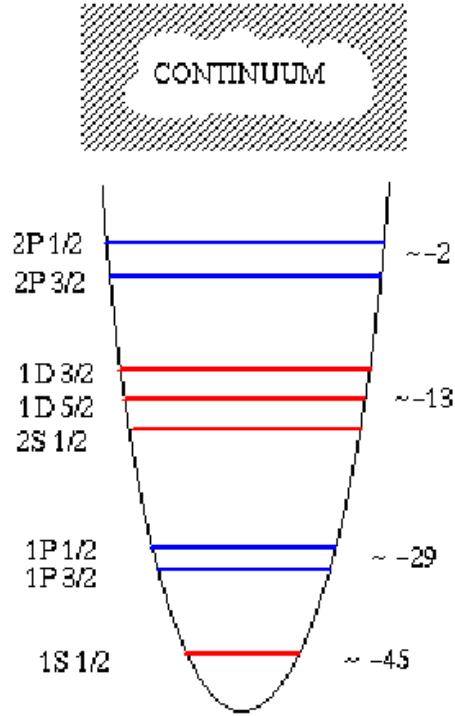,width=0.5\textwidth}
\end{center}
\caption{Schematic representation of the physical (2P) and unphysical 
(1S,1P,1D,2S) bound states and of the 
continuum states obtained solving the Schr\"odinger equation for the relative 
motion of the two clusters. They must be taken into account in the calculation
of the energy weighted molecular sum rule. The quantum numbers of 
the states are on the left, while on the right is indicated their energy 
with respect to the threshold (in MeV).}
\label{unphy}
\end{figure}

Limiting ourselves to the lowest few ones, as the ${7\over 2}^-$ and the 
${5\over 2}^-$, we have solved the Schr\"odinger equation for scattering 
states with a depth of the Woods-Saxon ($V_{WS}=-68.255$) and a spin-orbit
($V_{ls}=3.115$) adjusted to yield the two resonant states in the 
excitation spectrum just at the right energy.
Besides this resonant strength we observed a concentration of strength of
non-resonant character at the separation threshold, solely due to the 
weakly-bound nature of the $~^7$Li nucleus. This strength is small for 
multipolarities that have a resonance in the low-lying continuum, but it is
sizable when there are no resonances  (as in the $p$ cases). The widths of 
these two states are in reasonable agreement with experimental observations 
as shown in the last part of table 1.\\
\index{subject}{energy weighted sum rule}
\index{subject}{energy weighted molecular \\sum rule}
We have also compared our values with energy weighted sum rules as well as
with energy weighted molecular sum rules (EWMSR) \cite{Alha,Lang}, also 
called AGB sum rule from the initials of the three persons who introduced it
in the literature, that are particularly useful for molecular-like structures.
In light nuclei enhanced $E1$ 
transitions have been observed for which $B(E1)$ values may still be very small
in comparison with single-particle estimates. EWMSR have been introduced
as measures for these transitions and in the cases of dipole and quadrupole 
they read:
\be
S_I(E1,A_1+A_2)= \left( {9\over 4\pi} \right) {(Z_1A_2-Z_2A_1)^2 \over AA_1A_2}
\left( {\hbar^2 e^2\over 2m} \right)
\ee 
\be
S_I(E2,A_1+A_2)= \left( {25\over 2\pi} \right){1\over Z} 
\left(Z_1Z_2 +\Bigl(Z_1{A_2\over A}-Z_2{A_1\over A}\Bigr)^2 \right) S_0^2
\left( {\hbar^2 e^2\over 2m} \right)
\ee
where the notation means that the nucleus with mass $A$ and charge $Z$ is 
split in two clusters with masses $A_1$ and $A_2$, charges $Z_1$ and $Z_2$ and
neutron numbers $N_1$ and $N_2$.
$S_0$ is the equilibrium separation that may be simply calculated as the sum
of the radii of the two clusters (we have taken $S_0=3.63$ fm). 
We find that our dipole strength represents approximatively the 2.6\% of
the Thomas-Reiche-Kuhn sum rule, but it amounts to about 94\% of the energy 
weighted molecular dipole sum rule. Similarly the quadrupole strength is the
9.2\% of the energy weigthed quadrupole sum rule and about 42\% of the
EWMSR. 
Starting from the initial ground state ($2p_{3/2}$) we have included 
,in the calculation of the exhausted fraction
of sum rules, all the possible transitions to lower unphysical bound states
($1s_{1/2},2s_{1/2},1d_{5/2},1d_{3/2}$ for dipole and $1p_{3/2},1p_{1/2}$ for
quadrupole) and we have also included the quadrupole transition to the 
first excited state (see fig. \ref{unphy} for a schematic view).

\begin{table}[!t]
\caption{Exhausted fraction of the energy weighted sum rule  and of the 
energy weighted 
molecular sum rule (AGB sum rule) for dipole and quadrupole excitation.}
\label{ew}
\begin{center}
\begin{tabular}{|c|c|c|}
\hline
~&E1 & E2\\ \hline
EWSR&$2.62\%$ & $\sim 9\%$ \\
EWMSR&$94.22\%$ & $\sim 42\%$ \\\hline 
\end{tabular}
\end{center}
\end{table}

While the quadrupole transitions to bound states 
(both the physical $2p_{1/2}$ excited state and the two unphysical $1p$
levels) represent only a small part (about 2\%) of the total quadrupole 
strength (that amounts to $\sim 317 MeV e^2fm^4$), 
the total dipole strength is strongly modified: 
the dipole transitions to the continuum amount to $\sim 1.311 MeVe^2fm^2$
and the dipole transitions to unphysical states give a negative contribution
of about $\sim -0.645 MeVe^2fm^2$ to the total strength. While the result 
for molecular dipole energy weighted sum rule is an exact one, the other is 
rather robust with respect to small variations in the $S_0$ parameter. 
All these values are summarized in Table \ref{ew}

\section{Formalism and Form Factors}
\begin{figure}[!t]
\begin{center}
\begin{picture}(300,200)(0,0)
\psset{unit=1.pt}  
\pscircle*(280,100){20}   \rput(280,60){Target}
\pscircle(80,50){35}\rput(80,10){Core}
\pscircle(100,140){20}\rput(100,165){Cluster}
\psline{<->}(80,50)(100,140)
\psline{->}(280,100)(80,50)
\psline{->}(280,100)(100,140)
\psline{->}(280,100)(91,100) \rput(80,100){CM}
\rput(130,108){$\vec R$}
\psbezier[linewidth=0.5]{->}(40,180)(100,130)(30,140)(94,112)
\rput(35,190){$ f_1 \vec r$}
\psbezier[linewidth=0.5]{->}(10,60)(40,70)(20,98)(89,88)
\rput(00,50){$ -f_2 \vec r$}
\rput(180,140){$ \vec R+f_1 \vec r$}
\rput(180,60){$ \vec R-f_2 \vec r$}
\end{picture}
\end{center}
\caption{Coordinate system for the interaction between a dicluster nucleus
(white) and an external target (black).}
\label{coord}
\end{figure}
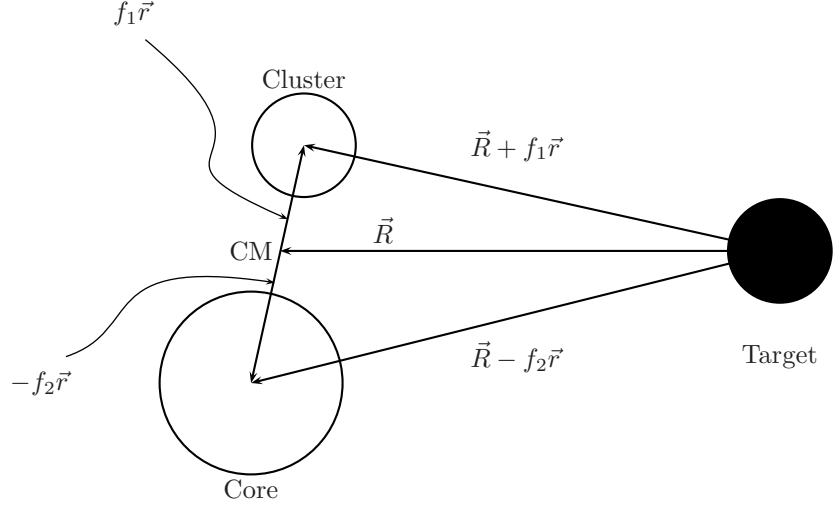

We wish now to move from the electromagnetic response, that is a structural 
feature of $^7$Li, to the study of a breakup reaction in which the dicluster
nucleus is used as a projectile on an heavy target.
\index{subject}{form factors}
The coordinate system for the interaction between a dicluster
nucleus and a target is depicted in fig. \ref{coord}.
The factors $f_1$ and $f_2$ are the ratios of the distances of the center of 
mass of each cluster from the common center of mass divided by the 
inter-cluster
distance $r$. We have named the two clusters as 'core' and 'cluster' to avoid
confusions even if the alpha particle has not a mass so large to justify the 
choice with respect to triton. Nonetheless it has spin $0$ that helps in 
simplifying formulae.

The wavefunction for the bound initial state with angular momentum quantum 
numbers $J,M$ is:
\be
\Psi_{JM}(\vec r,\xi_1,\xi_2) \=\sum_{\mu,m_2} \langle l\mu j_2 m_2 
\mid J M \rangle \varphi_{l\mu}(\vec r) \Phi_1^{j_1=0,m_1=0} (\xi_1) 
\Phi_2^{j_2m_2} (\xi_2) 
\ee 
while for the final scattering state with quantum numbers $J',M'$ it reads:
\be
\Psi_{J'M'}(\vec r,\xi_1,\xi_2)\= \sum_{\mu',m_2'} \langle l'\mu' 
j_2' m_2' \mid J' M' \rangle \varphi_{l'\mu'}(\vec r) \Phi_1^{j_1'=0,m_1'=0} 
(\xi_1) \Phi_2^{j_2'm_2'} (\xi_2) 
\ee
where $\xi_i$ are the internal coordinates of the two clusters with 
wavefunction $\Phi_i$ and intrinsic quantum numbers ${j_i,m_i}$.
Instead $\varphi_{l'\mu'}(\vec r)$ is the relative motion wavefunction, 
depending only on the relative coordinate.
For alpha particles we have $j_1=0$ and $m_1=0$, while the triton 
has $j_2=1/2$, and hence we may simplify both formulae.
The relevant interaction between the target and each component of
the projectile, $V$, may be split in two parts: 
$V_{\alpha-T}(\mid \vec R-f_2 \vec r\mid)+V_{t-T}(\mid\vec R-f_1 \vec r\mid)$. 
Furthermore each interaction consists in a nuclear and a coulomb part. 
The former has a Wood-Saxon parametrization of the standard type:
\be
V^N_{i-T} (\mid\vec r_i\mid) \= {V_{i-T} \over 
1+e^{\bigl[(\mid \vec r_i\mid-R_T)/a_{i-T}\bigr] } }
\ee
where the index $i$ refer to one of the two clusters and the notation
employed in the figure for the relative distances between each cluster 
and the target is shortened to $\mid \vec r_i \mid$.
The Coulomb interaction takes into account the extension of the 
charge distributions:
\begin{eqnarray}
V^C_{i-T} (\mid\vec r_i\mid) &=& \Phi_0 {R_T \over \mid \vec r_i \mid }
, \qquad \qquad \mbox{if $\mid \vec r_i \mid > R_T$} \nonumber \\
V^C_{i-T} (\mid\vec r_i\mid) &=& \Phi_0 \Biggl[ {3\over 2}-{\mid \vec r_i 
\mid ^2 \over 2 R_T^2}  \Biggr], \quad \mbox{if $\mid \vec r_i \mid < R_T$}
\end{eqnarray}
where $\Phi_0 = Z_{eff}Z_T e^2/R_T$.

\begin{figure}[!h]
\begin{center}
\epsfig{figure=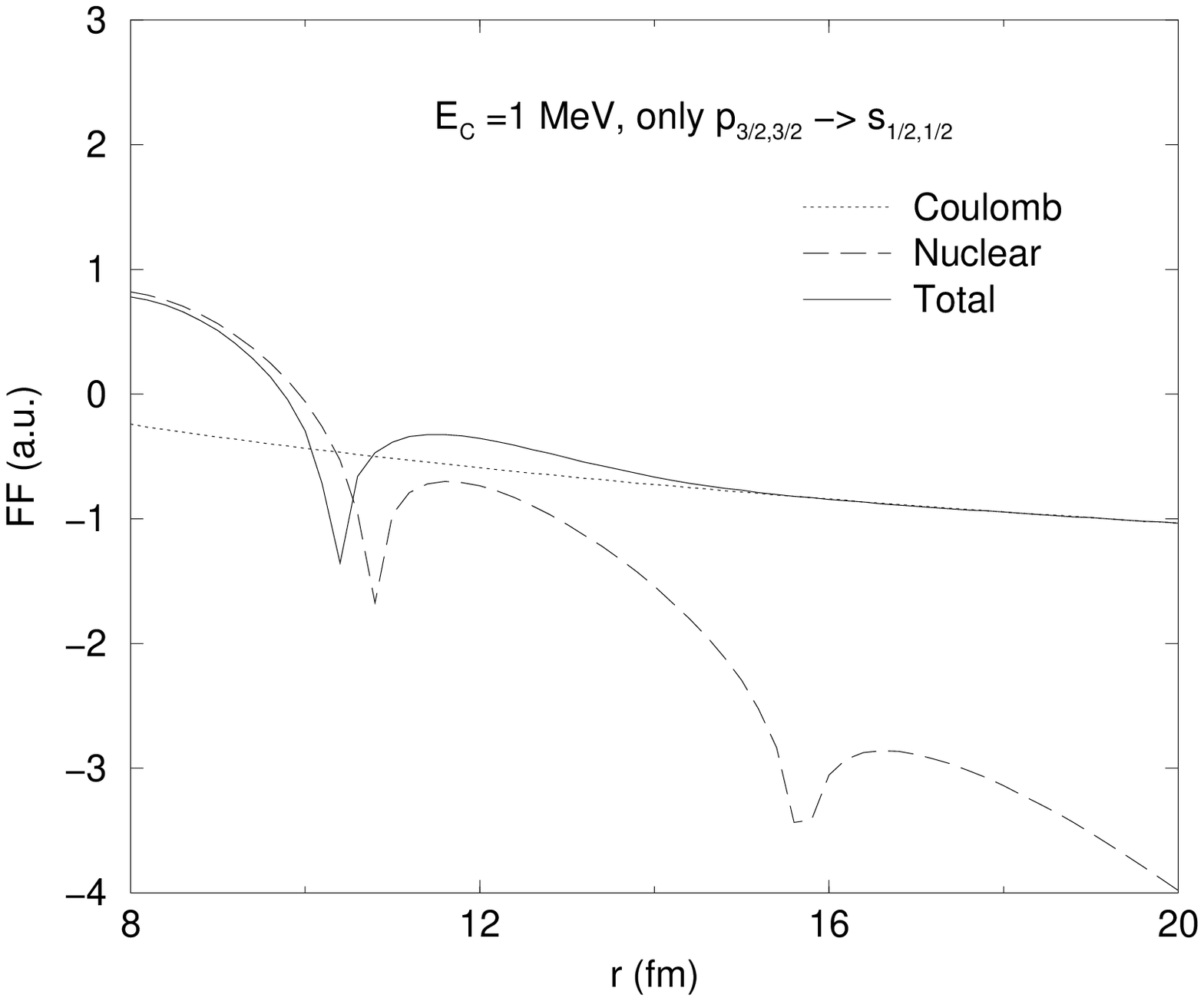,height=0.4\textheight}
\vspace{.5cm}
\epsfig{figure=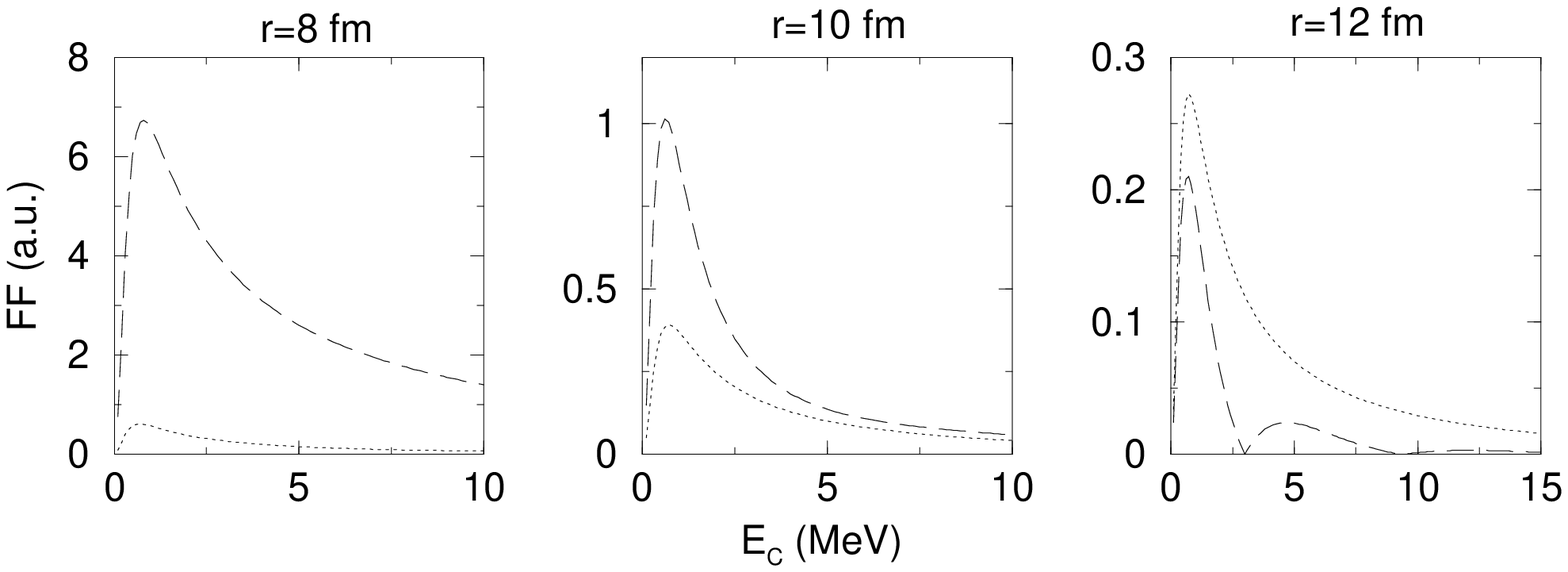,width=1.\textwidth}
\end{center}
\caption{Form factors (in arbitrary units) for a particular transition 
plotted against the distance, for a fixed energy in the continuum of 
$E_{C}=1 MeV$ (upper figure, logarithmic vertical scale) 
and against the energy in the continuum for
three fixed distances (lower row). Coulomb (dotted) and nuclear (dashed)
form factors are shown. See text for details.  }
\label{formfact}
\end{figure}

The formfactor depends only on
the $\vec R$ coordinate that is referred to the distance between the centers 
of mass of the two reactants and it has the form
$$
F(\vec R) \= \langle E J'M'\mid V(\vec R) \mid JM \rangle \=$$
\be
\int d\vec r d\xi_1 d\xi_2
\Psi_{J'M'}^{*}(\vec r,\xi_1,\xi_2) \Bigl( V_{1-T}(\mid \vec r_1\mid)+
 V_{2-T}(\mid \vec r_2\mid) \Bigr) 
\Psi_{JM}(\vec r,\xi_1,\xi_2)
\ee
that can be further reduced inserting the wavefunctions above and
integrating over the internal degrees of freedom. Using also some
properties of the Wigner 3J and 6J coefficients and of the spherical 
harmonics one ends with the formula:
$$F(\vec R)_{JMl\rightarrow J'M'l'} \= $$
$$\= \sqrt{\pi(2J+1)(2J'+1)}   
\sum_{L,\Theta} (-1)^{1/2-M'} 
\Biggl( {J \atop -M} {J' \atop M'} {L \atop \Theta} \Biggr) 
\Biggl( {J \atop 1/2} {J' \atop -1/2} {L \atop 0} \Biggr) $$
\be
\Biggl[ \int_0^\infty r_i^2 dr_i\int_{-1}^1 du R^*_l(r_i)R_{l'}
V(\sqrt{R^2+r_i^2-2r_iRu}) P_L(u)\Biggr] Y_{L\Theta} (\hat R) 
\label{fofa}
\ee
where ${L,\Theta}$ are the change in orbital angular momentum and its 
third component due to the transition and $u$ is the cosine of the angle 
between the two vectors $\vec R$ and $\vec r_i$. Since the index $i$ is 
present in the square bracket there are two similar terms referring to the 
the two clusters. The above formula refers to a cluster with intrinsic spin
quantum number $1/2$ and a more general one is provided in the appendix.

We show some results in fig. (\ref{formfact}) where 
the Coulomb and nuclear form factors, in the same arbitrary units, are plotted
for a dipole transition between the $p_{3/2,3/2}$ 
(second index refers to third component of angular momentum) 
ground state and the 
$s_{1/2,1/2}$ state in the continuum. From the upper panel, that shows
the dependence upon $r$ when $E_C =1$ MeV, it is evident that the nuclear field
dominates at smaller distances, while the coulomb one dominates at larger 
distances. This is once again displayed in the next three figures where three
different distances have been kept constant and the dependence upon $E_C$
has been calculated. 
The nuclear contribution is still very important at a distance of $12-14$ fm
that is far beyond the geometrical sum of the radii of the two systems.
In halo systems close to the drip lines, where the wavefunctions are 
much more extended than in this case, this effect is even magnified.
It is seen that the nuclear form factor starts to show
radial nodes when the distance is increased as already noticed in \cite{Das1}. 
Similar figures may be obtained for all the possible transitions and all
display similar features.

\section{Cross section}
The formfactors obtained in the last section contain all the elements to 
build up elastic and inelastic cross-sections and Q-value distributions. The
reaction amplitudes can be calculated in a semiclassical coupled-channel 
approach.
\index{subject}{semiclassical coupled channel}

\begin{figure}[!t]
\begin{center}
\epsfig{figure=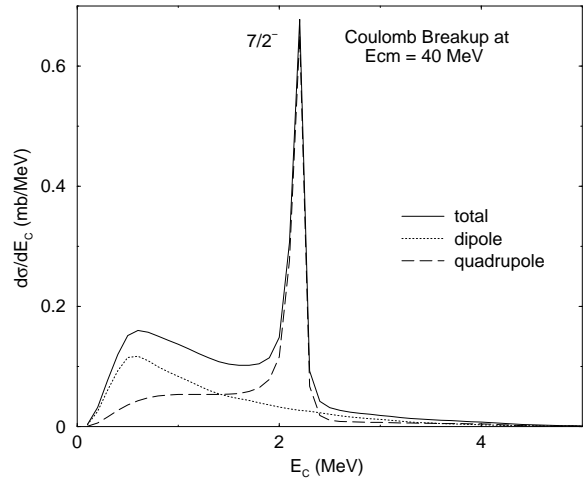,height=0.33\textheight}
\end{center}
\caption{Q-value distribution for Coulomb breakup of $^7$Li on $^{165}$Ho at
$E_{cm}=40 MeV$. Dipole (dotted) and quadrupole (dashed) 
contributions are shown together with their sum (full line). The $7/2^-$ 
resonance is marked, while the $5/2^-$ around 4 MeV is very small.}
\label{crossE}
\end{figure}
\begin{figure}[!h]
\begin{center}
\epsfig{figure=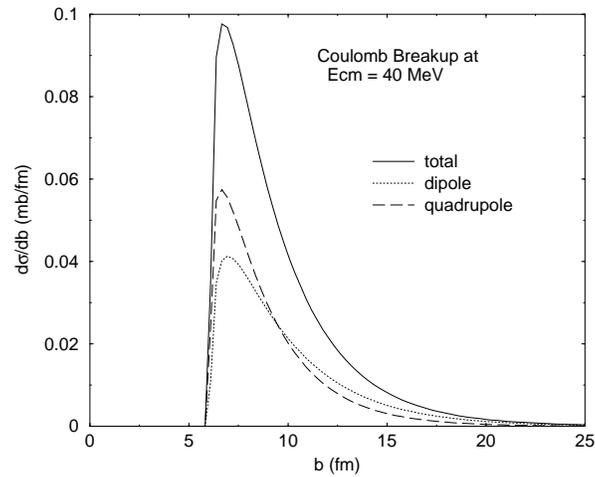,height=0.33\textheight}
\end{center}
\caption{Differential Coulomb breakup cross sections as a function of 
the impact parameter with the same data of the last figure. Again different
multipolarities are shown separately and one may notice the different 
behaviour at large impact parameter, that is dominated by the dipole 
contribution.}
\label{crossb}
\end{figure}     

We implement here a scheme analogous to the one presented in \cite{Das1}, 
that follows a standard way of construction of elastic and inelastic 
cross sections \cite{Brog}.
For each partial wave ($\ell$) we have a set of coupled first 
order differential equations for the amplitudes:
\be
\dot a_i^\ell = -i\hbar \sum F_{i,j}[\vec R (t)] 
e^{-i(\epsilon_i-\epsilon_j)t/\hbar} a_j^\ell
\ee 
where $\vec R (t)$ represents the trajectory of relative motion, while
$\epsilon_k$ is the energy of a particular channel.
The integration in time is done assuming a standard parametrization for 
the ion-ion potential with an imaginary part that yields a first-order 
elastic shift.
The energy in the continuum is divided in a suitable number of intervals, 
treated as different channels. The reaction amplitudes obtained from this
system of equations are used to calculate cross sections for the excitation 
of a given channel and differential cross sections as a function of the 
energy in the continuum.
The calculations of total cross sections (as well as differential ones)
must take into account the fact that one can have many choices for the 
initial magnetic substate (over which we must average) that are simplified
since $\sigma_{p,j,m\rightarrow lj'}=\sigma_{p,j,-m\rightarrow lj'}$. 

\begin{figure}[!t]
\begin{center}
\epsfig{figure=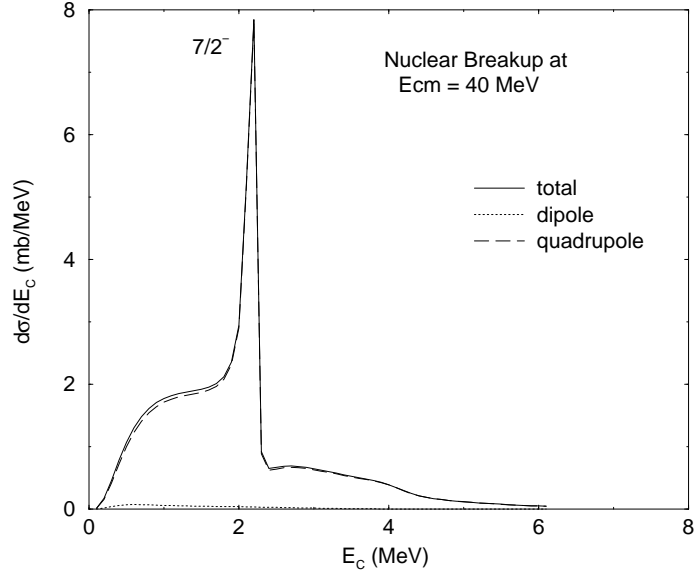,height=0.4\textheight}
\end{center}
\caption{Q-value distribution for nuclear breakup of $^7$Li on $^{165}$Ho at
$E_{cm}=40 MeV$. Dipole (dotted) and quadrupole (dashed) 
contributions are shown together with their sum (full line). The $7/2^-$ 
resonance is marked.}
\label{crossE_N}
\end{figure}    

Fig. \ref{crossE} is very interesting since displays the expected
Q-value distribution for Coulomb breakup ( the contributions of the dipole 
and quadrupole transitions are separately shown, together with their sum).
It is worthwhile to notice that the strength in the two peaks arise from 
different mechanisms: the peak at $1$ MeV is mostly build up with transition
to the continuum due to the matching between initial and final wavelengths
that we have already discussed, while the peak at $2.186$ MeV has a real
resonant nature ($7/2^-$).  Also in the case of quadrupole the non-resonant
strength is seen  just above the threshold, but its relative magnitude is 
small compared to the dipole one.
The same informations about the relative importance of various 
multipolarities may be deduced from figure \ref{crossb}. In this picture
one can appreciate the role of the absorption at small impact parameters,
due to the implementation of a transmission factor obtained integrating the
imaginary part of the potential.
Moreover one can see the different behaviour of the two tails:
the quadrupole contribution goes to zero in a faster way with respect to the
dipole. Consequently at large impact parameters, that correspond to forward
angles, the Coulomb breakup cross sections are mostly due to dipole transitions
to the continuum. One can thus expect to measure dipole breakup preferably
at forward angles.

\begin{figure}[!t]
\begin{center}
\epsfig{figure=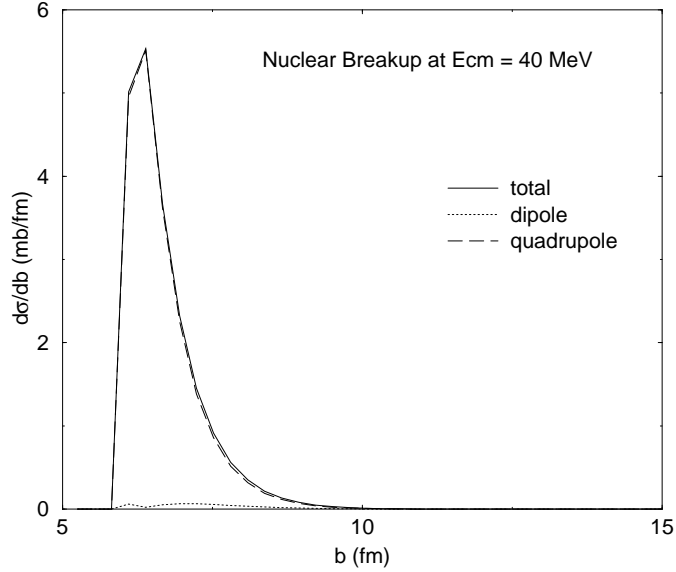,height=0.4\textheight}
\end{center}
\caption{Differential nuclear breakup with respect to the impact parameter.}
\label{crossE_N}
\end{figure}  
\begin{figure}[!h]
\begin{center}
\epsfig{figure=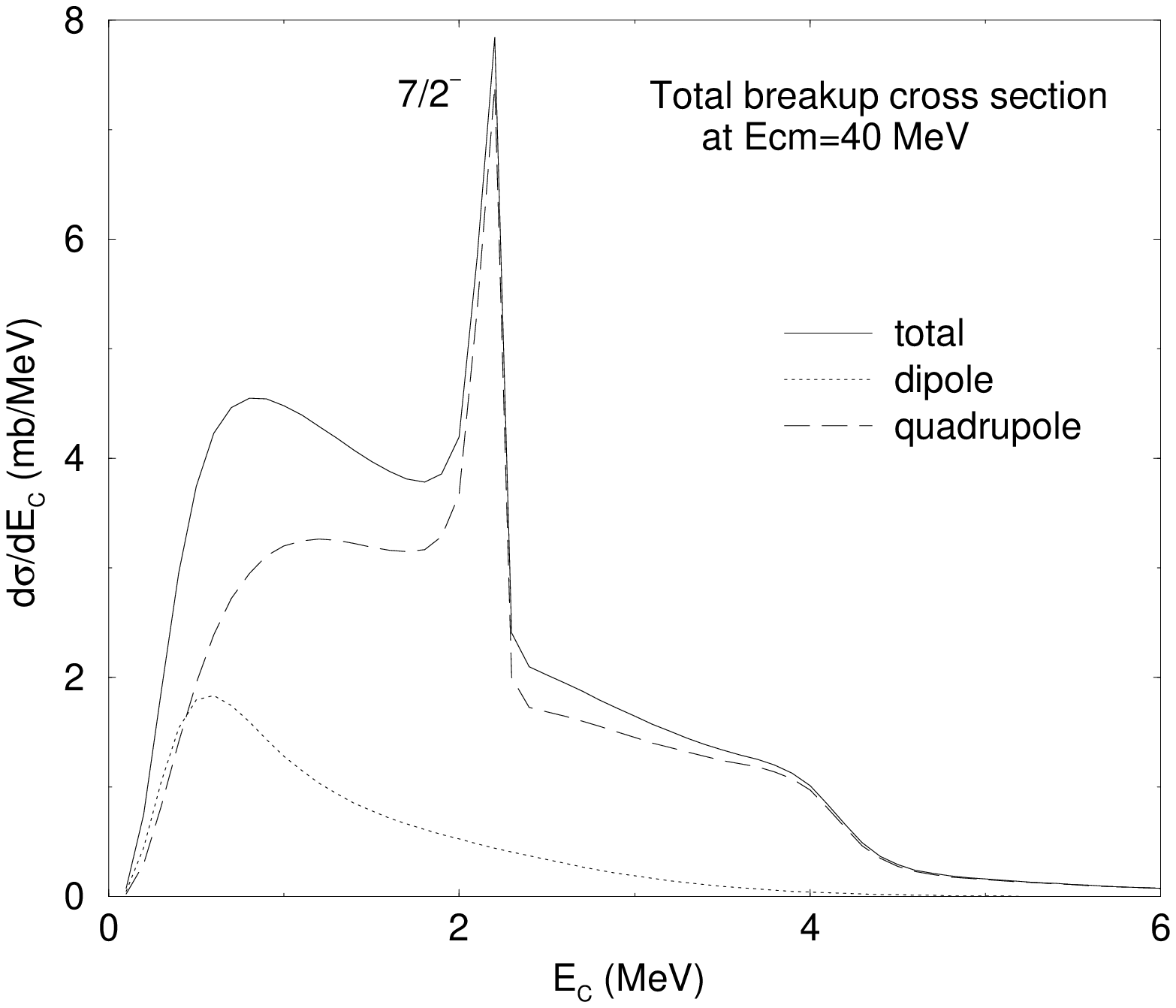,height=0.4\textheight}
\end{center}
\caption{Q-value distribution for total (interference of Coulomb and nuclear)
 breakup of $^7$Li on $^{165}$Ho at
$E_{cm}=40 MeV$. Dipole (dotted) and quadrupole (dashed) 
contributions are shown together with their sum (full line). The $7/2^-$ 
resonance is marked.}
\label{crossE_I}
\end{figure}    
\begin{figure}[!h]
\begin{center}
\epsfig{figure=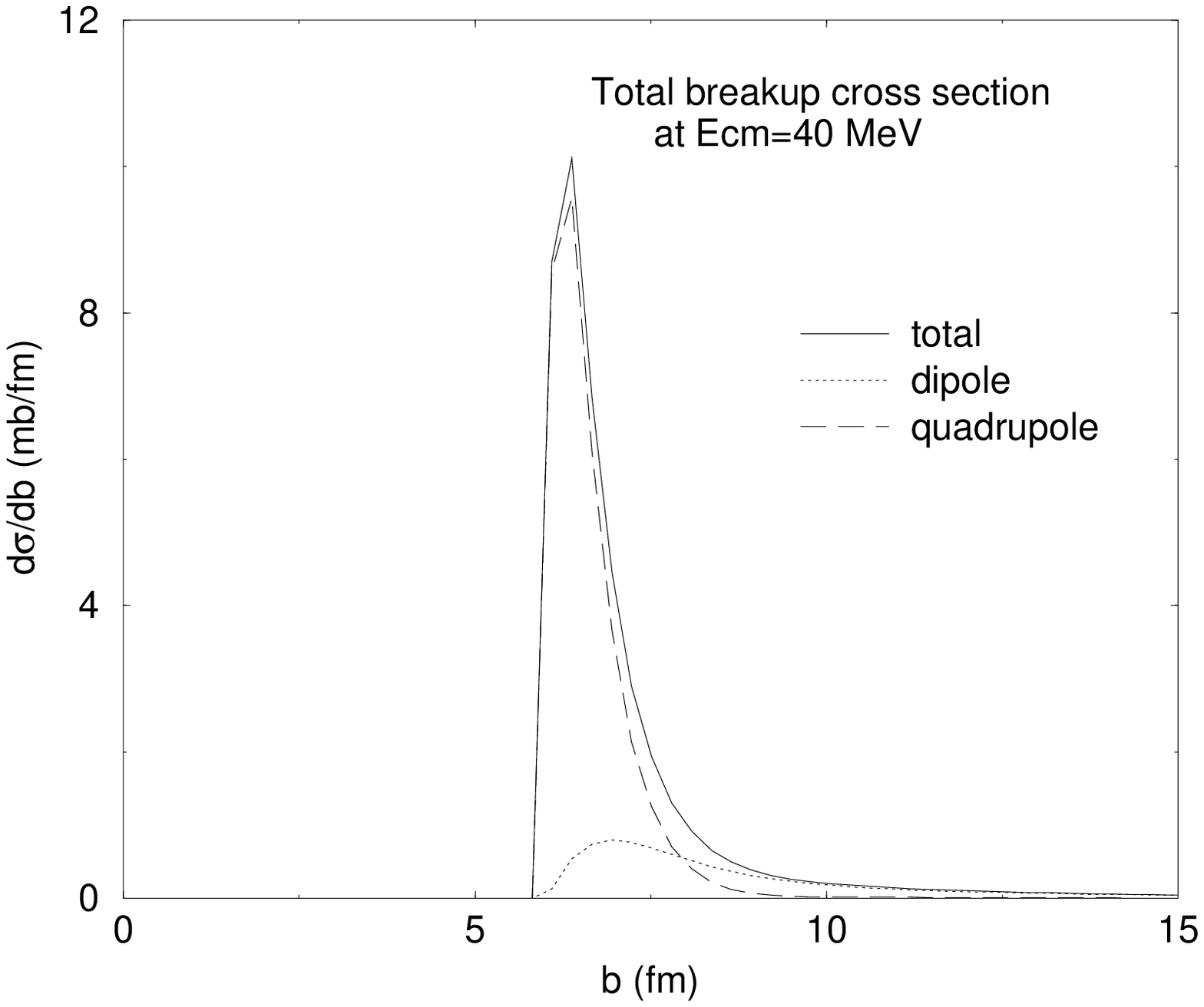,height=0.4\textheight}
\end{center}
\caption{Differential total breakup with respect to the impact parameter for 
the interference of Coulomb and nuclear fields.}
\label{crossb_I}
\end{figure}  
The total Coulomb cross section at $E_{cm}=40$ MeV is around $0.48$ mb, 
that could be further analyzed separating the contribution of the different 
multipolarities: $0.20$ mb for the dipole and $0.28$ mb for the quadrupole. 

The nuclear breakup at the same energy has a very similar shape for the 
Q-value distribution (depicted in fig. \ref{crossE_N}), being the total
integrated cross-section about $5.78$ mb.  
At a variance with previous findings the dipole contribution to this
cross-section ($0.14$ mb) is now much smaller that the quadrupole 
one ($5.64$ mb).

The comparison with fig. \ref{crossE} shows that, while the resonances are 
always due to the quadrupole component, the peak at low energy is mostly due
to dipole transitions for the Coulomb field and to quadrupole transitions
for the nuclear field and has a different shape. 

Summing up these two considerations results in a 
rather complicated mixing of the various components excited by the cumulative 
effect of the two interactions. 
The coulomb and nuclear cross sections have been displayed explicitly only 
to discuss their features.
The interference of the two fields give
a final Q-value distribution depicted in fig. \ref{crossE_I} and a 
corresponding curve as a function of the impact parameter in fig.
 \ref{crossb_I}. 
The total cross sections in this case is about $11.9$ mb. 
The dipole contribution amounts to $2.5$ mb while the quadrupole 
is about $9.4$ mb. This result is rather large if compared with the separated 
Coulomb and nuclear cross sections and it is due to the interference.
In order to check the validity of these estimates we have performed a simple
and independent calculation with another code that evaluates Coulomb 
and nuclear
excitation from the ground state to a sharp state (we did it for the $7/2^-$
resonance) given its energy and the deformation length calculated on the basis 
of the B(E2) value (that is the integral of the curve in figure 
\ref{quadrupole}
corresponding to the resonance). The result gives $0.34$ mb for Coulomb
excitation. This number is roughly consistent with the value of the quadrupole 
component of the Coulomb cross section (half of which may be attributed to the
resonance).

\section{Parallel work on $^6$Li}
\index{subject}{$^6$Li}
Besides the detailed discussion on $^7$Li a parallel investigation on $^6$Li
has been carried on with the purpose of comparing the results with other 
theoretical predictions and experiments \cite{Mazz1,Kelly}. We briefly
report here on the status of this research line.

\begin{figure}[!t]
\begin{center}
\includegraphics[height=.85\textheight]{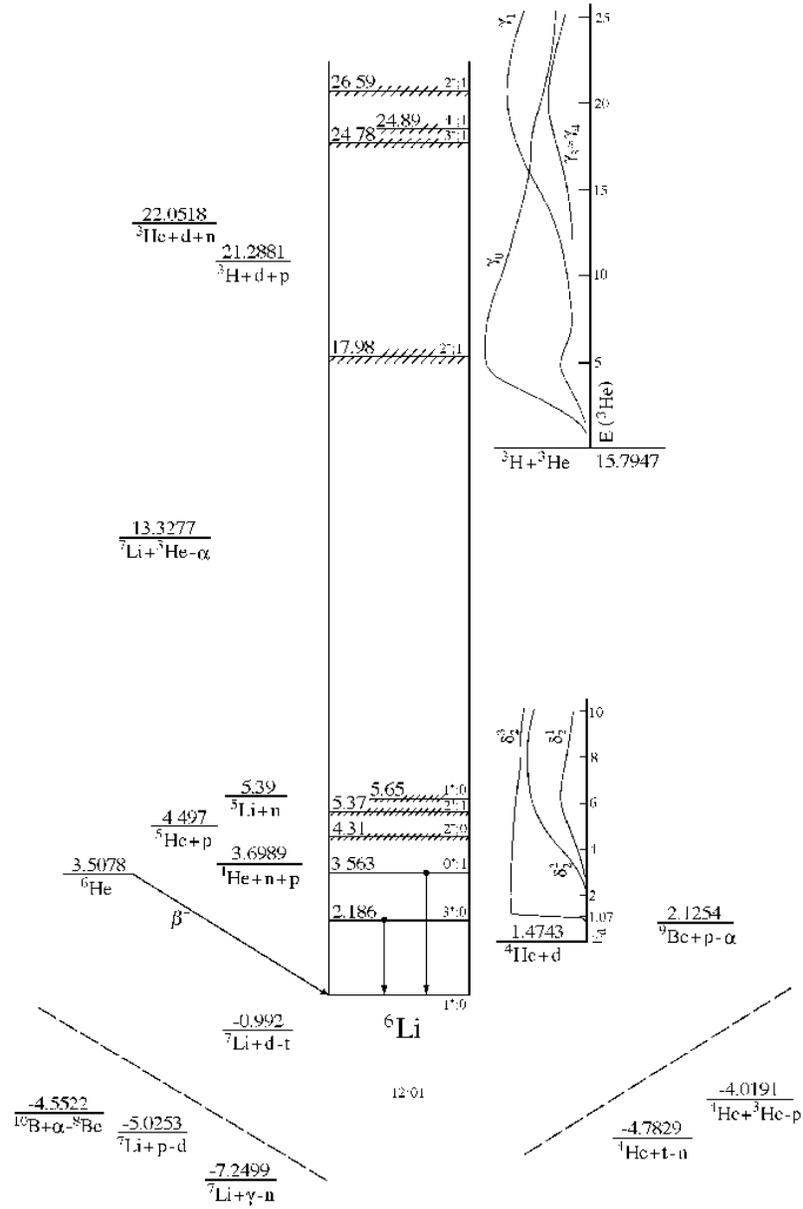}
\end{center}
\caption{Level scheme of $^6$Li from TUNL website \cite{TUNL}. The data
displayed here come from the compilation of Fay Ajzenberg-Selove.}
\label{6Lischeme}
\end{figure}

The level scheme of $^6$Li is reported in fig \ref{6Lischeme} for the sake of 
completeness. The ground state is the only bound state below the threshold 
for separation into an $\alpha$ particle plus a deuteron. Its binding energy 
is about 1.5 MeV. This makes of $^6$Li a very weakly-bound nucleus. Three 
resonances
with $T=0$ isospin quantum number are seen in the low-lying continuum with
$J^\pi \= 3^+,2^+,1^+$ interpreted as the coupling of the spin of the clusters
with a d-wave relative motion. Resonances with $T=1$ are not included in the 
present analysis.

The same physical idea of the description of $~^6$Li as a 
dicluster nucleus (alpha plus deuteron) has been retained and 
the Schr\"odinger equation has been
solved for the relative motion both for the bound and the continuum states, 
adjusting the depth of the Saxon-Woods potential to obtain the correct 
binding or resonance energies. 
The relative motion has then been treated along a classical trajectory in a 
semi-classical coupled channel approach that has the advantage to be very
straightforward to solve numerically and to require lower CPU times than other 
commonly used approaches.
This allows a finer subdivision of the continuum (for example we have done the
calculations with energy bins of 0.1 MeV). The results for the
cross sections and excitation energy distribution, although not including 
continuum-continuum couplings, are very similar to the one obtained with
other methods, like the continuum discretized coupled channel method. 
At 39 MeV bombarding energy our total breakup cross section 
(integrated over the continuum energy) is about 85 mb, close to the value of 
104 mb obtained with CDCC procedure.\\
\index{subject}{CDCC}

\begin{table}
\begin{center}
\caption{Comparison between $^6$Li breakup cross section evaluated by
semi-classical calculation and  CDCC, and experimental data. All cross
sections are in mb. It is seen that both theories slightly overestimate the 
experimental values. [Data are due to the courtesy of M.Mazzocco].}
~~~~~~\\~~\\
\begin{tabular}{lccc}
\hline\noalign{\smallskip}
 Beam & Semi-classical    & CDCC & Experimental   \\
 energy & calculation   &  & data   \\
\noalign{\smallskip}\hline\noalign{\smallskip}
\hline\noalign{\smallskip}
 31 MeV & 54.9 & 48.76 & 41.3(0.7) \\
 33 MeV & 69.3 & 65.43 & 59.2(1.2) \\
 35 MeV & 77.9 & 80.91 & 63.0(2.6) \\
 39 MeV & 84.5 & 102.32 & 76.5(5.3) \\
\noalign{\smallskip}\hline
\end{tabular}
\end{center}
\end{table}

\begin{figure}[!t]
\begin{center}
\epsfig{figure=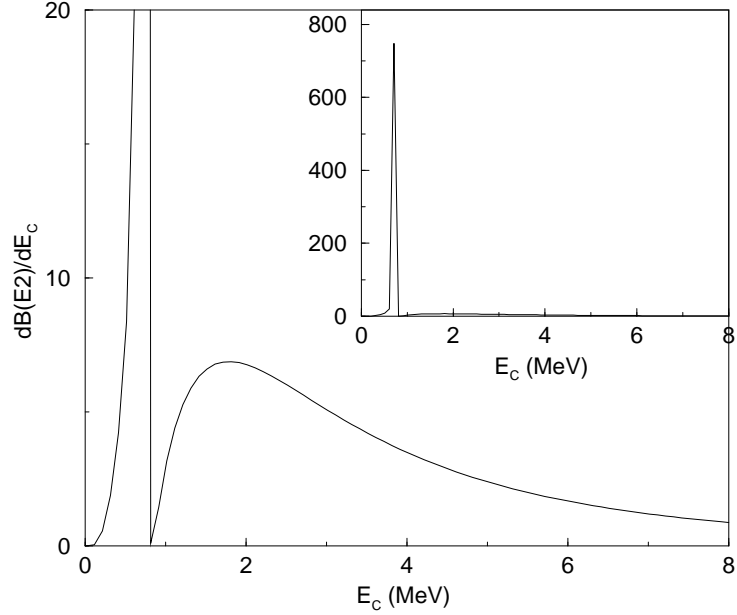,width=0.8\textwidth}
\end{center}
\caption{Differential B(E2) values (in $e^2 fm^4/MeV$) for transitions 
from the ground state to the continuum for the breakup of $^6$Li. 
In the inset we display the full
vertical scale. Energies are in MeV, referred to the 
threshold for break-up into the $\alpha-d$ channel.}
\label{6li}
\end{figure}

The most important qualitative difference is the proper inclusion of 
non-resonant continuum states \cite{Cata} that arise from 
the weakly-bound nature of the $^6$Li nucleus \cite{Das2}. 
We report in fig. \ref{6li} the differential B(E2) values to the low-lying 
continuum (in the case of $^6$Li the dipole vanish identically in a cluster 
picture because the effective charge is zero for an alpha plus deuteron 
cluster configuration).
The transitions to the non-resonant continuum are strong enough to smear 
the 2$^+$ and 1$^+$ resonances, thus introducing a slight qualitative 
difference between the two lineshapes.
This is consistent with the trends of the data sets and a finer 
energy resolution would be advisable to settle the problem.
Instead the 3$^+$ resonance is not modified 
appreciably because, at a variance with the CDCC calculations, we find a 
narrow and strong peak. This difference may be due to the different way
of subdivision of the continuum into energy bins. \\
As it appears from the table below, however, both approaches slightly 
overestimate the integrated experimental cross sections.

\section{Conclusions}
We have illustrated a general method to treat dicluster nuclei, showing
how to obtain their structure features and we have implemented a semiclassical 
calculation to perform breakup reaction studies.\\
The main idea was developed in the papers by Dasso, Lenzi ans Vitturi, who 
showed how the non-resonant continuum affects the excitation properties
and the breakup of neutron and proton halos in weakly-bound nuclei. 
Our aim was to extend this picture to cluster states and to show the 
importance of transition to low-lying continuum in weakly-bound dicluster 
systems.\\
We have not attempted a detailed comparison with other methods,
but we based our arguments on the fact that the cluster description 
of light nuclei is widely accepted.
We expect that the semiclassical implementation of a coupled channel scheme
is not inferior to some, in principle more correct, fully quantistic 
coupled channel calculations. Our scheme is simpler and faster and allows 
a binning of the continuum that may be as fine as one wishes. Other methods
are constrained to a rougher subdivision of the energy.  
Nonetheless we argue that the effects of the non-resonant strength 
on the final cross sections are twofold: they modify the lineshape and 
increase the magnitude of the cross sections in a sizable way.\\
This work may turn out to be relevant to
studies on  complete and incomplete fusion reactions:
in a nuclear reaction with a weakly-bound projectile, the field of the 
target may act on the projectile in such a way to break it before the 
collision, and the breakup may be followed by fusion of both fragments, of 
only one of the two or of none. In this respect the effect of break-up 
on fusion processes is among the topics of higher current interest 
for the nuclear physics community.

\section{Appendix to the chapter}
We would like to give here a formula for the calculation of the form factors 
with the same purpose of (\ref{fofa})
for a generic cluster with intrinsic spin quantum number $j_2$:
$$ F(\vec R)_{JMl\rightarrow J'M'l'} = \sqrt{\pi(2J+1)(2J'+1)
(2\ell+1)(2\ell'+1)}$$
$$ \sum_{L,\Theta} (-1)^{3j_2-M'}
\Biggl( {\ell' \atop 0} {\ell \atop 0} {L \atop 0} \Biggr) 
\Biggl( {J \atop -M} {J' \atop M'} {L \atop \Theta} \Biggr) 
\Biggl\{ {J \atop \ell} {J' \atop \ell'} {L \atop j_2} \Biggr\} $$
$$
\Biggl[ \int_0^\infty r_i^2 dr_i\int_{-1}^1 du R^*_l(r_i)R_{l'}
V(\sqrt{R^2+r_i^2-2r_iRu}) P_L(u)\Biggr] Y_{L\Theta} (\hat R) 
\label{fofa}$$
where the symbols have the same meanings of the text. The simplifications used 
to calculate the form factor for $^7$Li no longer apply when the cluster has 
not $j_2=1/2$.


\chapter{Summary and epilogue}

\section{Summary}
We have tried to give a unitary exposition of the researches that we 
undertook during the three years of study and work in Padova, sewing
together apparently different subjects, that nevertheless have a common
basic goal: the understanding of collective modes and of the continuum 
in which these modes are usually embedded, especially in connection
with novel phenomena seen in nuclei far from the stability line.
We attacked this problem from many different sides taking particular care 
for the role of exotic nuclei either as a subject or as a tool to
understand a particular issue.
Not only the presence of clustering, halos and skins have been considered 
as 'exotic' phenomena, affecting the properties of excitation of dripline 
nuclei, but also 'exotic' (in the sense of unconventional or less frequently 
discussed) collective modes have been discussed, as for example the Giant
Pairing Vibrations excited with unstable nuclei.

The themes covered in the present thesis are: a detailed study of double phonon
giant resonances in normal nuclei, the enhancement of the excitation of
Giant Pairing Vibrations using weakly-bound nuclei, an extension of the 
Steinwedel-Jensen model for Giant Dipole Resonances to treat surface effects 
and presence of neutron skin and, finally, a model on breakup reactions in 
weakly-bound dicluster nuclei.

The common idea underlying these four chapters, and in some way its mortar,
is the interpretation of nuclear phenomena by means of collective models
that have the virtue of being simple and well characterized from a physical
point of view. Needless to say it has been also useful to look at the 
many-body problem in terms of microscopic theories, if need be.
Another cementing feature of our arguing has been the profitable resort to 
semiclassical reaction theory, as far as reaction models were concerned.

The keystone of many problems that we had encountered was to solve
 ordinary differential equation or systems of 
ODE's, from the Schr\"odinger equation in dicluster states to the equations of 
motion in the semiclassical theory nuclear reactions, 
or from the wave equation in 
two-fluids acoustical models to coupled channels reaction amplitudes. 
This task has been accomplished 
analitycally, when possible, or numerically, implementing throughout the 
whole thesis computer codes that exploit a very versatile subroutine to
solve systems of first ordinary differential equations.

~\\~\\~\\
\section*{Epilogue}
There would be no better conclusion to this thesis than to quote again
the same Master that we cited at the beginning of this work, because, once
again, his words are of the utmost relevance:
\begin{quote} 
"He! he! he! --he! he! he! --yes, the Amontillado. But is it not getting late? 
Will not they be awaiting us at the palazzo, the Lady Fortunato and the rest? 
Let us be gone."\\
"Yes," I said, "let us be gone." \\ 
- {\it The cask of Amontillado \index{subject}{amontillado}} - Edgar Allan Poe.
\end{quote}


\printindex{subject}{Subject Index}


\begin{thebibliography}{000}
\bibitem{harak} M. N. Harakeh and A. van Woude, {\it Giant Resonances}
                   (2001 Oxford University Press, New York).
\bibitem{Bothe} W. Bothe and W.Gentner, {\it  Z. Phys.} {\bf 106} (1937) 236.
\bibitem{Bald} G.C.Balwin and G.Klaiber, {\it Phys.Rev.} {\bf 71} (1947) 3.
\bibitem{GT} M.Goldhaber and E.Teller, {\it Phys.Rev.} {\bf 74} (1948) 1046.
\bibitem{SJ} H.Steinwedel and J.H.D.Jensen, {\it  Z.Nat.} {\bf 52} (1950) 413.
\bibitem{Pitth} A.Pitthan and Th.Walcher, {\it  Phys. Lett.} {\bf 318} 
(1971) 563.
\bibitem{Lewis} M.B.Lewis and F.E.Bertran, {\it  Nucl.Phys.} {\bf A196}
(1972) 337.
\bibitem{BroBer} G.F.Bertsch and R.A.Broglia, {\it Oscillation in finite 
quantum systems}, (Cambridge Univerity Press, 1994).
\bibitem{Frasca} N Frascaria, Nucl. Phys. {\bf A687} (2001) 154c-161c.
\bibitem{rev} H. Hemling, {\it Prog. Part. Nucl. Phys.} {\bf 33} (1994) 729;
      T. Aumann, P. Bortignon, and H. Emling,
                 {\it  Annu. Rev. Part. Sci.} vol. {\bf 48} (1998) 351.
\bibitem{Chom}  Ph. Chomaz, and N. Frascaria, {\it Phys. Rep.} {\bf 252} 
(1995) 275.
\bibitem{Ko79} M.A.Kovash {\it et al.}, {\it Phys.Rev.Lett.} {\bf 42} 
(1979) 700.
\bibitem{LAND} LAND collaboration, {\it  Phys.Rev.Lett.} {\bf 70} (1993) 1767.
\bibitem{Eml} H.Emling {\it et al.}, {\it  Nucl. Phys.} {\bf A569} (1994) 141c
\bibitem{wam} S. Nishizaki and J. Wambach, {\it   Phys. Lett.} {\bf B349}
                   (1995) 7;  ibid. {\it   Phys. Rev.} {\bf C57} (1998)1515.
\bibitem{pon} V. Yu. Ponomarev, P. F. Bortignon, R. A. Broglia,
   E. Vigezzi and V. V. Voronov, {\it Nucl. Phys.} {\bf A599} (1996) 341c.
\bibitem{ber} C. A. Bertulani and V. Yu. Ponomarev, {\it  Phys. Rep.} {\bf 321}
                  (1999) 139.
\bibitem{cat}  F. Catara, Ph. Chomaz and N. Van Giai, {\it Phys. Lett.} {\bf
                 B233} (1989) 6.
\bibitem{pon2} V. Yu. Ponomarev, P. F. Bortignon, R. A. Broglia,
     and V. V. Voronov, {\it Phys. Rev. Lett.} {\bf 85} (2000) 1400.
\bibitem{lan} E. G. Lanza, M. V. Andr\'es, F. Catara, Ph. Chomaz and
                 C. Volpe,  {\it Nucl. Phys.} {\bf A613} (1996) 445; 
{\it ibid.} {\it Nucl. Phys.} {\bf A654} (1999) 792c.
\bibitem{vol}  C. Volpe, F. Catara, Ph. Chomaz, M. V.  Andr\'es and
                  E. G. Lanza,  {\it Nucl. Phys.} {\bf A589} (1995) 521;
            {\it ibid.}{\it Nucl. Phys.}  {\bf A599} (1996) 347c.
\bibitem{bor} P. F. Bortignon and C. H. Dasso,  {\it Phys. Rev.} {\bf C56}
                  (1997) 574.
\bibitem{brax} D. Brink, PhD Thesis, Oxford University, 1955,
                  unpublished; P. Axel,  {\it Phys. Rev.} {\bf 126} (1962)
                  671.
\bibitem{car} B. V. Carlson et al.,  {\it Ann. Phys.} (NY) {\bf 276} (1999)
                 111; {\it  Phys. Rev.} {\bf C60} (1999) 014604.
\bibitem{wei} J. Z. Gu and H. A. Weidenm\"uller,  {\it Nucl. Phys.} {\bf
                  A690} (2001) 382.
\bibitem{cat87} F. Catara, Ph. Chomaz and A. Vitturi, {\it Nucl. Phys.}
{\bf A471} (1987) 661. 
\bibitem{lan2} E. G. Lanza, M. V. Andr\'es, F. Catara, Ph. Chomaz and
                 C. Volpe,  {\it Nucl. Phys.} {\bf A636} (1998) 452.
\bibitem{and} M. V. Andr\'es, F. Catara, E. G. Lanza, Ph. Chomaz,
  M. Fallot anf J. A. Scarpaci,  {\it Phys. Rev.} {\bf C65} 
(2001) 014608.
\bibitem{can} L. F. Canto, A. Romanelli, M. S. Hussein and A. F. R. de
                 Toledo Piza,  {\it Phys. Rev. Lett.} {\bf 72} (1994) 2147.
\bibitem{ber96} C. A. Bertulani, L. F. Canto, M. S. Hussein and A. F. R. de
                 Toledo Piza,  {\it Phys. Rev.} {\bf C53} (1996) 334.
\bibitem{ald} K. Alder and  A. Winther, {\it Electromagnetic Excitation}  
                  (North-Holland, Amsterdam, 1975).
\bibitem{win} R. A. Broglia and A. Winther, {\it Heavy Ion Reactions} 
Vol.I (1981 The Benjamin/Cummings Publishing Company).
\bibitem{land} S. Landowne and A. Vitturi, in {\it Treatise on Heavy Ion
                 Science},  Ed. D. A. Bromley, vol. 1, p.355.

\bibitem{VOV} W. von Oertzen and A. Vitturi,  {\it Rep. Prog. Phys.} 
{\bf 64} (2001) 1247-1338  
\bibitem{BB} R.A. Broglia and D.R. Bes,  {\it Phys. Lett.} {\bf B69} (1997) 
129 \\
M.W Herzog, R.J. Liotta and T Vertse,  {\it Phys. Lett.} {\bf B165} (1985) 35.
\bibitem{WvO} G.M. Crawley et alt., {\it Phys. Rev.} {\bf C23} (1981) 589. \\
W. von Oertzen and al.,  {\it Z.Phys.} {\bf A313} (1983) 371-372.
\bibitem{BM} A. Bohr and B.R. Mottelson, {\it Nuclear Structure}, 
Vol. II, Benjamin, Reading (1975)\\
R.A. Broglia, O. Hansen and C. Riedel,  {\it Adv. Nucl. Phys.} {\bf 6} 
(1973) 287.
\bibitem{Dal} {\it Collective Aspects in Pair Transfer Phenomena}, 
ed. C.H. Dasso and A. Vitturi, Societ\`a Italiana di Fisica, Editrice 
Compositori, Bologna 1987, Vol. 18.
\bibitem{RPA} D.R. Bes and R.A. Broglia, {\it Phys. Rev.} {\bf C3} (1971) 2349.
\bibitem{MJ}  M.G.Mayer and J.H.D.Jensen, {\it Elementary Theory of
Nuclear Shell Structure}, Willey, New York, N.Y. (1955) 
\bibitem{Da} C.Dasso, private communication
\bibitem{AfF} J.Blomqvist and S.Wahlborn, {\it Arkiv f\"or Fysik}
 {\bf 16} (1960) 545.
\bibitem{US} R.A.Uhrer and R.A.Sorensen,  {\it Nucl Phys.} {\bf A86} (1966) 1.
\bibitem{DaLi} C.H.Dasso and R.J.Liotta,  {\it Phys.Rev. } {\bf C36} 
(1987) 448.
\bibitem{Sch} G.R. Satchler, Direct Nuclear Reactions, Clarendon press, Oxford
1983
\bibitem{DP} C.H.Dasso and G.Pollarolo, {\it Phys. Lett.} {\bf 155B} 
(1985) 223. 
\bibitem{DV} C.H.Dasso and A.Vitturi, {\it Phys. Rev. Lett.} {\bf 59} 
(1987) 634. 
\bibitem{BPW} R.A.Broglia, G.Pollarolo and A.Winther, {\it  Nucl Phys.} 
{\bf A361} (1981) 307
\bibitem{vOe2} W.von Oertzen in {\it Heavy Elements and Related Phenomena} 
Vol.I, ed. R.K.Gupta and W.Greiner (1999) World Scientific (Singapore).
\bibitem{COM} M.Rhoades-Brown, M.McFarlane and S.Pieper, {\it Phys. Rev.}
 {\bf C21} (1980) 2417;  {\it Phys. Rev.} {\bf C21} (1980) 2436.
\bibitem{Aky} R.A.Broglia and A.Winther, {\it Heavy Ion Reactions} Vol.I
(1981) The Benjamin/Cummings Publishing Company.
\bibitem{GLN}  S.Gustavson, I.L.Lamm, B.Nilsson and S.G.Nilsson,
 {\it Arkiv f\"or Fysik} {\bf 36} (1967) 613.
\bibitem{Lot} P.Lotti {\it et al.}, {\it Phys.Rev.} {\bf C40} (1989) 1791.
\bibitem{skin} T.Suzuki {\it et al.}, {\it Phys.Rev.Lett.} {\bf 75} (1995) 
3241;\\ I.Hamamoto {\it et al.}, {\it Phys.Rev.} {\bf C52} (1995) R2326;
{\it ibid.} {\bf C53} (1996) 765;{\it ibid.} {\bf C54} (1996) 2369; \\
S.Mizutori  {\it et al.}, {\it Phys.Rev.} {\bf C61} (2000) 044326;\\
H.Sagawa  {\it et al.}, {\it Phys.Rev.} {\bf C65} (2002) 064312;\\
S.Karataglidis  {\it et al.}, {\it Phys.Rev.} {\bf C65} (2002) 044306. 

\bibitem{VanI} P.Van Isacker, M.A.Nagarajan and D.D.Warner, {\it Phys.Rev}
 {\bf C 45} (1992) 45-49.
\bibitem{cat22} F.Catara {\it et al.}, {\it Nucl.Phys.} {\bf A624} (1997) 449;
 {\it Nucl.Phys.} {\bf A614} (1997) 86;
\bibitem{Lipp} E.Lipparini and S.Stringari,  {\it Phys.Rep.}
 {\bf 175} (1989), issues 3-4,  103-261.
\bibitem{Migd} A.Migdal,  {\it J.Phys.} (Moscow) {\bf 8} (1944), 331.
\bibitem{BF} B.L.Berman and S.C.Fultz,  {\it Rev.Mod.Phys.} {\bf 47}, 
(1975), 713.
\bibitem{MS} W.D.Myers and W.J.Swiatecki,  {\it Ann.Phys. }(N.Y.) {\bf 55}, 
(1969), 395.
\bibitem{Jone} G.A.Jones, {\it Rep. Prog. Phys.} {\bf 33} (1970), 645-689
\bibitem{Grei} W.Greiner and J.A.Maruhn, {\it Nuclear Models}, 
Springer (1996). 
\bibitem{SteMon} Stefano Montagnani, degree thesis, University of Padova 
(2002) [in italian].


\bibitem{Das2} C.H.Dasso, S.M.Lenzi and A.Vitturi, {\it Nucl. Phys. }
{\bf A639} (1998), 635-653. 
\bibitem{Das1} C.H.Dasso, S.M.Lenzi and A.Vitturi, {\it Nucl. Phys. } 
{\bf A611} (1996), 124-138. 
\bibitem{Naga} M.A.Nagarajan, S.M.Lenzi, A.Vitturi, to be published.
\bibitem{TUNL} http://www.tunl.duke.edu/nucldata/
\bibitem{Wall} H.Walliser and T.Fliessbach, {\it Phys. Rev. }C{\bf 31} 
(1985), 2242.
\bibitem{Bohr} A.Bohr and B.Mottelson, {\it Nuclear Structure} W.A.Benjamin, 
Inc.(1969) New York,Amsterdam.
\bibitem{Till} D.R.Tilley et al. {\it Nucl. Phys. }A{\bf 708} (2002), 3-163.
\bibitem{Voel} H.G.Voelk and D.Fick, {\it Nucl. Phys. }A{\bf 530} (1991), 
475-489.
\bibitem{Toki} Y.Tokimoto et al., {\it Phys. Rev. }C{\bf 63} (2001), 035801.
\bibitem{Alha} Y.Alhassid, M.Gai and G.F.Bertsch, {\it Phys. Rev. Lett.}
{\bf 49} (1982), 1482.
\bibitem{Lang} H.J.Assenbaum, K.Langanke and A.Weiguny, {\it Phys. Rev. }C
{\bf 35} (1987), 755.
\bibitem{alfa} C.M.Perey and F.G.Perey, {\it At. Data Nucl. Data Tab.}{\bf 17}
(1976), 1. 
\bibitem{trit} R.P.Ward and P.R.Hayes, {\it At. Data Nucl. Data Tab.}{\bf 49}
(1991), 316. 
\bibitem{Brog} R.A.Broglia and A.Winther, Heavy ion reactions 
(Addison-Wesley, Redwood City,1991).
\bibitem{Mazz1} C.Signorini {\it et al.}, submitted.
\bibitem{Kelly} G.R.Kelly {\it et al.}, {\it Phys. Rev. } {\bf C63} (2000), 
024601.
\bibitem{Cata} F.Catara,C.H.Dasso and A.Vitturi,  {\it Nucl. Phys.} 
{\bf A602} (1996), 181.

\begin{center}
------------------------------------------------------------------------------
\end{center}

\bibitem{Io1} C.H.Dasso, L.Fortunato, E.G.Lanza and A.Vitturi,  {\it Nucl. 
Phys.} {\bf A724} (2003), issue 1-2, 85-98. 
\bibitem{Io2} L.Fortunato, H.M.Sofia, W.von Oertzen and A.Vitturi,  {\it 
Eur.J.Phys.} {\bf A 14}, (2002), issue 1,  37-42.
\bibitem{Io3} L.Fortunato,  {\it Phys. At. Nucl.} (Yadernaya Fizika) {\bf 66}
 (2003),issue 8, 1445-1449, and Procedings of HIP02, Dubna, Russia (2002) 
(World Scientific).
\bibitem{Io3b} L.Fortunato, in "Theoretical Nuclear Physics in Italy", 
Proceedings of the 9th Conference on Problems in Theoretical Nuclear Physics,
S.Boffi {\it et al.} ed., (2003) (World Scientific). 
\bibitem{Io4} L.Fortunato, S.Montagnani and A.Vitturi, in preparation.
\bibitem{Io5} L.Fortunato and A.Vitturi, in preparation.
\bibitem{Io6} L.Fortunato, Heavy Ion Physics ( {\it Heavy Ion Phys.}, 
{\it in press}) and Proceedings of 284th Heraeus seminar, Marburg, Gemany 
(2002) (World Scientific). 
URL: http://arxiv.org/abs/nucl-th/0209013
\bibitem{Io7} L.Fortunato and A.Vitturi, {\it Nucl. Phys.} {\bf A722} (2003) 
85c-91c and Proceedings ISPUN02 (Elsevier, 2003).
\bibitem{Io8} L.Fortunato and A.Vitturi, Proceedings CM2002.
\bibitem{Io9} C.Signorini, P.Scopel, M.Mazzocco, L.Fortunato, F.Soramel,
I.J.Thompson, A.Vitturi, M.Barbui, A.Brondi, M.Cinausero, D.Fabris, E.Fioretto,
G.La Rana, M.Lunardon, R.Moro, A.Ordine, G.F.Prete, V.Rizzi, L.Stroe, M.Trotta
and E.Vardaci, Eur.J.Phys. {\it accepted}

\end{thebibliography}
\end{document}